\documentclass[fleqn,usenatbib]{mnras}

\usepackage{newtxtext,newtxmath}
\usepackage[T1]{fontenc}
\usepackage{ae,aecompl}

\usepackage[final]{graphicx}	
\usepackage{amsmath}	
\usepackage{amssymb}	
\usepackage{ltxtable}
\usepackage{subcaption} 
\title[A survey of TiO$\lambda$567~nm absorption in solar-type stars]{A survey of TiO$\lambda$567~nm absorption in solar-type stars}

\author[Fatemeh Azizi and Mohammad Taghi Mirtorabi]{Fatemeh Azizi$^{1}$\thanks{E-mail:
f.azizi@pnu.ac.ir (F.A.)} and{ Mohammad Taghi Mirtorabi$^{2}$\thanks{E-mail:
torabi@alzahra.ac.ir (M.T.M.)}}\\
$^{1}$Department of Physics, Payame Noor University (PNU), P.O.BOX 19395-3697, Tehran, Iran\\
$^{2}$Department of Physics, Alzahra University, P.O. BOX 1993893973, Tehran, Iran}

\date{Accepted XXX. Received YYY; in original form ZZZ}

\pubyear{2015}

\begin{document}
\label{firstpage}
\pagerange{\pageref{firstpage}--\pageref{lastpage}}
\maketitle
%
%
%
%
\begin{abstract}
Molecular absorption bands are estimators of stellar activity and spot cycles on magnetically active stars. We have previously introduced a new color index that compare absorption strength of the titanium oxide ($TiO$) at 567 nm with nearby continuum. In this paper we implement this index to measure long-term activity variations and the statistical properties of the index in a sample of 302 solar-type stars from the HARPS planet search program. The results indicate a pattern of change in star's activity, covers a range of periods from 2 years up to 17 years.

\end{abstract}

\begin{keywords}
 Activity cycles -- Stars: Solar-type -- Stars: active -- Spectroscopic: visual band. 
\end{keywords}

\section{Introduction}
Solar-type stars are $F8V-K2V$ main-sequences with $B-V$ color index between $0.48 \leq B-V \leq 0.8$. They covers more than 10 percent of nearby stars. Their spectral manifestation in visual region are full of spectral lines including molecular absorption bands. Molecular absorption is a fair indicator of cool regions like spots on the surface of active stars \citep{wal}. They can be used as a proxy to stellar oscillations, granulation, magnetic activity, plage and activity cycles. 

Activity in solar-type stars is relatively old subject in astrophysics which has been addressed by many authors \citep{wil1, wil2, saar, santos, diaz, boisse, cin, Lovis, gomes}. In late 60's \citet{wil1,wil2} introduced an index, {\it S}, to measures the $CaII~H\&K$ lines emission.  \citet{diaz} have found $NaI~D1\&D2$ lines useful to investigate activity in atmosphere of late-type dwarf stars. He observed a series of $F6$ to $M5.5$ dwarfs and found that these two lines can be used as activity indicator in stars with high level of activity and exhibit the Balmer lines in emission. \citet{gomes} have computed  activity in a sample of late $M$ stars using indicators which was defined based on variability in a series of absorption lines including $CaII$, $NaI$ and $H_\alpha$. They found significant level of correlation between $CaII$ and $NaI$ absorption and confirm that  $NaI~D1\&D2$ lines can be used as an activity indicator in stars with low level of activity. While \citet{gomes} asserted that index manipulating $HeI$ variability is a poor indicator of stellar activity, \citet{saar} claimed,  they have found correlation between $HeI$ D3 and $CaII~H\&K$ variation.  

Solar-type spectra are also covered with molecular absorption. These are wide absorption bands which widely spread on the infra-red part of the spectrum. \citet{ber} showed that many molecular bands including Titanium Oxide ($TiO$) are indicators of solar magnetic fields concentrated in sunspots. This sensitivity can also be implemented as an indicator of stellar activity. Most of the previous investigations on molecular absorption bands was focused on near-infrared and longer wavelengths in coolest red giants and dwarf stars. Technical advancements in sensitivity and resolution of spectrographs make it possible to detect wide band absorption features made by temperature sensitive molecules like $TiO$ in visual regions.      

The near-infrared spectrum of cool late-type giant and dwarf stars are covered by wide and pervasive $TiO$ absorption bands. 
$TiO$ can survive in stellar atmosphere with surface temperature cooler than 4000 $K$. In hotter atmospheres $TiO$ absorption can be used as a proxy to detect cool features like star spot. Long-term observation of $TiO$ absorption in stellar spectrum can reveal stellar activity cycle. The idea that $TiO$ bands could be used to measure spot's characteristics was first stated by \citet{ram}. They observed the active star V711 Tau (HR1099) near its photometric minimum and found $TiO$ absorption in its spectrum.
They claim that the $TiO$ absorption observed in the photosphere of the K1 IV star must be produced in a cooler region like a spot.  Further observations by  \cite{nef,onel} revealed $TiO$ bands in spectra of other active stars and let them to estimate the spot's area and temperature. \citet{onell} have used absorption bands of titanium oxide to study dark and cool stellar spots on magnetically active stars. 

As an indicator of $TiO$ absorption in near-infrared, \citet{win} introduced a photometric system consist of three filters, which measure the $TiO\left ( \gamma ,0,0 \right )$ R-branch band head at 719 nm. He used this system to determine near-infrared magnitudes, near-infrared color index and spectral type of a sample of cool giant stars. He also recalibrate the color temperature relation of $M$ type sub classes with his $TiO$-index. By careful investigation on the continuum band passes in Wings system in high resolution spectra of cool stars, \cite{fat} have found that a  small shift on continuum band passes (filters B and C) to shorter wavelengths will make the continuum less contaminated with weak absorption of $VO$ bands and yields a better correlation between the index and surface temperature of the calibration stars \footnote{The Wing three color photometric system consists of three filters. Filter A (FWHM = 11 nm) at 719 nm appropriately set to measure the strongest absorption bands of $TiO$ in near infrared. Filter B (FWHM = 11 nm) with central wavelength at 754 nm  and filter C  (FWHM = 42 nm) with central wavelength of 1024 nm both are located in continuum region.}.

\citet{mir} used Wing photometric system to study the correlation between chromospheric activity and $TiO$ absorption strength in the variable star $\lambda$  Andromeda. They found an estimated activity cycle of about 4-14 years and an anti correlation between average $TiO$ absorption (representing coverage of spots) and  average visual brightness; similar to what is observed in the Sun. They also found that $TiO$ molecules can survive even when the star is brightest, which means that part of the active regions might be distributed evenly on the surface of the star with no observable manifestation on the visual light curve.

We previously introduced a new color index (B-index) to measure the $TiO$ absorption band strength centered at 567 nm \citep{bahar}. In that paper we showed that there is a clear correlation between the B-index and surface temperature of cool M giants where cooler atmospheres exhibits higher B-index. The index can be implemented as temperature indicator up to $4200~K$ where the TiO molecule breaks down and absorption bands disappear from the spectrum. Solar-type stars are too hot to have TiO molecules in their atmosphere but like the Sun if stellar activity is in close connection with appearance of cool spots in surface of the star then an active solar type star can exhibit significant B-index in spite of its hot atmosphere. In this work  we implement our visual B-index to search for stellar activity in a sample of solar-type stars. The sample has been observed by the  High Accuracy Radial Velocity Planet search Spectrograph (HARPS) at European Southern Observatory (ESO). This sample was already searched for stellar activity identification based on $CaII~H\&K$ lines by \citet{Lovis}.

\section[]{The visual Absorption of Titanium Oxide}

The B-index implements two narrow band filters. The first filter (D) is designed to measure $TiO$ absorption band centered at 567 nm with full width at half-maximum (FWHM) of 16 nm. The depth of this absorption band is compared with the nearby continuum with the second filter (E) at 610 nm with FWHM equal to 10 nm \citep{bahar}. The continuum filter was appropriately chosen to be free from molecular absorption. The central wavelengths and  bandpass at these filters are listed in Table~\ref{tbl:1}. By the way, the B-index is defined as 

\begin{table}
\caption{The B-index Visual Filters.} 
 \label{tbl:1}
 \begin{tabular}{@{}cccc@{}}
 \hline  
Filter     &  Region Measured     & Central  & Bandpass  \\
              &                &Wavelength &FWHM \\
    &        &(nm) &(nm) \\
\hline D & TiO$\lambda$567 nm band &567 &16 \\
E  &Continuum &610 &10 \\
\hline 
 \end{tabular}
 \end{table}
 
\begin{equation}
\label{eqn:1}
B-index=-2.5 ~log~ \frac{\int_{}^{}F_{D}(\lambda)~ S_{D}(\lambda)~d\lambda }{\int_{}^{}F_{E}(\lambda) ~S_{E}(\lambda)~d\lambda}
\end{equation}
where $F_D{(\lambda)}$ and $F_E{(\lambda)}$ are the integrated spectral flux in wavelength $\lambda$ and $S_D{(\lambda)}$ and $S_E{(\lambda)}$ are appropriate filter response function.

\section[]{The Stars Sample}
We used a sample of 302 stars in $FGK$ spectral type from the HARPS high-precision planet search program \citep{mayor}. This sample was already used by \citet{Lovis} to study magnetic activity cycles and its correlation with HARPS precise radial velocities in solar-type stars by using $CaII~H\&K$ activity index. They implement a volume limited selection of stars closer than 56 pc, brighter than $V=9.5$ mag and with a rotational velocity $vsini \leq 2~km.s^{-1}$ in the solar neighborhood. These stars are slowly rotating and non binary or multiple.  Thus, this sample represents an old single, solar-type stars. All stars have been observed by the 3.6 m telescope at European Southern observatory in {\it La Silla \/} using the HARPS  spectrograph (R$\simeq$115,000). The original sample used by Lovis had 304 stars, we discard two, because their spectra were noisy and very sparse in time. The spectra are accessible from the ESO archives \footnote{The HARPS webpage: http://www.eso.org./instruments/harps/}. 

We now implement an extended sample with more data which were taken during 2011 to 2014 as part of the HARPS program extension. We use our B-index to determine any signs of activation periodicity in the time series. All of stars in our sample at least have 4 measurements and a total time span of observations covers at least 750 days.  We divided the whole observation periods of each star to bins of 150 days then averaged all data points within each bin.

The fundamental stellar parameters ($vsini$, spectral type, $T_{eff}$, log $g$, and [Fe/H]) were derived by \citet{sousa} using the same  HARPS spectra that we used here. Absolute magnitudes and luminosities  derived from the Hipparcos catalogue.
Table ~\ref{tbl:03} lists the sample stars. The first three columns of table ~\ref{tbl:03}  show the star's name (HD name), the number of observations and their observation time span, respectively.

\section{HARPS B-index Properties}

\subsection{Distribution of the mean B-indices}
\begin{figure}
	\includegraphics[width=\columnwidth]{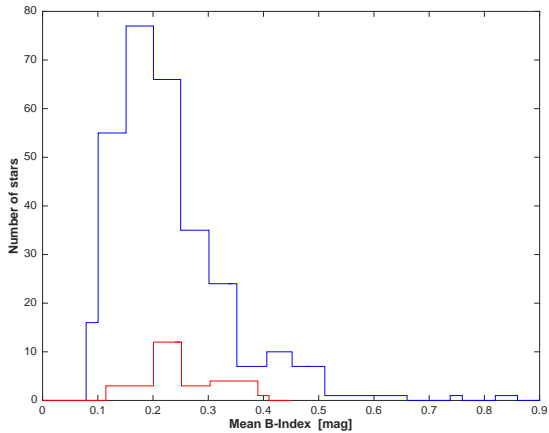}
 \caption{Distribution of the mean B-index values for 302 stars in the sample (blue). The same plot for 28 sub giants in the sample (red).}
\label{fig:02}
\end{figure}

Figure \ref{fig:02} shows distribution of mean B-index, calculated using equation~\ref{eqn:1}, for 302 stars in the sample. The fourth and fifth columns of Table \ref{tbl:03} shows the mean and standard deviation of B-index for all stars, respectively. The coolest star in the sample has a effective temperature of $4600 \  K$ which according to \cite{bahar} calibration is associated with a photospheric B-index of less than 0.05. Figure \ref{fig:02} clearly shows that all of the stars in the sample exhibit  high level of TiO absorption in the photospheres  which are too hot too sustain TiO molecules. \cite{mir} have associated this TiO excess to cool spots which could be assumed as a tracer of activity, similar to the sun.     The distribution displays  a large peak around 0.2 and a smaller one in the tail between 0.4 and 0.5. Most of the stars have B-index between 0.15 - 0.3 with an average of 0.22. According to calibration this level of TiO is associated with cool regions with $T \sim 4000\  K$ or $\sim 1200\  K$ cooler than average effective temperature of the sample. This is another evidence to confirm this TiO is coming from stellar spots. At the tail of the distribution there is a smaller peak which represents a smaller sample of stars with relatively higher level of activity with spots as cool as  $3800~K$. The low level of the second peak with  respect to the first one indicates that, as \citep{Lovis} asserts, the sample mostly contains native solar-type stars with less active ones. The lower (red)  distribution in the figure \ref{fig:02} shows a subsample of the stars which contain sub giants. Searching the SIMBAD we found 28 sub giants in the sample. The average value of B-index in subgiants are 0.23 which is higher than dwarf. This might be due to cooler atmosphere of these star which are more prone to sustain TiO molecules than dwarfs.

\subsection{Impact of metallicity on the mean B-index}
\begin{figure}
\includegraphics[width=\columnwidth]{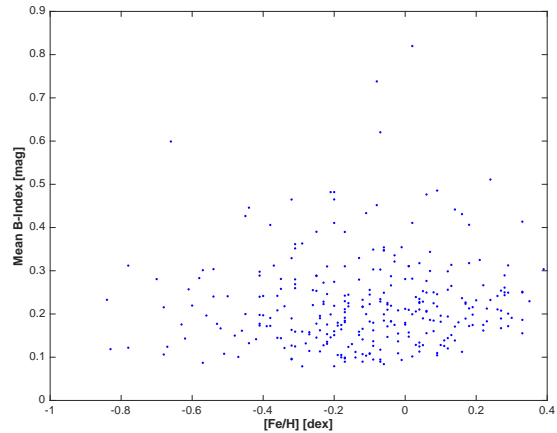}
 \caption{Mean B-index as a function of stellar metallicity.}
\label{fig:03}
\end{figure}

Figure \ref{fig:03} shows the mean B-index values as a function of stellar metallicity for all stars in the sample. The metallicity of these stars have been calculated by \citet{sousa} implementing the same  HARPS spectra that we used here. No evidence of correlation between mean B-index and $[Fe/H]$ can be recognized from figure \ref{fig:03}. This result is consistent with previous findings that TiO absorption strength is independent of metallicity \citep{white}.  This is also in agreement with results of \citet{Lovis}. 

\subsection{B-Index dispersion and noise}
\begin{figure}
\includegraphics[width=\columnwidth]{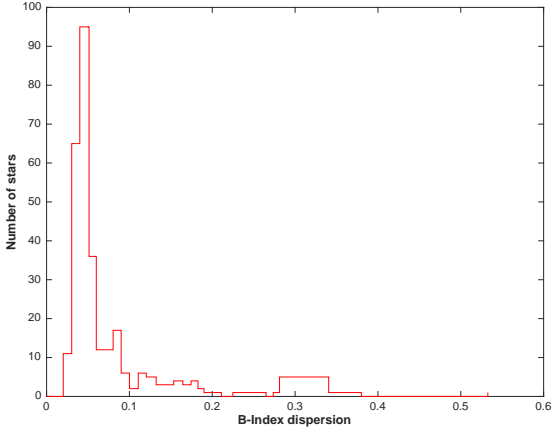}
\caption{Distribution of B-index dispersions for all stars in the sample.}
\label{fig:04}
\end{figure}

Figure \ref{fig:04} shows distribution of standard deviations (dispersion) of the B-index for 302 stars in the sample. Apart from large peak at 0.05 that might be produced by the observational noise, the distribution shows a smaller peak at about 0.3. Highly dispersed TiO absorption in long-term observation could be interpreted as periodic reappearance of cool, active regions in atmosphere of star. This small peak indicates that at least part of this sample might have experienced a period of stellar activity during last ten years. 

\begin{figure}
\includegraphics[width=\columnwidth]{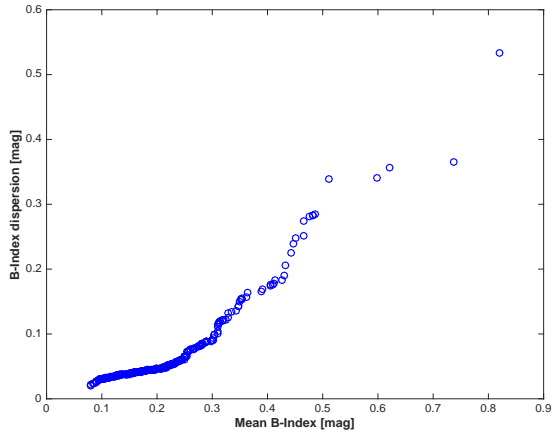}
 \caption{B-index dispersion as a function of mean B-index.}
\label{fig:33}
\end{figure}

To clarify that B-index has capability to resemble stellar activity we plot B-index dispersion versus mean B-index. Figure ~\ref{fig:33} shows that stars with higher B-index show more index variations. Higher B-index in an indication of existence of  active regions like spots which regularly occurs when star is active. These stars are prone to have activity cycles and must exhibit higher value for B-index dispersion.
B-index variability could also caused by rotation of cool regions around the star which regularly resembles short period variation in B-index. We eliminate this type of variability by avoiding very low periods (less than two years).

\section[]{Method and Analysis}
Indices based on $TiO$ absorption band have shown their potential to reveal active regions in surface of stars like $\lambda$ Andromeda \citep{mir}. The large data set in hand, provides us a unique opportunity to check this capability in a more reliable and statistically approved way. Our main goal is to search for significant long-term periodicities in the sample stars, according to their $TiO\lambda567$ nm variation.

\subsection{Periodicities}
We derived the periodicity on the basis of results from calculations of power spectra of the time series. The most commonly used method of calculating the spectrum of non-uniformly spaced data is periodogram analysis \citep{lom,sca}.
It is equivalent to least-squares fitting of sine and cosine  waves, in form of $y=a\cos \omega t+b\sin \omega t$. 
The normalized periodogram analysis yields the best sinusoid that fits the unevenly sampled data, and the significance of the period is inferred from it's false alarm probability ({\it FAP\/}) \citep{peter}.


In order to search for activity cycles, we computed generalized Lomb-Scargle periodogram ({\it GLS\/}) for each star. The {\it GLS\/} is an extension to the Lomb-Scargle periodogram which takes into account the measurement of errors and also is more suitable for time series with none zero average. {\it GLS\/} tries to fit the equation  $y=a\cos \omega t+b\sin \omega t+C$ to the time series and find the spectrum for frequencies \citep{zech}.

\begin{figure}
  \includegraphics[width=\columnwidth]{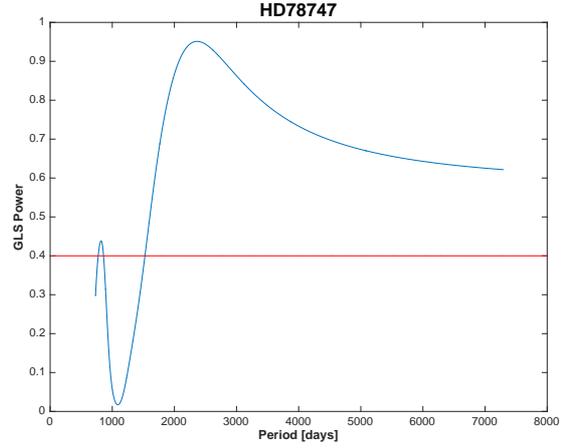}
 \caption{GLS periodogram of the B-index for HD78747, The red line denotes the 1$\%$ false-alarm probability. }
\label{fig:05}
\end{figure}

\begin{figure}
	\includegraphics[width=\columnwidth]{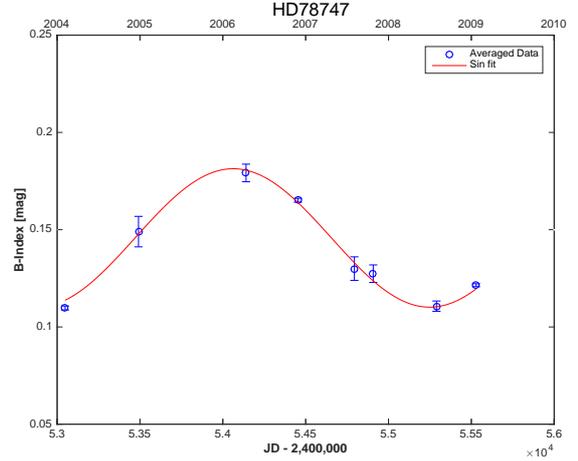} 
 \caption{Time series of the B-index with the fitted activity cycle for HD78747.}
\label{fig:06}
\end{figure}

We consider a given periodogram peak, derived from {\it GLS\/} significant when it exceed the  one present "false alarm probability" level  ({\it FAP\/}), which means  there is  99$\%$ confidence that the peak is real and could not be simulated by Gaussian noise. {\it FAP\/} levels are calculated by performing random permutations of the data with similar times of observations. To confirm that the period represented by the peak is appropriate, we rechecked the period by fitting a rescaled sin wave with that period to the data. Figure \ref{fig:05} and \ref{fig:06} show the periodogram and fitted sin wave for the star HD78747. Both figures indicates a significant activity cycle at a period of 2371 days.  

 To reduce the risk of inclusion of rotational variables from the final sample of active stars,   we select a lower limit for the calculated peaks in periodograms. Peaks with period less than 2 years were discarded and we focused  here on activity cycles in the 2 - 20 yr range. The 2 years time scale would be a relevant minimum for separating two variability regimes. It is more than 30 times larger than the rotational period of the sample and about five times smaller than activity period of the Sun. Main sequence stars with rotational periods higher than 2 years are rare and this minimum might appropriately disentangle any sign of rotational variability from final results. A time scale of 20 years was selected as an upper limit for activity because typical time span of observations were limited to $\sim$10 - 11 years.

Using Mont Carlo simulation we calculate confidence intervals for each measured period. A Gaussian distribution was used to generate different noise realization for each individual data points, with the same standard deviation taken from computed short-term scatter. By recomputing of {\it GLS\/} periodograms a confidence interval was calculated from distribution of derived periods.

\subsection{Result} 
The results of the B-index analysis for determination of activity cycle are given in table \ref{tbl:03} for all stars in the sample. 156 stars show significant activity cycle. The sixth and seventh columns of table \ref{tbl:03} tabulates the activity period and the cycle semi-amplitude calculated from B-index. For those stars that have shown cycles in both indices we have also added their $R' _{HK}$ period in the last column. 

\subsubsection{Activity cycles}
Among 302 FGK stars, we find 156 stars with a detected cycle. There are 146 stars with no significant periodic variation. In the active subsample there are 153 stars with a cycle semi-amplitude larger than 0.01, where we assigned them as large-amplitude cycles. This is more than $51\%$ of the sample. This is significantly greater than what \citet{Lovis} obtained by searching activity cycle in the variation of $CaII~H\&K$ emission. No clear correlations between $R' _{HK}$ index used by Lovis and B-index was found. $61 \%$  of Lovis active stars were also active in our B-index results. 

\subsubsection{Periods and amplitudes}
\begin{figure}
\includegraphics[width=\columnwidth]{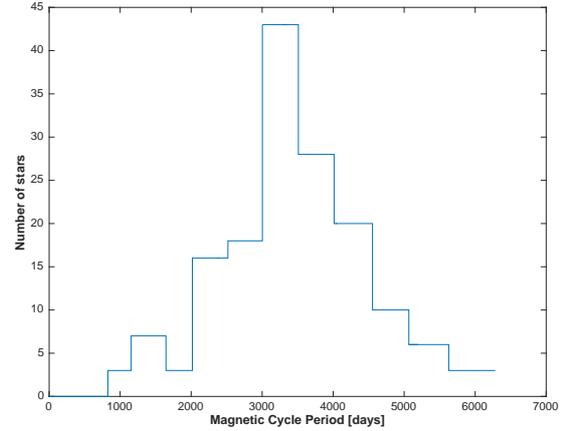}
 \caption{Distribution of activity cycle periods for 156 stars exhibiting long-term modulations in B-index.}
\label{fig:07}
\end{figure}

Figure  \ref{fig:07} shows the distributions of activity cycle periods. The histogram represents a maximum at 3000-4000 days and a smooth decrease for longer periods, unlike Lovis results that has steeper decrease. This might be due to that Lovis survey was not long enough to properly contain periods beyond $\sim$4000 days.  About 47$\%$ of the  stars in the sample show an activity period similar to the Sun. 

\begin{figure}
\includegraphics[width=\columnwidth]{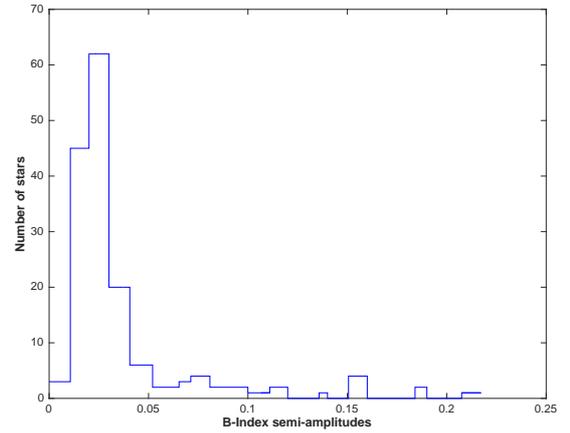}
 \caption{Distribution of activity cycle semi-amplitude for the 156 stars exhibiting  long-term modulations in B-index.}
\label{fig:08}
\end{figure}

The distribution of activity cycle semi-amplitude is shown in figure \ref{fig:08}. The sharp peak at $\sim0.03$ belongs to those stars with semi-amplitude smaller than 0.01. They are mostly cycles that are not strong  enough to result in a period with high confidence level or may be due to inactive stars or stars with very long activity cycle with no significant variation in the observation time span. 

\begin{figure}
 \includegraphics[width=\columnwidth]{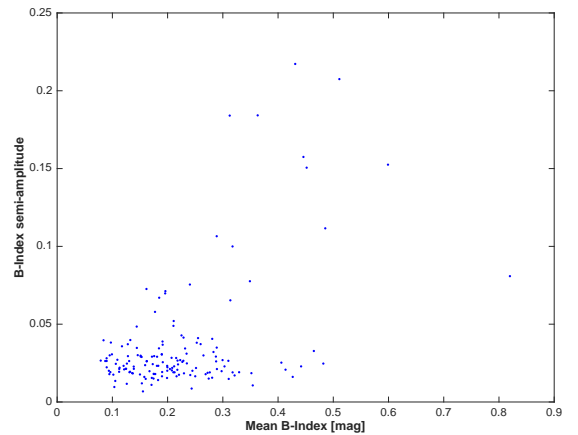}
 \caption{Activity cycle semi-amplitude as a function of mean B-index.}
\label{fig:09}
\end{figure}

Figure \ref{fig:09} shows plot of activity cycle's semi-amplitude as a function of mean activity level (mean B-index). Although most of the stars with lower activity level ($ 0.1 \leq$ B-index $\geq 0.25$) are concentrated in lower region with semi-amplitude less than 0.05, a few with higher B-index display a trend to have larger amplitude. The trend is disperse and not obvious. These are stars with activity period as short as they can illustrate a complete period in the course of whole observation like HD78747 shown in figure \ref{fig:06}.   

\section{Discussion and Conclusion}
The present survey has allowed us to measure activity cycles in a sample of solar-type stars by using molecular absorption bands.
We are using a temperature sensitive index which measuring strength of an absorption band of the TiO$\lambda$567~nm as a proxy to the activity cycle. This molecule is normally appear in cool and active regions of star surface.       

In this work we have performed a similar study like \citet{Lovis} in a sample of old solar-type stars. The data sample was taken from high-precision spectrograph, HARPS, \citep{mayor}. It spans 10 - 11 years of observation, compared to 7 - 8 years of the \citet{Lovis} survey. The longer time span allowed us to detect cycles with periods up to $\sim$17 yr. Among all, 51.7 $\%$ of stars show a meaningful activity cycle. Since the B-index is high when the star is highly active. The lack of activity in some stars may be due to that they are in their quiet stage.

The distribution of activity cycle periods show a peak between 8-11 years, and smoothly decrease in both side of  longer and shorter periods, compared to \citet{Lovis} who have calculated a sharper decrease towards longer periods. This sharp decrease may be due to insufficient number of stars with long observation time span to reveal a possible long-period cycles. Our results are roughly compatible with the results of \citet{bal} (see their figure 3). In their studies a few of the stars are showing activity cycles with periods greater than 15 years. This confirms that our lower time duration than \citet{bal} does not significantly affect the derived result for activity cycles.

No clear correlations between $R' _{HK}$ index used by Lovis and B-index was found. We have four stars (HD 15337; HD 20794; HD125881; HD215152) with quiet B-index and significantly variable $R'_{HK}$. It seems hard to justify why two indices give different cycles. The sample of spectra we got from HARPS normally are sparse. The spectra are in groups of few tens taken in few days and groups are few months apart.  Because few days are two short in our cycle investigation, we did average all the B-index inside a group. The error bars in the time series plots (figure \ref{fig:06})  are calculated from relative dispersion of B-indices in a group. Appendix A shows that although the B-index varies significantly over the years which could be assumed as stellar activity but since averaged points are two sparse, a more continuous and evenly spread in time observation needed to find out whether these two indices are really correlated or not.     

Apart from the lack of correlation between results from two indices, this work illustrates that molecular bands can also be used as a proxy to stellar activity and reveals the long period variation of stellar characteristics.

\section*{Acknowledgments}

This research is based on observations collected with the {\it HARPS \/}  spectrograph on the 3.6-m telescope at {\it La Silla \/} Observatory,  European Southern Observatory, Chile, Science Archive Facility. We thank the anonymous referee for the helpful comments and suggestions. M.T.M. is financially supported by the Alzahra university office of research and technology. FA is financially supported by the Payame Noor University.

\appendix

%


\onecolumn
\begin{longtable}{llllllll}
\caption{The Results} \label{tbl:03} \\

\hline
   $Star$ & $No.$  & $Observed$ & \multicolumn{2}{c}{$B-Index$} &$Period$& $Semi-$ & $Period $\\ 
  $HD\ name$& $Obs.$  & $time (days)$ &$ mean$  & $\sigma$  & $(days)$ & $amplitude$ & $R'_{HK}$ $(days)$  \\

    \hline  
\endfirsthead

\multicolumn{8}{l}%
{\tablename\ \thetable{} -- continued from previous page} \\
\hline
$Star$ & $No.$  & $Observed$ & \multicolumn{2}{c}{$B-Index$} &$Period$&$Semi-$& $Period ~$ \\  
 $HD\ name$& $Obs.$  & $time (days)$ &$ mean$ & $\sigma$  & $(days)$ &$amplitude$ & $R'_{HK}$$(days)$    \\
    \hline 
\endhead

\hline \multicolumn{8}{l}{{Continued on next page}} \\ 
\endfoot

\endlastfoot
55    & 60    & 3647  & 0.599& 0.249 &$2730_{-208}^{+208}$&0.152 \\
361&58&3685&0.0897& 0.0436 &$1729_{-12}^{+10}$&0.0223\\
967&64& 2556& 0.216& 0.100& $2624_{-386}^{+382}$&0.0283\\
1320  & 20    & 3230  & 0.115& 0.0455   \\
1388&59&3398&0.0936&0.0372&$2032_{-1237}^{+1237}$&0.0201\\
1461&190&3688& 0.174&0.0821& $2317_{-295}^{+295}$&0.0151 &$3754_{-565}^{+807}$ \\
1581& 118   & 3242  & 0.175& 0.103 \\
3569&121&3677&0.0951&0.0366& $3897_{-1022}^{+1022}$ &0.0300\\
3823  & 69    & 2226  & 0.132& 0.0422\\
4307&93&3676&0.223& 0.177& $2948_{-2091}^{+2089}$&0.159\\
4308&182&3358&0.218&0.0336& $5066_{-500}^{+448}$&0.0266\\
4915&44&1770& 0.199& 0.118&$2140_{-750}^{+850}$&0.0154 &$1863_{-211}^{+1951}$\\
6348  & 21    & 3328  & 0.196& 0.0291&\\
6673  & 20    & 1142  & 0.255& 0.122\\
6735&31&3654&0.0840&0.0495&$2927_{-120}^{+120}$ &0.0397\\ 
7134&22&3663&0.0789&0.0439& $4125_{-136}^{+159}$&0.0266\\
7199&116&3639&0.289&0.0490&$5873_{-6}^{+2}$&0.0212 &$2760_{-319}^{+416}$ \\
7449& 108 &3346& 0.104& 0.0347& $3060_{-383}^{+394}$&0.00963 \\
8638& 50    & 3293  & 0.191& 0.0830&   $2399_{-362}^{+336}$&0.0389 \\
8828&55& 3704& 0.218&0.0666&$1365_{-4}^{+0.3}$&0.0207&$2163_{-234}^{+300}$ \\
9246&49&3328& 0.177& 0.0766&$3296_{-2324}^{+2324}$&0.0579 &$2747_{-387}^{+540}$ \\
9782  & 30    & 2217  & 0.104& 0.0311\\
9796  & 22    & 2855  & 0.253& 0.142\\
10002& 30    & 3331  & 0.189& 0.0727&  $2487_{-1238}^{+1268}$&0.0307\\
10180& 192&3387&0.160& 0.0384&$4312_{-587}^{+646}$&0.0149&$2737_{-318}^{+414}$ \\
10700 & 192   & 2560  & 0.167& 0.0481 \\
11226 & 44    & 1766  & 0.134  & 0.0399\\
11505& 24& 3325  & 0.131  & 0.0423&  $4033_{-1583}^{+1583}$&0.0191 \\
11964 & 159   & 3677  & 0.250 & 0.0534 &$5403_{-475}^{+414}$&0.0165\\
12345 & 43    & 3574  & 0.146  & 0.0557 & $4115_{-867}^{+842}$&0.0301\\
12387 & 95    & 3246  & 0.196& 0.0844 &$2357_{-83}^{+87}$ &0.0855\\
12617 & 94    & 3246  & 0.196& 0.0847 & $2658_{-4}^{+13}$&0.0712\\
13060 & 92    & 3689  & 0.282  & 0.0443 &\\
13724 & 36    & 3621  & 0.183  & 0.0451 \\
13808&190&3672&0.274&0.0433& $5897_{-832}^{+1130}$&0.0150&$3715_{-562}^{+807}$ \\
14374 & 19    & 2089  & 0.219 & 0.0579 \\
14747 & 20    & 2379  & 0.172& 0.0452  &$2739_{-246}^{+243}$&0.0291\\
15337& 46    & 3643  & 0.250 & 0.0381 \\    
16297 & 60    & 3651  & 0.354& 0.0845 \\
16417 & 184   & 3729  & 0.151  & 0.0411&   $6282_{-4419}^{+4369}$&0.0292 \\
16714 & 185   & 3729  & 0.151 & 0.0411&   $6003_{-936}^{+867}$&0.0297\\
17970 & 35    & 3690  & 0.200  & 0.0596 \\
19034 & 30    & 3660  & 0.150 &  0.0450&  $3494_{-83}^{+102}$&0.0177\\
19467 & 51    & 3774  & 0.112  & 0.0467& $3736_{-940}^{+940}$&0.0194 \\
20003&184&3695&0.210&0.0319&$4854_{-110}^{+110}$&0.0191 &$3149_{-414}^{+562}$ \\
20407 & 65    & 3693  & 0.446& 0.366& $3364_{-222}^{+222}$&0.212\\
20619&36& 2138 &0.153& 0.0317& $2043_{-4}^{+9}$&0.0120 &$1687_{-151}^{+184}$ \\
20781&167& 3734&0.200&0.0527&$4136_{-11}^{+32}$&0.0223 &$8000_{-4249}^{+\infty}$ \\
20782&165&3734& 0.199& 0.0529&$4033_{-10}^{+30}$&0.0232 &$1150_{-58}^{+65}$ \\
20794 & 168& 763   & 0.242 & 0.126 & \\
20807&150& 3692& 0.228&0.183&$2413_{-3}^{+11}$&0.0265&$1133_{-65}^{+1090}$ \\
21019 & 47    & 3752  & 0.173 & 0.0342& $2865_{-153}^{+153}$&0.0161\\
21209 & 44    & 3259  & 0.426 & 0.122 \\
21693&190&3412&0.212&0.0325&$3842_{-2214}^{+2214}$&0.0229 &$2483_{-256}^{+322}$ \\
21938 & 22    & 3341  & 0.101  & 0.0368 \\
22879 & 100   & 3778  & 0.119  & 0.0757 \\
23249 & 177   & 3777  & 0.314 & 0.120& $3974_{-9}^{+30}$ &0.0653\\
23356 & 33    & 2228  & 0.310 & 0.0308 \\
23456 & 28    & 3429  & 0.0963 & 0.0373& $4033_{-268}^{+258}$&0.0192\\
24331 & 11    & 2700  & 0.280  & 0.0512 \\
24892 & 27    & 3428  & 0.163  & 0.0322 \\
25673 & 23    & 3427  & 0.241  & 0.0357&  $3179_{-413}^{+431}$&0.0173\\
26965&136&3368&0.269& 0.177&$3282_{-7}^{+20}$&0.0179 &$3352_{-543}^{+805}$ \\
27063&45& 1576 & 0.190& 0.0387& $1352_{ 155}^{160 }$ &0.0165 &$1316_{-124}^{+153}$ \\
28471 & 30    & 3378  & 0.210 & 0.0861 \\
28701 & 23    & 3223  & 0.190  & 0.0403 \\
28821 & 26    & 3665  & 0.141  & 0.0406 \\
30278 & 29    & 1514  & 0.242 &  0.0450 \\
30306 & 22    & 2905  & 0.224  & 0.0571 \\
31527 & 27    & 3612  & 0.211 & 0.0931&  $2658_{-493}^{+500}$&0.0433 \\
31822 & 192   & 3612  & 0.105  & 0.0376 \\
32724 & 33 &3673& 0.0973& 0.0455&   $3465_{-15}^{+15}$ &0.0382\\
33725 & 25    & 3671  & 0.184 & 0.0443&    $3759_{-567}^{+545}$&0.0397 \\
34449 & 61    & 3791  & 0.0946& 0.0362 \\
34688 &22 & 3383& 0.219& 0.0418& $3974_{-9}^{+29}$&0.0242 &$2026_{-172}^{+208}$ \\
35854 & 42    & 3816  & 0.279  & 0.0410 \\
36003 & 196   & 3411  & 0.482 & 0.0303 \\
36108 & 25 &3467  & 0.0720 & 0.0411 & $3480_{-872}^{+872}$ &0.0247\\
36379 & 95    & 3810  & 0.0896  & 0.0422&$4312_{-10}^{+36}$&0.0281\\
37986 & 20    & 3340  & 0.211 & 0.0471&  $3603_{-41}^{+41}$ &0.0489 \\
38277 & 20    & 3386  & 0.117 & 0.0430&    $2582_{-536}^{+573}$&0.0358 \\
38858&176&3763&0.136&0.0435&  $4580_{-14}^{+38}$&0.0186 &$3406_{-428}^{+570}$ \\
38973 & 30    & 3345  & 0.128  & 0.0476& $3702_{-372}^{+372}$&0.0353 \\
39194 & 178   & 2915  & 0.257& 0.135 \\
40105 & 52    & 3675  & 0.276 & 0.0433 & $3350_{-360}^{+356}$ &0.0190\\
40307 & 170   & 3010  & 0.362& 0.0325 \\
40397 & 40    & 3824  & 0.159 & 0.0326\\
44120 & 43    & 3707  & 0.120  & 0.0390&  $3130_{-387}^{+404}$&0.0232 \\
44420 & 22    & 3375  & 0.172& 0.0460 \\
44447 & 60    & 3676  & 0.246 & 0.190 \\
44573 & 73    &       & 0.620  & 0.252 \\
44594 & 40    & 3290  & 0.181  & 0.0417& $2495_{-12}^{+28}$&0.0232\\
45184 & 166&   3772  &0.125  &0.0514 &$3753_{-1173}^{+1233}$&0.0213 &$1595_{-108}^{+125}$ \\
45289 & 92    & 3771  & 0.109  & 0.0459& $3788_{-173}^{+195}$&0.0271\\
45364&85&3636&0.191&0.0579& $4825_{-13}^{+44}$&0.0256 &$8000_{-6958}^{+\infty}$ \\
47186 & 135   & 3761  & 0.213 & 0.116& $4740_{-369}^{+408}$&0.0191 \\
48611 & 17    & 1225  & 0.242 & 0.0589 \\
50590 & 20    & 3074  & 0.310 & 0.0482  & $2959_{-1084}^{+1075}$&0.0264\\
50806 & 74    & 1819  & 0.206 & 0.0410  \\
51608&172& 3715& 0.229& 0.116&$3480_{-727}^{+708}$&0.0414 &$2481_{-256}^{+322}$ \\
 55693&38&3434& 0.312&0.282&  $3106_{-18}^{+43}$&0.184 &$2403_{-218}^{+266}$ \\
52919 & 4     & 1021  & 0.390 & 0.0286 \\
59468 & 190   & 3756  & 0.239 & 0.0440\\
59711 & 47    & 3771  & 0.107 & 0.083 \\
63765&53&2303&0.225 & 0.0578& $2284_{-864}^{+859}$&0.0172 &$2219_{-1051}^{+304}$ \\
65277&27&3716&0.352&0.0360& $3652_{-802}^{+802}$&0.0185 &$3791_{-565}^{+2048}$ \\
65562 & 46    & 3629  & 0.223 & 0.0579& $3378_{-124}^{+137}$&0.0269 \\
65907&61&3466&0.125&0.0404&$4244_{-316}^{+301}$&0.0188&$8000_{-2216}^{+\infty}$ \\
66221 & 50    & 3629  & 0.227  & 0.0602 &$3771_{-100}^{+119}$&0.0258\\
67458 & 48    & 3463  & 0.124 & 0.0534 \\
68607&50&1210&0.299& 0.0462&$899_{-80}^{+78}$&0.0269 &$863_{-130}^{+\infty}$ \\
68978&126&3759&0.113& 0.0381&   $3571_{-8}^{+23}$&0.0221 &$1021_{-310}^{+58}$ \\
69655 & 51    & 3465  & 0.165 & 0.239 \\
69830&193& 3670  & 0.249& 0.0731\\      
70889 & 20    & 1580  & 0.139 & 0.0282 &$1232_{-439}^{+446}$&0.0213\\
71334 & 42    & 3724  & 0.126 & 0.0450&$3619_{-43}^{+56}$&0.0247\\
71479&73&3387&0.511& 0.341 & $2002_{-4}^{+8}$ &0.0475&$1525_{-114}^{+1993}$ \\
71835&78& 3731&0.183&0.0425&$3685_{-17}^{+17}$&0.0232 &$3222_{-418}^{+565}$ \\
72579 & 39    & 2975  & 0.232 & 0.117 &$2843_{-59}^{+75}$&0.0214\\
72673 & 162   & 3001  & 0.240 & 0.0437 \\
72769 & 32    & 3385  & 0.191  & 0.0664 & $4942_{-1167}^{+1145}$&0.0368\\73121 & 136   & 3420  & 0.110  & 0.0525 \\
73524&159&3420& 0.112&0.0541 & $3217_{-31}^{+57}$&0.0212 &$2801_{-322}^{+984}$ \\
74014 & 25    & 3069  & 0.175  & 0.0419&$4685_{-452}^{+428}$&0.0181\\
78429 & 68    & 3790  & 0.246 & 0.182 \\
78558 & 24    & 3810  & 0.133  & 0.0660 &$3350_{-107}^{+107}$&0.0398\\
78612 & 22    & 3071  & 0.113 & 0.0515 \\
78747&53& 2554&0.124&0.0344&$2371_{ -147 }^{ +147 }$ &0.0213 &$2389_{-250}^{+316}$ \\
81639 & 24    & 3109  & 0.181 & 0.0481&     $2767_{-224}^{+213}$&0.0291 \\
82342 &52& 3720&0.240& 0.136&$3256_{-80}^{+80}$ & 0.0755 &$3434_{-428}^{+571}$  \\
82516&71&3719&0.311&0.0466&$2703_{-612}^{+634}$&0.0148&$6065_{-4074}^{+\infty}$ \\
83529 & 28    & 3685  & 0.176 & 0.165&  $2855_{ 1464 }^{  1463 }$&0.0295\\
85390&8&3682&0.234 & 0.0879&$2835_{-140}^{+163}$ &0.0311 &$2476_{-255}^{+322}$ \\
85512&105&3467&0.465&0.0482& $3130_{-248}^{+248}$&0.0328 &$3793_{-566}^{+806}$ \\
86065 & 28    & 3110  & 0.289  & 0.0473\\
86140&43&3098&0.347&0.0995&$3003_{-5}^{+17}$&0.0350 &$2798_{-1427}^{+418}$ \\
86171 & 121   & 1960  & 0.390 & 0.0571\\
88084 & 24    & 3071  & 0.109  & 0.0447 \\
88218 & 61    & 3813  & 0.0994 & 0.0355 \\
88742 & 49    & 3435  & 0.155  & 0.0325 \\
89454 & 59    & 2218  & 0.299& 0.225\\
90156&152&3434&0.162&0.0725&$4013_{-107}^{+69}$&0.0263 &$1279_{-72}^{+81}$ \\
90711& 27    & 3339  & 0.233 & 0.0445 \\
90812&29&3339&0.208&0.0465&$3510_{-12}^{+18}$&0.0219&$8000_{-2419}^{+\infty}$ \\
92588 & 120   & 2343  & 0.235  & 0.0754 & $2017_{-3}^{+7}$&0.0248\\
92719 & 67    &       & 0.108  & 0.0451 \\
93083&106&3434&0.486&0.281&$3378_{-137}^{+114}$ &0.112&$3453_{-550}^{+806}$ \\
93385 & 177   & 3788  & 0.102 & 0.0441& $3619_{-828}^{+842}$&0.0176\\
94151&55&3385&0.318&0.166& $4538_{-9}^{+26}$& 0.999 &$3792_{-566}^{+806}$ \\
95456 & 166   & 3748  & 0.431  & 0.356 & $3510_{-5}^{+25}$ &0.199\\
95521 & 24    & 3383  & 0.130 & 0.0410&$4554_{-1487}^{+1478}$&0.0232 \\  
96423 & 69    & 3722  & 0.128  & 0.0526 & $3243_{-3}^{+21}$&0.0295\\
96700&167&3696&0.105&0.0458&$3083_{-1020}^{+1065}$&0.0135 &$830_{-27}^{+\infty}$ \\
97037& 87 & 3693  & 0.0897& 0.0331 &$4768_{-2451}^{+2451}$ &0.0263\\ 
97343 & 77    & 4160  & 0.185 & 0.111&    $5403_{-18}^{+55}$&0.0670 \\
97998 & 25    & 3384  & 0.142  & 0.0518 \\
98281&66&3687&0.221&0.0381&$1401_{-4}^{+9}$&0.0175&$2861_{-396}^{+1354}$ \\
100508 &32&2284&0.304&0.0422\\
101930&103&3436& 0.265&0.0654& $3230_{-349}^{+381}$&0.0299&$8000_{-2453}^{+\infty}$ \\
102117 & 119   & 3787  & 0.261  & 0.0637 \\
102438 & 87    & 3790  & 0.159& 0.0403& $4053_{-9}^{+32}$&0.0237 \\  
104006 & 91 & 4128  & 0.122 & 0.0992 & $4289_{-564}^{+532}$&0.0890\\
104067 & 88    & 2271  & 0.347 & 0.0413 \\
104263&32 & 2245  & 0.199 & 0.0600 & $2123_{-478}^{+495}$&0.0201 \\
104982 & 47  & 3043  & 0.173& 0.0659 &   $2487_{-851}^{+892}$&0.0243\\
105837 & 26    & 3049  & 0.108 & 0.0902 \\
106116&109&3675&0.171&0.0401&$3269_{-8}^{+18}$ &0.0110 &$3194_{-417}^{+564}$ \\
106275&27&3010&0.303&0.0587& $2317_{-368}^{+368}$&0.0229 &$3187_{-417}^{+564}$ \\
108309&49& 3333&0.144&0.0478& $5368_{-59}^{+141}$&0.0347 &$1552_{-574}^{+3005}$ \\
109200 & 137   & 2966  & 0.260  & 0.0373 \\
109409 & 63 & 3333  & 0.155 & 0.0444 & $3142_{-585}^{+600}$&0.00680\\
110619 & 36    & 3273  & 0.289 & 0.274 &    $3243_{-137}^{+137}$&0.0474\\
111031 & 40    & 3337  & 0.167 & 0.0351&   $3014_{-5}^{+17}$&0.0198 \\
112540 & 12    & 3275  & 0.230 & 0.0489 \\
114613 & 195   & 3560  & 0.216 & 0.0529 \\
114747 & 39    & 1473  & 0.325& 0.0361\\
114853 & 62    & 3783  & 0.128 & 0.0462 \\
115585 & 33    & 3015  & 0.230 & 0.0529\\
115617&164& 2918&0.137& 0.0382\\
115674&39&2252&0.204&0.0271&$3378_{-1321}^{+1318}$&0.0212 &$1947_{-588}^{+236}$ \\
116920 & 14    & 2305  & 0.272 & 0.0402 & $2063_{-195}^{+195}$ &0.0187\\
117105 & 46    & 3337  & 0.363 & 0.339&  $3060_{-136}^{+124}$&0.209\\
117207 & 63    & 3779  & 0.162  & 0.0730 \\
119638&27&3324&0.0945&0.0404&$2992_{-1233}^{+1233}$&0.0297 &$6000_{-2000}^{+\infty}$ \\
119782 & 19    & 2636  & 0.303 & 0.0369 \\
122862 & 50    & 3328  & 0.166 & 0.0520 \\
123265 & 35    & 4127  & 0.198  & 0.0749 \\
124292 & 31    & 3037  & 0.151 & 0.0383 \\
124364 & 18& 3031  & 0.153 & 0.0417&    $3897_{-1676}^{+1676}$ &0.0289\\
125072 & 44    & 3653  & 0.406& 0.0460\\
125184 & 151   & 3749  & 0.189 & 0.0444& $4033_{-325}^{+325}$&0.0305 \\
125455 & 132   & 3639  & 0.0997 & 0.0898\\
125881 & 196   & 3603  & 0.476& 0.282\\         
129642&64&1510&0.354&0.0283&$828_{-33}^{+34}$&0.0107 &$1419_{-149}^{+436}$ \\
126525 & 104   & 3675  & 0.182  & 0.0780 \\
128674 & 22    & 3028  & 0.172  & 0.0385\\
130930 & 21    & 2882  & 0.311& 0.0413\\
130992 & 23    & 3266  & 0.329& 0.0492&  $1922_{-70}^{+70}$&0.0191 \\
132648 & 40    & 3002  & 0.312  & 0.153\\
134060 &172& 3748  & 0.139 & 0.0312 & $3230_{-712}^{+712}$ &0.0228\\
134606 & 174   & 3602  & 0.208 & 0.0314 & $6282_{-507}^{+478}$&0.0203 \\
134664 & 33    & 1067  & 0.182  & 0.0319\\
134985 & 23    & 3030  & 0.220 & 0.0374 \\
136352&177&3495&0.121&0.0389& $3435_{-71}^{+77}$&0.0209 &$1041_{-97}^{+581}$ \\
136713&43& 2203  & 0.344& 0.0375\\
136894 & 50    & 2044  & 0.256  & 0.0806 \\
137303 & 44    & 2044  & 0.258 & 0.0803 \\
137388&91&2942&0.318&0.0458&$3974_{-3120}^{+3120}$&0.0190&$2770_{-389}^{+541}$ \\
138549&23&2942&0.176&0.0461&$3105_{-40}^{+33}$ &0.0180&$2180_{-205}^{+253}$ \\
140901 & 47    & 1429  & 0.217  & 0.0546 \\
142709 & 81    & 3329  & 0.281 & 0.132 \\
143114 & 21    & 3113  & 0.177  & 0.0583 \\
144411 & 16    & 2481  & 0.329 & 0.0433 \\
144585 & 168   & 2904  & 0.186 & 0.0759 \\
144628&47&2105&0.298&0.0334&$2697_{-736}^{+729}$&0.0193&$4103_{-804}^{+1321}$ \\
145598 & 44    & 2107  & 0.312  & 0.533 \\
145666 & 26    & 3970  & 0.266  & 0.0822 \\
145809 & 34    & 2906  & 0.145  & 0.0605 \\
146233 & 200   & 1147  & 0.228& 0.0232 \\
147512 & 54    & 1734  & 0.452  & 0.284&   $1722_{-2}^{+6}$&0.201 \\
148303&72&3215&0.321&0.0512& $3540_{-620}^{+602}$&0.0171&$2416_{-1274}^{+936}$ \\
150433 & 91    & 3204  & 0.145 & 0.0419 \\
151504 & 21    & 2261  & 0.205  & 0.0201 \\
154088&187&2701&0.244&0.0445&$4659_{-56}^{+80}$&0.00864&$8000_{-6487}^{+\infty}$ \\
154363 & 126   & 2999  & 0.143  & 0.0388 \\
154577&109&3390&0.281&0.0307&$3824_{-119}^{+139}$&0.0156 &$2702_{-316}^{+412}$ \\   
 157338 & 200   & 2936  & 0.820 & 0.177 \\
157172&186&2891&0.126&0.0855& $1354_{-40}^{+36}$ &0.0117&$8000_{-2581}^{+\infty}$ \\
157347 & 132   & 3566  & 0.738  & 0.153 &$4013_{-287}^{+307}$&0.0809\\
157830&67&3024&0.178&0.0415&$3435_{-252}^{+229}$&0.0180&$1613_{-119}^{+140}$ \\   
 161612 & 56    & 4021  & 0.237 & 0.174 \\
161098&116&2696&0.158&0.0376&$3480_{-53}^{+86}$&0.0161 &$8000_{-2791}^{+\infty}$ \\
162236 & 43    & 3496  & 0.233 & 0.0727 \\
162396 & 83& 3351  & 0.144  & 0.0643&$3954_{-1562}^{+1551}$&0.0435 \\
165920 & 55    & 3660  & 0.201  & 0.0886 \\
166724 & 71    & 3347  & 0.349  & 0.142 &$2641_{-31}^{+31}$&0.0981\\
168871 &125 & 3079  & 0.137  & 0.0352 &$3014_{-1314}^{+1314}$&0.0184\\
170493 & 31& 3237  & 0.442 & 0.0582 &  $4335_{-589}^{+572}$&0.0229\\
171665 & 20    & 2512  & 0.171 & 0.0366 &$3060_{-934}^{+929}$&0.0155\\
171990 & 73    & 1797  & 0.207 & 0.0528&  $1382_{-226}^{+223}$&0.0284\\
172513 & 44    & 1081  & 0.222& 0.0331 \\
174545 & 35    & 3081  & 0.266 & 0.0610 \\
176986 & 170& 3867  & 0.281 & 0.0887 &   $2845_{-758}^{+758}$&0.0406\\
177409 & 34 & 3221  & 0.183  & 0.0251 &$3897_{-2419}^{+2419}$&0.0141\\
177565&74&3641&0.185&0.0452 &$6003_{-2357}^{+2298}$&0.111 &$2016_{-262}^{+355}$ \\
177758 & 165   & 3867  & 0.283 & 0.0867& $3003_{-68}^{+68}$&0.0321 \\
180409& 22 & 3257& 0.0992  & 0.0456 &$3243_{-1485}^{+1483}$&0.0306\\
181433 & 190   & 3877  & 0.413  & 0.0516 &$4094_{-197}^{+197}$&0.0208\\
183658 & 54    & 3571  & 0.129 & 0.0540\\
183783 & 26    & 3678  & 0.411 & 0.0893 \\
185615 & 19    & 2947  & 0.225  & 0.0588 \\
187456 & 21    & 3639  & 0.411 & 0.0346 \\
188559 & 33    & 796   & 0.433  & 0.0340 \\
188748 & 20  & 2133  & 0.216 & 0.0270&   $2048_{-622}^{+617}$&0.0173 \\
189567 & 198   & 3640  & 0.190  & 0.0451 &$5789_{-22}^{+61}$&0.0193\\
189625 & 57    & 1758  & 0.172& 0.0325 \\
190248 & 193   & 2911  & 0.251  & 0.0440 &$3935_{-19}^{+19}$&0.0218\\
190954 & 24    & 3062  & 0.198& 0.0524 \\
192031 & 22    & 3376  & 0.233  & 0.206 \\
192310&196&3287&0.287&0.154&$3392_{-1786}^{+1781}$&0.0292&$3792_{-566}^{+806}$ \\
193193 & 52    & 2889  & 0.147  & 0.0420 \\
195564 & 63    & 3869  & 0.156  & 0.0360 \\
196761 & 68    & 3869  & 0.159  & 0.0598 \\
197210 & 19    & 2831  & 0.175  & 0.0492 \\
197823 & 23    & 1056  & 0.256& 0.0433 \\
199288 & 198   & 3963  & 0.176  & 0.0215 \\
199960 & 150   & 3294  & 0.250 & 0.169 \\
203384 & 29    & 2834  & 0.242  & 0.0900 \\
203432 & 73    & 1405  & 0.249& 0.0325\\
203850 & 22& 3281  & 0.106 & 0.0315 & $3435_{-256}^{+256}$&0.0279 \\
204313 & 158   & 3277  & 0.213& 0.122 \\
204385 & 21    & 2890  & 0.186 & 0.0402\\
204941 & 176   & 3462  & 0.180  & 0.0773 \\
205536 & 60    & 3260  & 0.226& 0.0432 \\
207129 & 186   & 1402  & 0.179  & 0.0243\\
207700 & 60    & 3260  & 0.226  & 0.0432 \\
209100 & 198   & 956   & 0.465  & 0.119 \\
209742 & 26    & 3610  & 0.244  & 0.0368 \\
210752 & 27& 3657  & 0.0871 & 0.0423 &$5628_{-331}^{+331}$&0.0263\\
210918 & 79    & 3889  & 0.180  & 0.0551 \\
211038 & 185   & 3887  & 0.288  & 0.0448&$4554_{-505}^{+468}$&0.0261\\
212708 & 53    & 3693  & 0.254  & 0.0430 \\
213628 & 29    & 4172  & 0.147  & 0.0402 \\
213941 & 29    & 2218  & 0.161  & 0.0459 \\
215152 & 193   & 3630  & 0.301 & 0.0483 \\
215456 & 196   & 3892  & 0.162 & 0.0478 &$4713_{-13}^{+42}$&0.0261 \\
219249 & 21    & 2901  & 0.197  & 0.0654 \\
220256 & 26    & 3680  & 0.245  & 0.0306& $3480_{-748}^{+792}$ &0.0201\\
220507 & 89    & 3601  & 0.157 & 0.0465 \\
221356 & 34    & 4221  & 0.0791  & 0.0369 \\
221420 & 66    & 3684  & 0.250 & 0.0308 \\
222237&37&3690&0.406&0.0360&$4244_{-191}^{+191}$&0.0253 &$2168_{-424}^{+2049}$ \\
222422 & 8     & 2522  & 0.217& 0.0989 \\
222595 & 33    & 1687  & 0.210 & 0.0366&  \\
222669 &62& 3656  & 0.229 & 0.0447 \\
223121 & 26 & 2252  & 0.253 & 0.0536 &$4136_{-2466}^{+2466}$&0.0381\\
223171 & 180   & 3893  & 0.194  & 0.0337 \\
224619 & 23    & 3280  & 0.160 & 0.0485\\
224789 & 35    & 2833  & 0.336  & 0.0482\\
hip22059 & 36    & 2584  & 0.127  & 0.150 \\
hip26542 & 5     & 1078  & 0.304 & 0.157 \\

 \hline

\end{longtable}




\newpage
\appendix

\section[]{Typical examples of GLS periodogram and time series for stars with a significant activity cycle. }
Here we present typical periodogram and time series for those stars that exhibit significant activity cycle. Column one shows the periodogram. The red horizontal line in the periodogram plot represents FAP level. Column two shows the time series for that star which fitted with a sinusoidal curve with period that maximize the periodogram. Column three and four is the same as one and two.

%

\newpage
\begin{figure}
        \begin{subfigure}[b]{0.25\textwidth}
                \includegraphics[width=\linewidth]{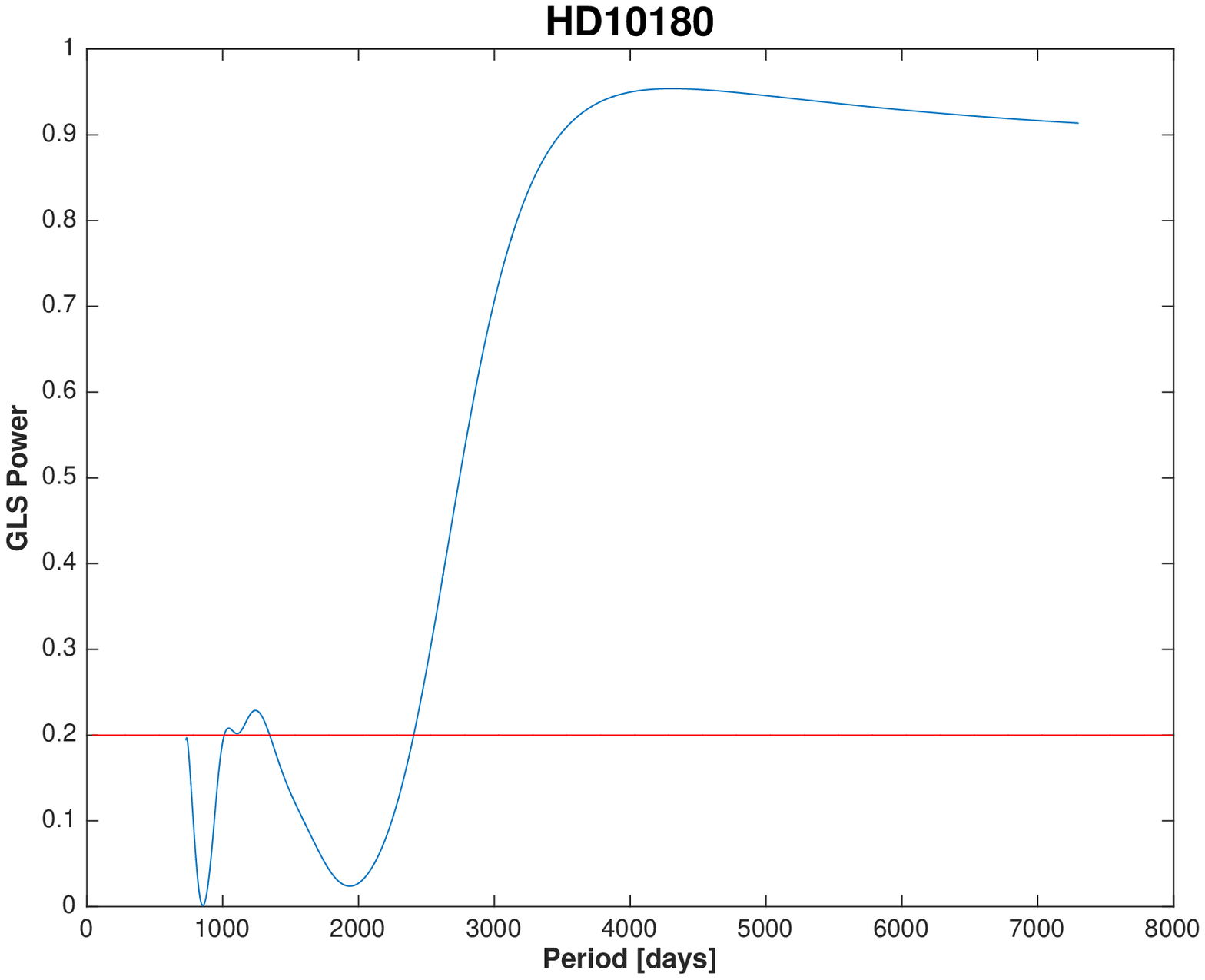}
        \end{subfigure}%
        \begin{subfigure}[b]{0.25\textwidth}
                \includegraphics[width=\linewidth]{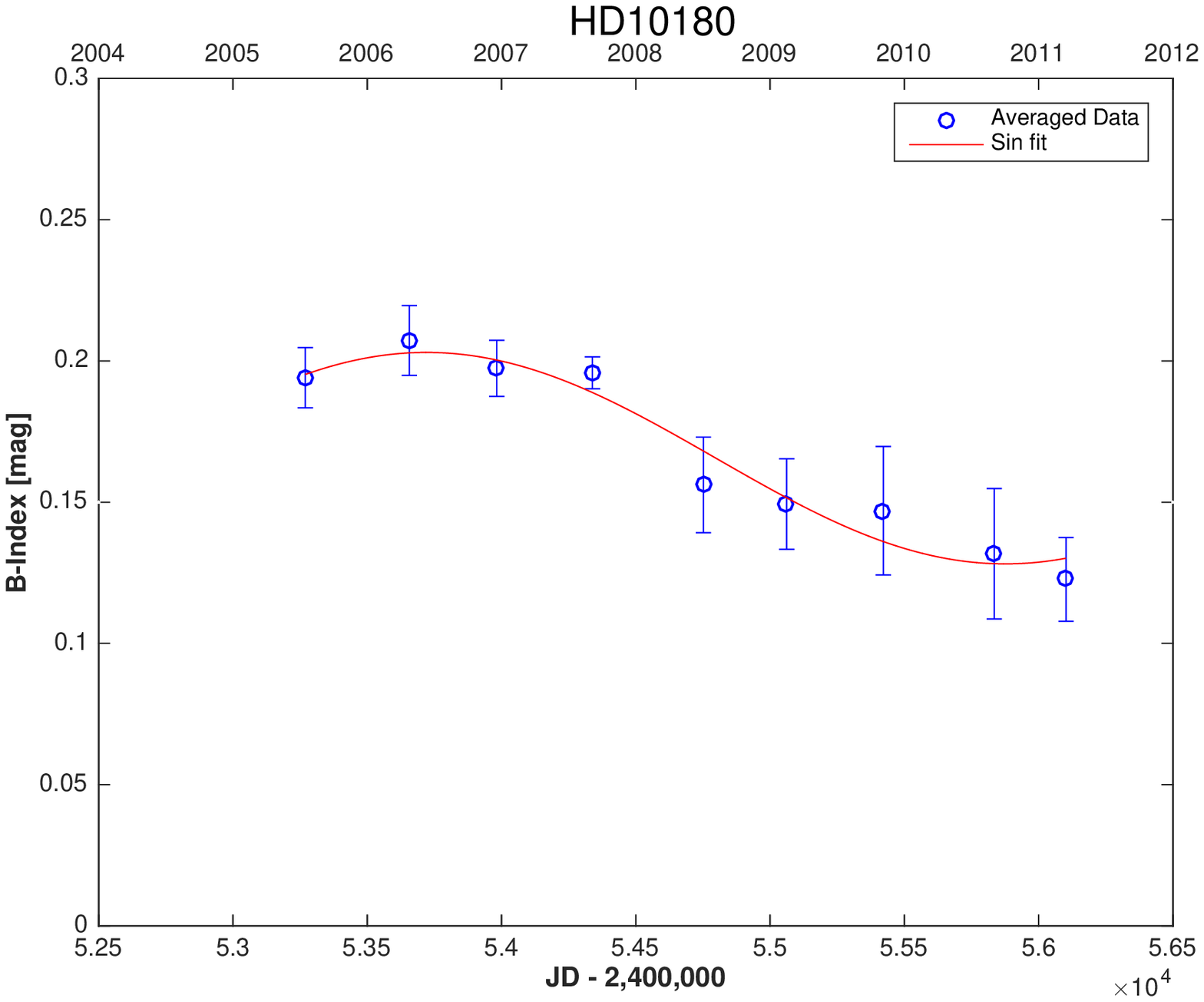}
        \end{subfigure}%
        \begin{subfigure}[b]{0.25\textwidth}
                \includegraphics[width=\linewidth]{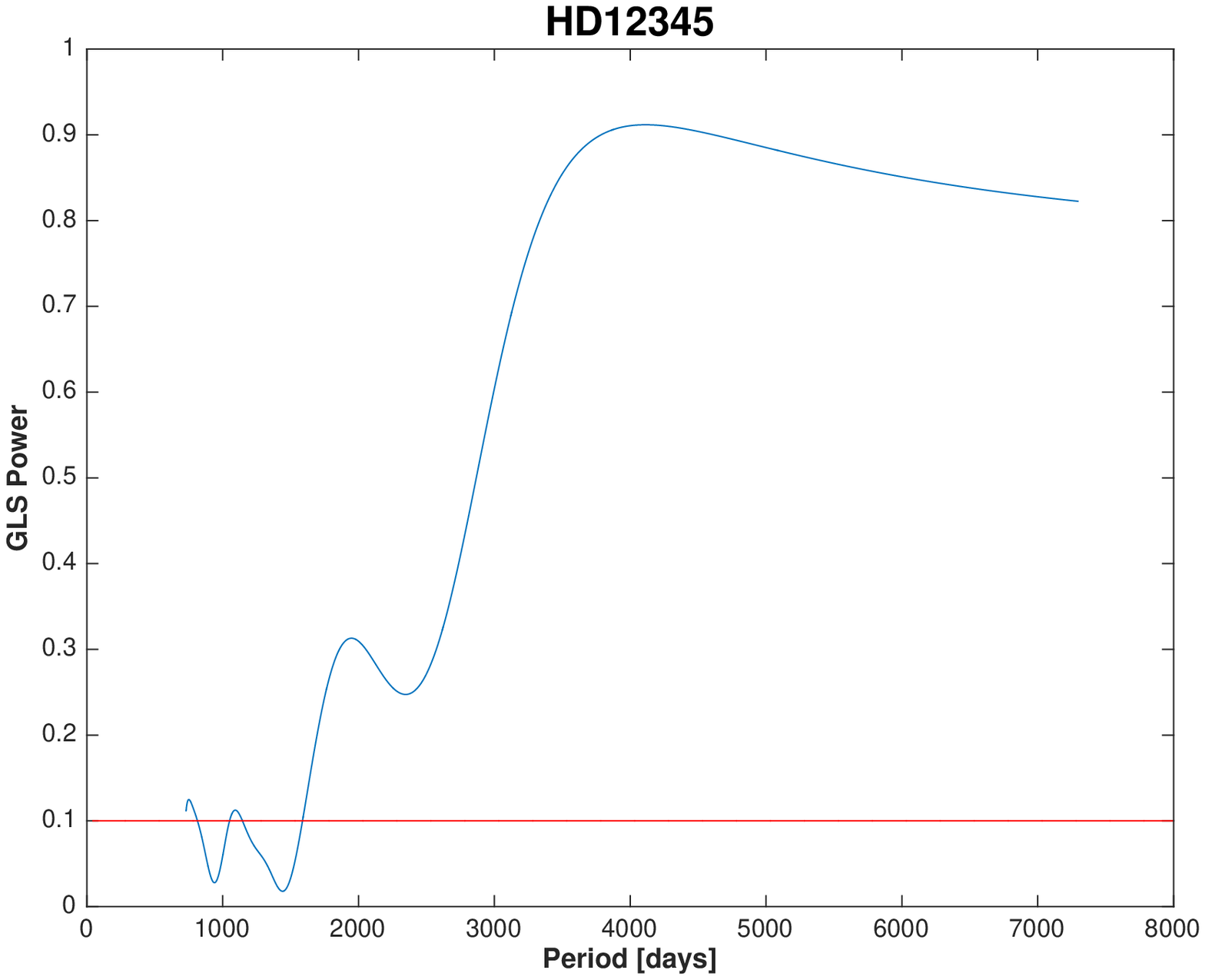}
        \end{subfigure}%
        \begin{subfigure}[b]{0.25\textwidth}
                \includegraphics[width=\linewidth]{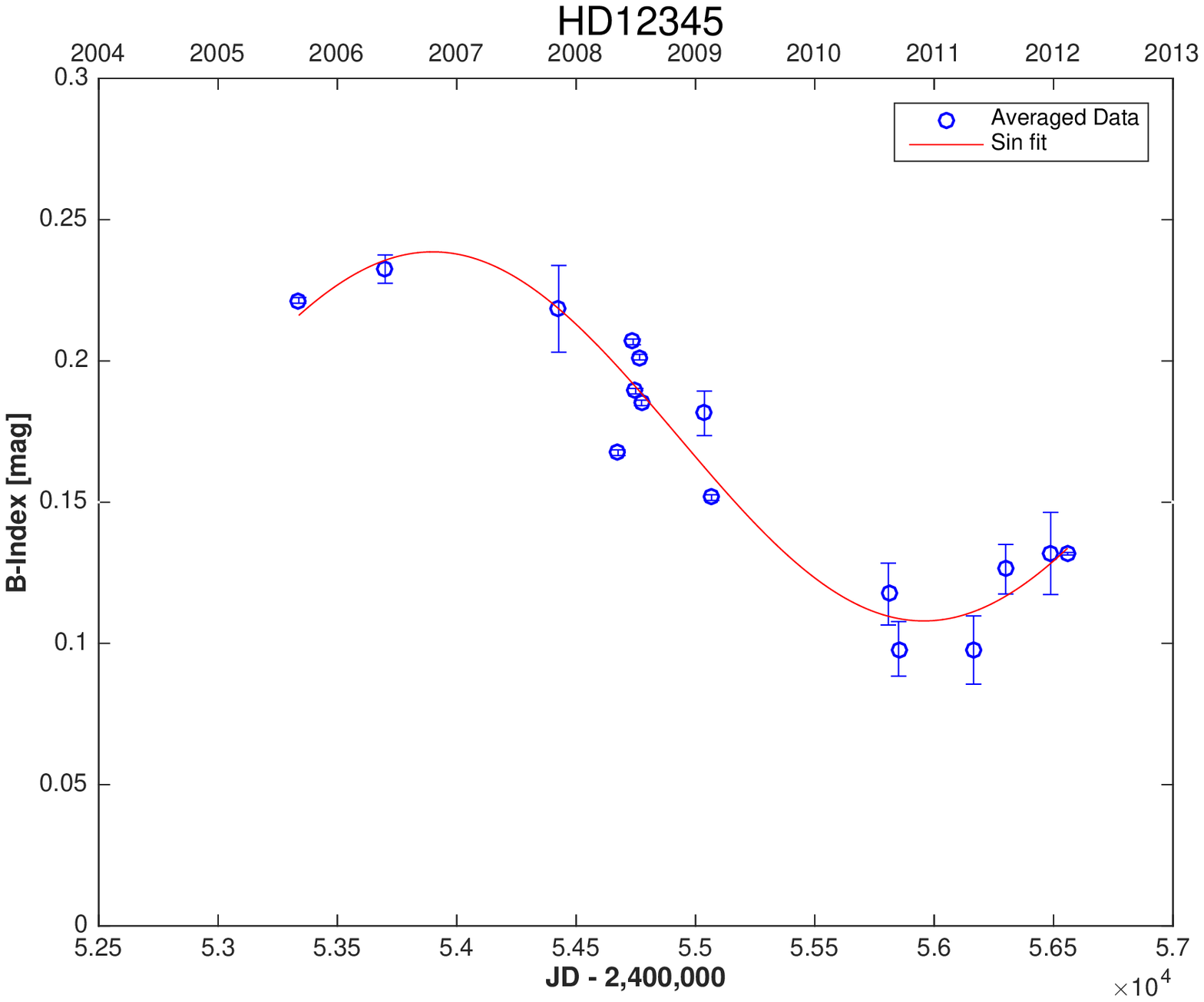}
        \end{subfigure}
        
         \begin{subfigure}[b]{0.25\textwidth}
                \includegraphics[width=\linewidth]{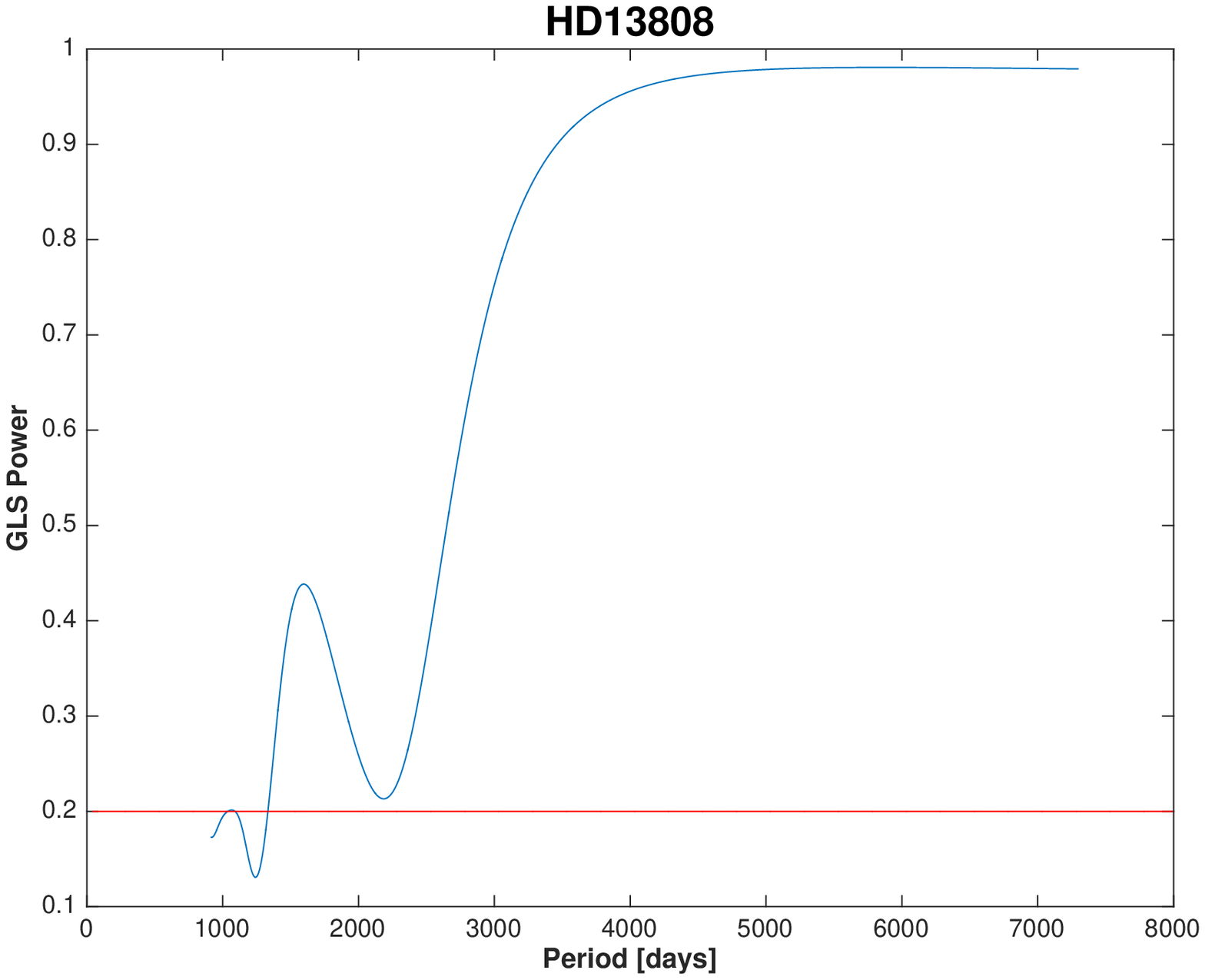}
        \end{subfigure}%
        \begin{subfigure}[b]{0.25\textwidth}
                \includegraphics[width=\linewidth]{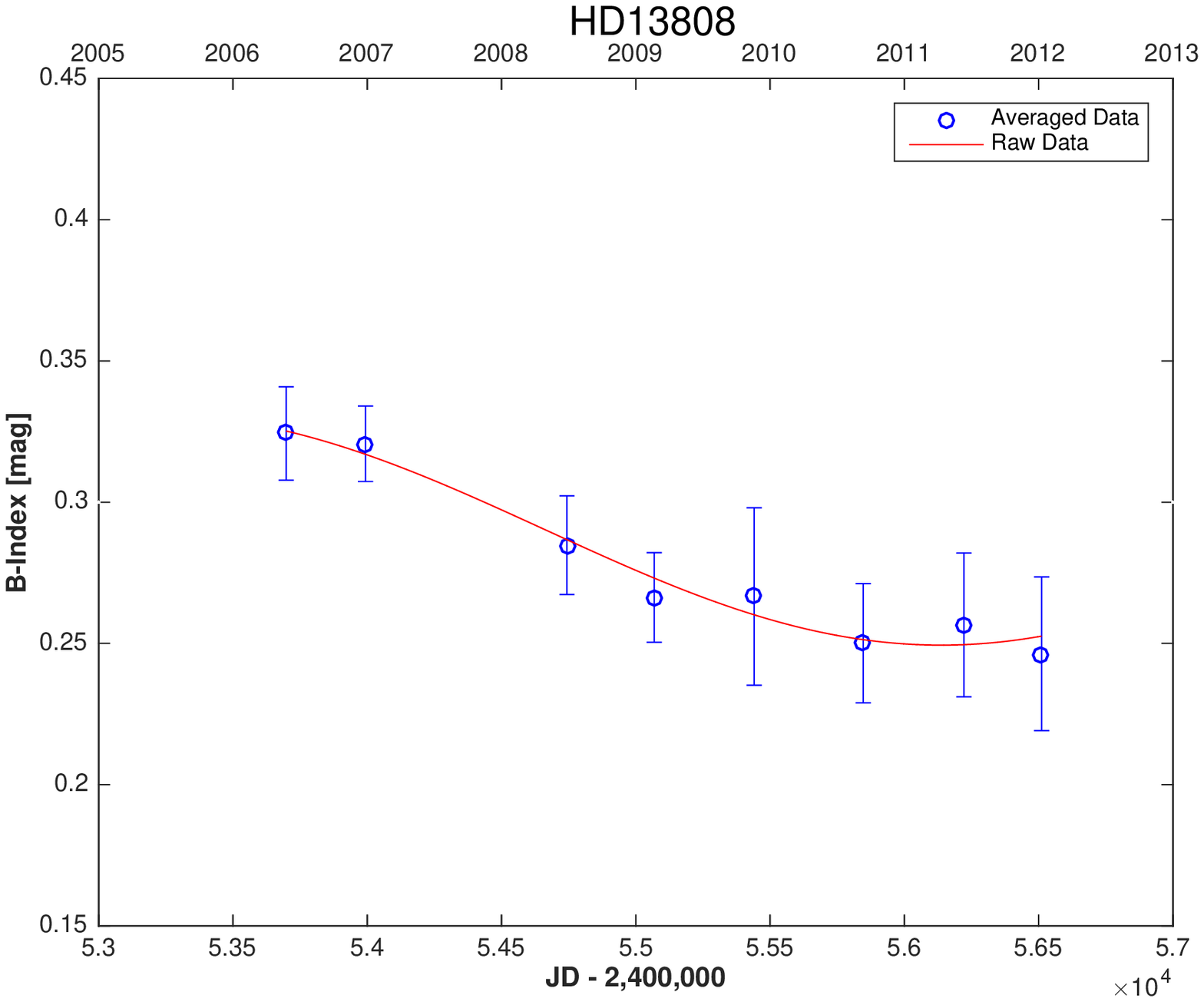}
        \end{subfigure}%
        \begin{subfigure}[b]{0.25\textwidth}
                \includegraphics[width=\linewidth]{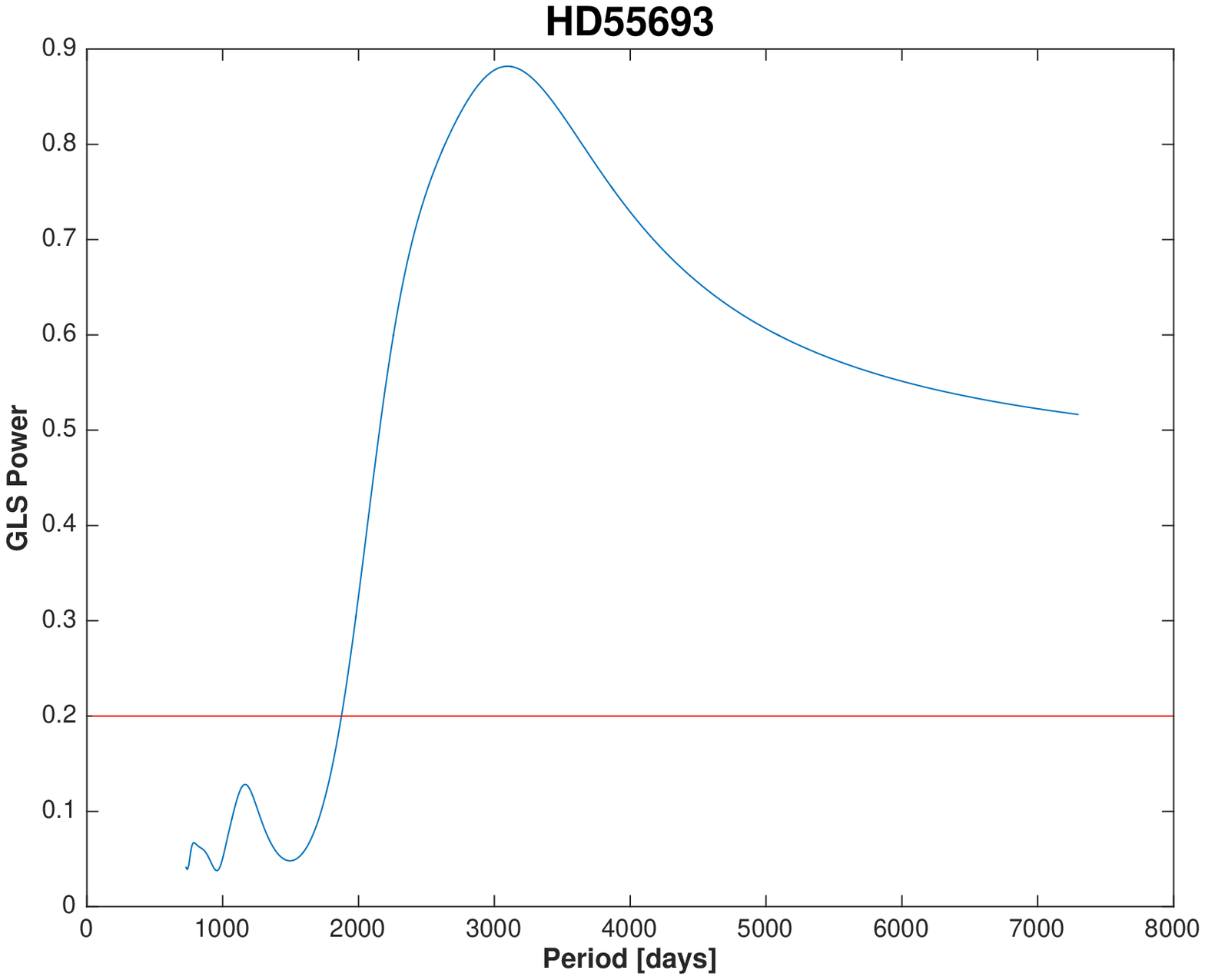}
        \end{subfigure}%
        \begin{subfigure}[b]{0.25\textwidth}
                \includegraphics[width=\linewidth]{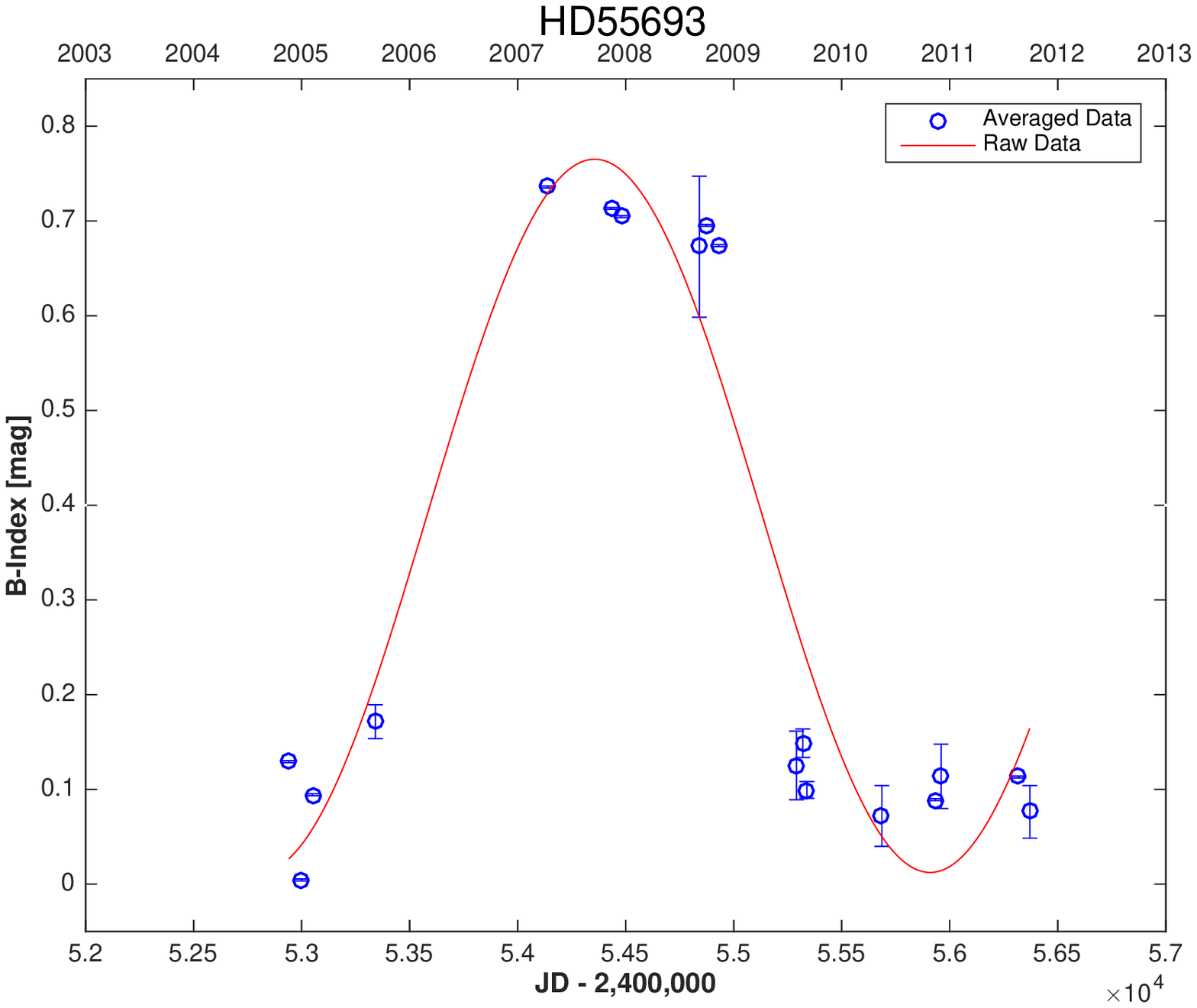}
        \end{subfigure}

  \begin{subfigure}[b]{0.25\textwidth}
                \includegraphics[width=\linewidth]{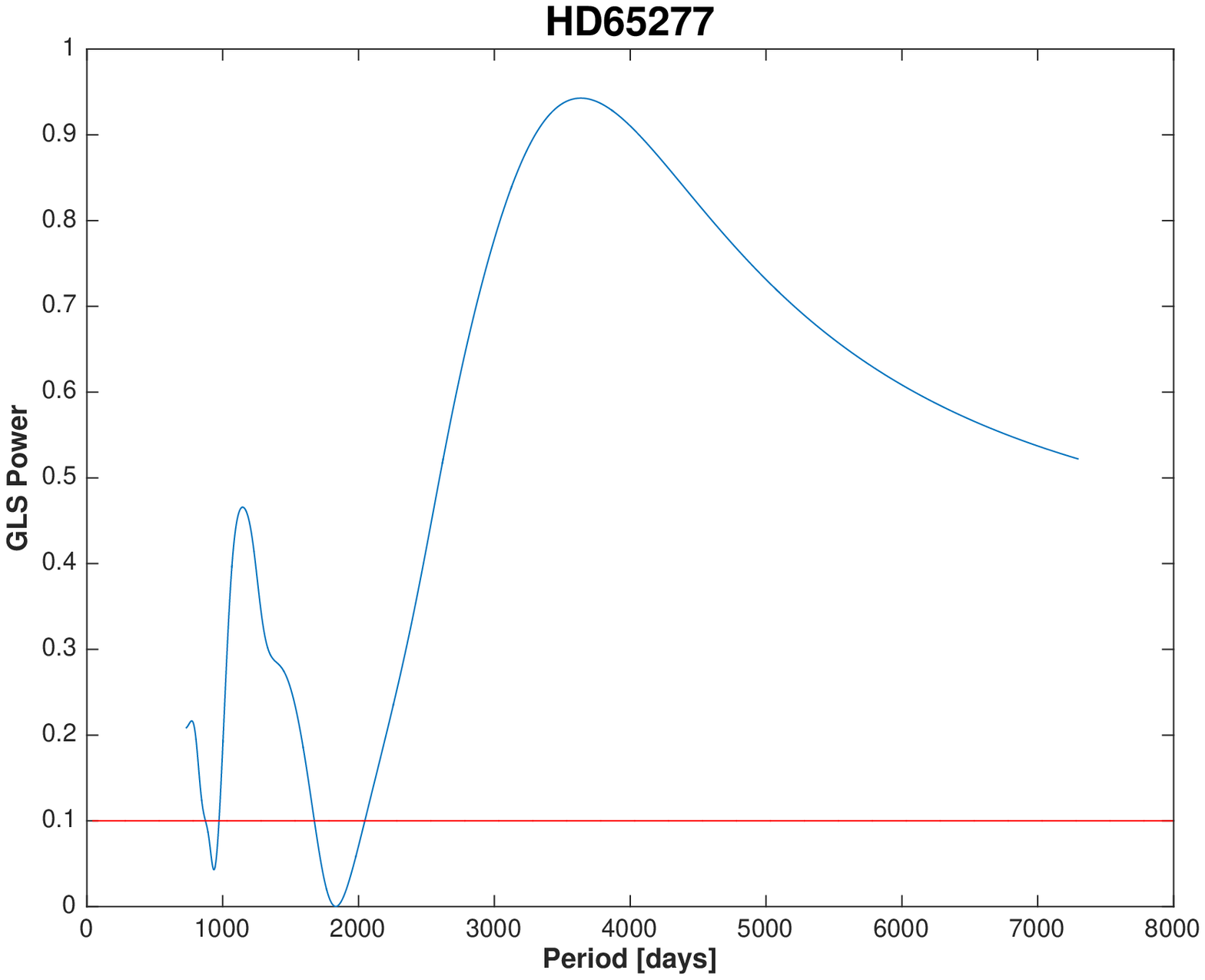}
        \end{subfigure}%
        \begin{subfigure}[b]{0.25\textwidth}
                \includegraphics[width=\linewidth]{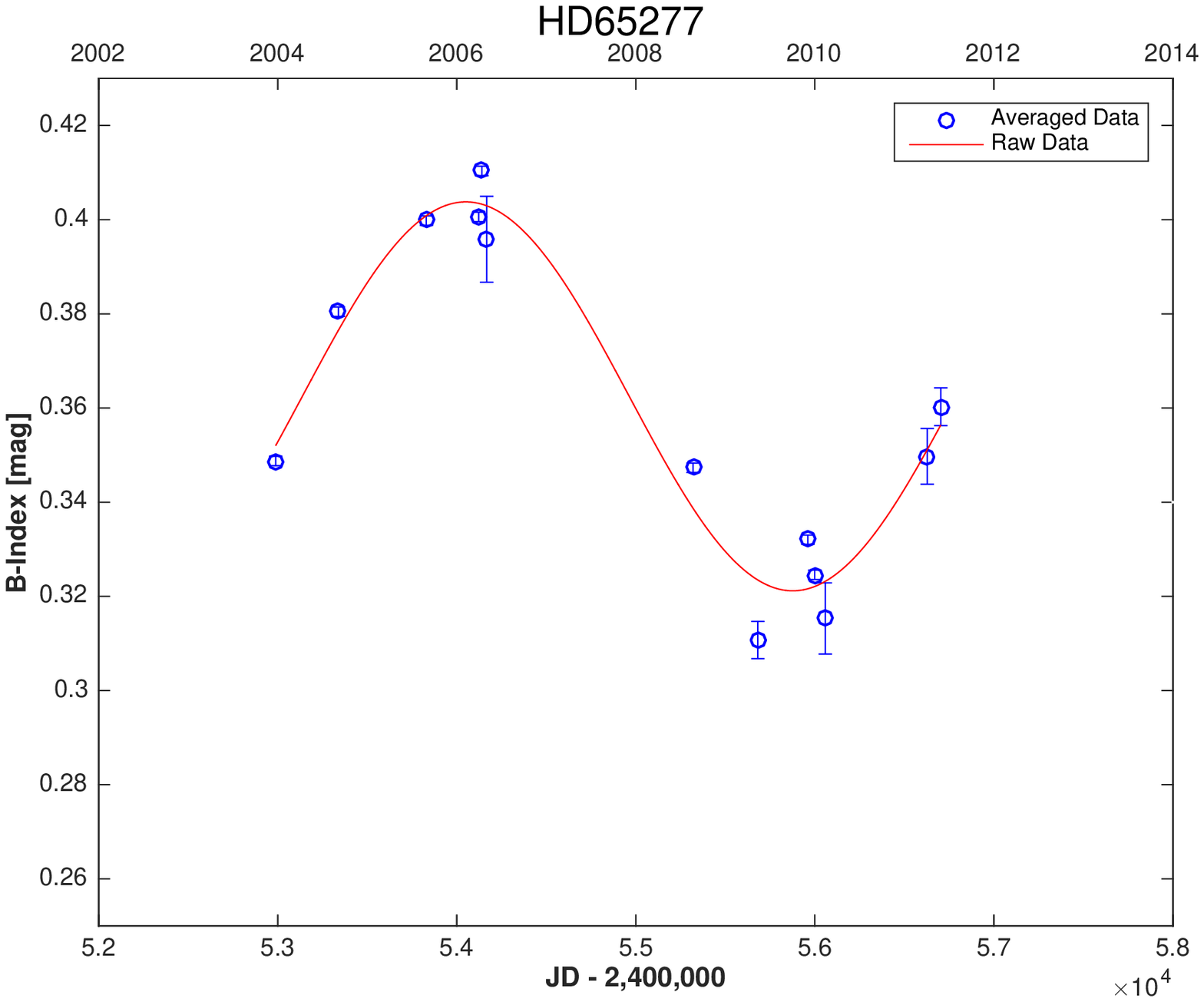}
        \end{subfigure}%
        \begin{subfigure}[b]{0.25\textwidth}
                \includegraphics[width=\linewidth]{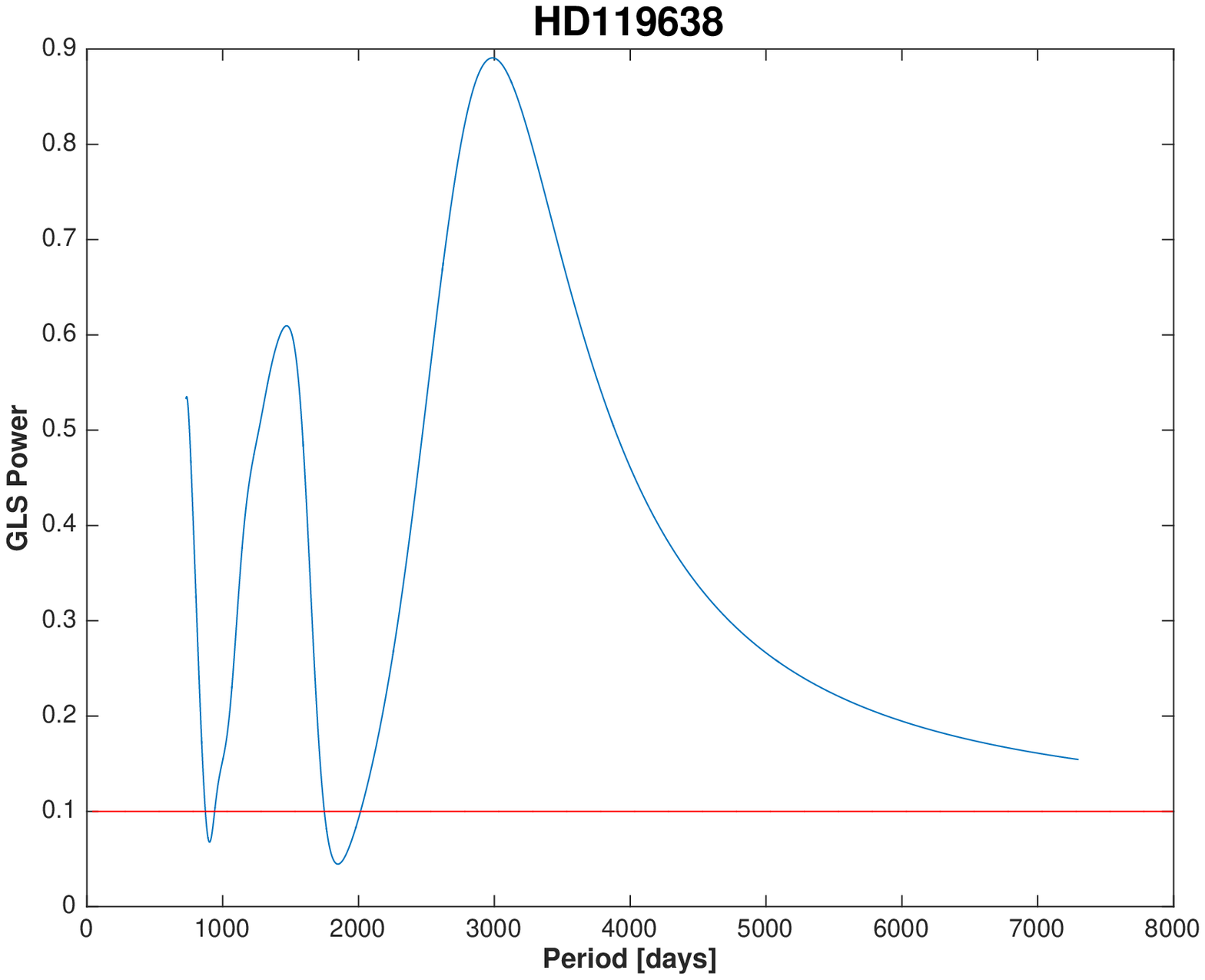}
        \end{subfigure}%
        \begin{subfigure}[b]{0.25\textwidth}
                \includegraphics[width=\linewidth]{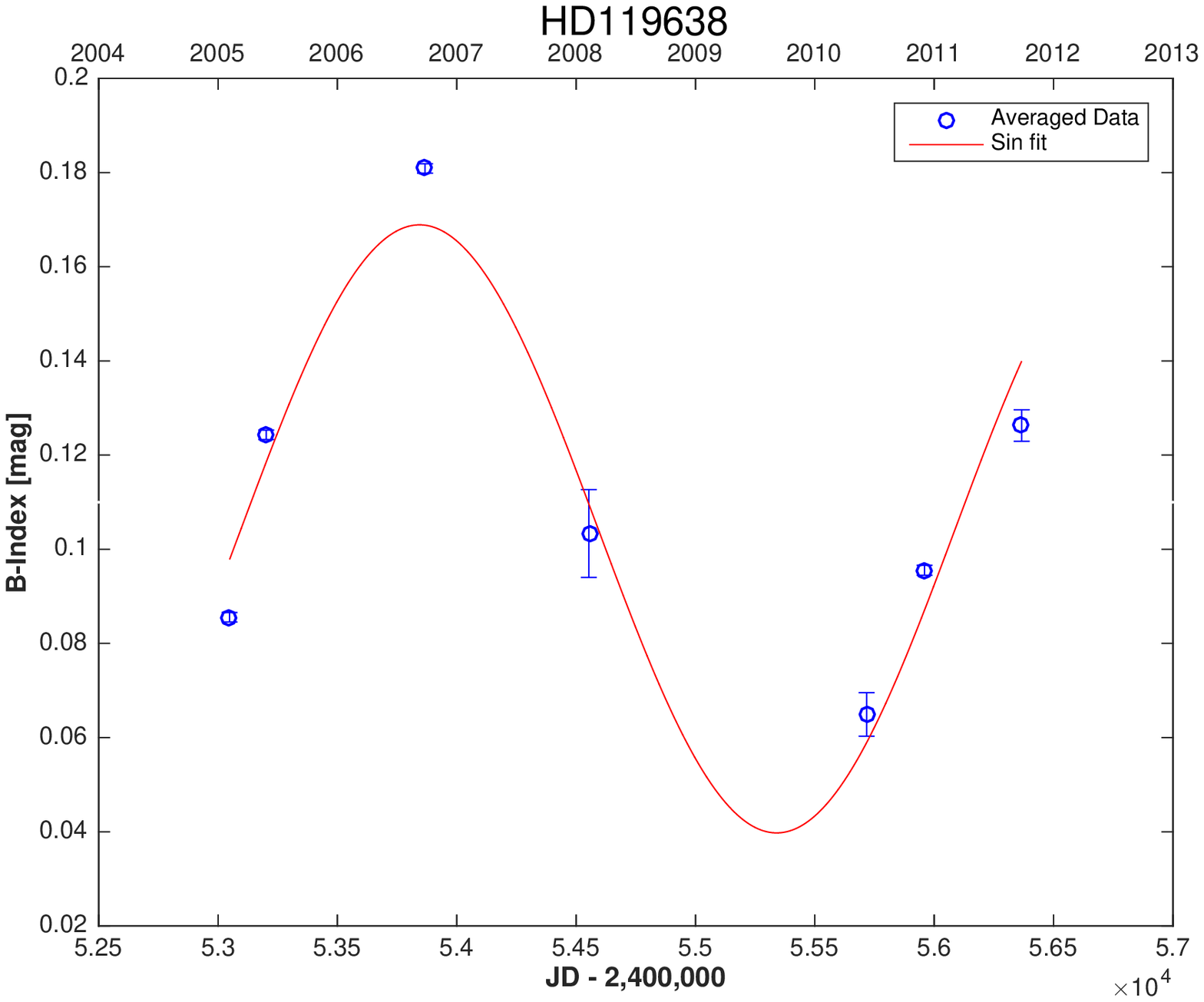}
        \end{subfigure}

       \begin{subfigure}[b]{0.25\textwidth}
                \includegraphics[width=\linewidth]{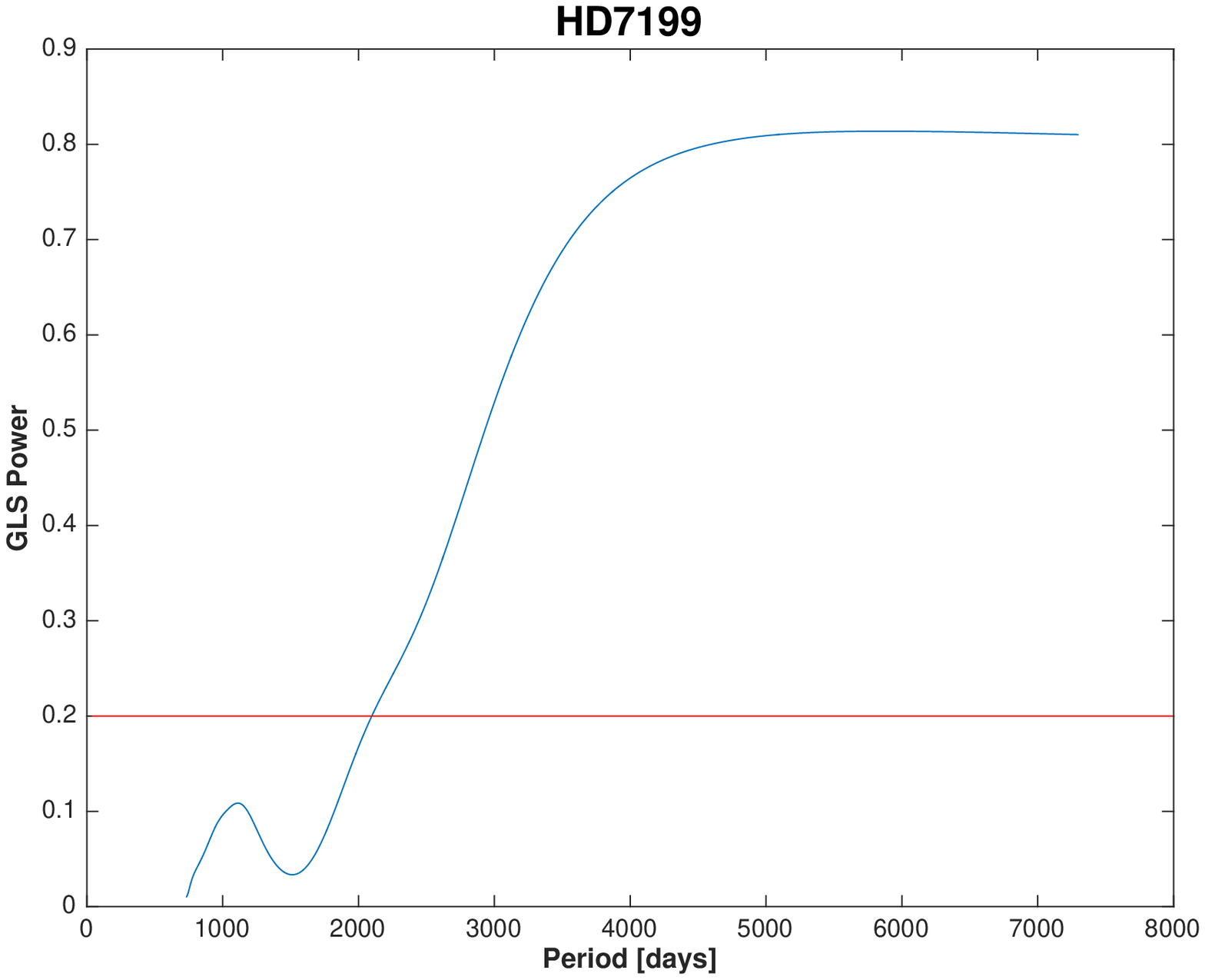}
        \end{subfigure}%
        \begin{subfigure}[b]{0.25\textwidth}
                \includegraphics[width=\linewidth]{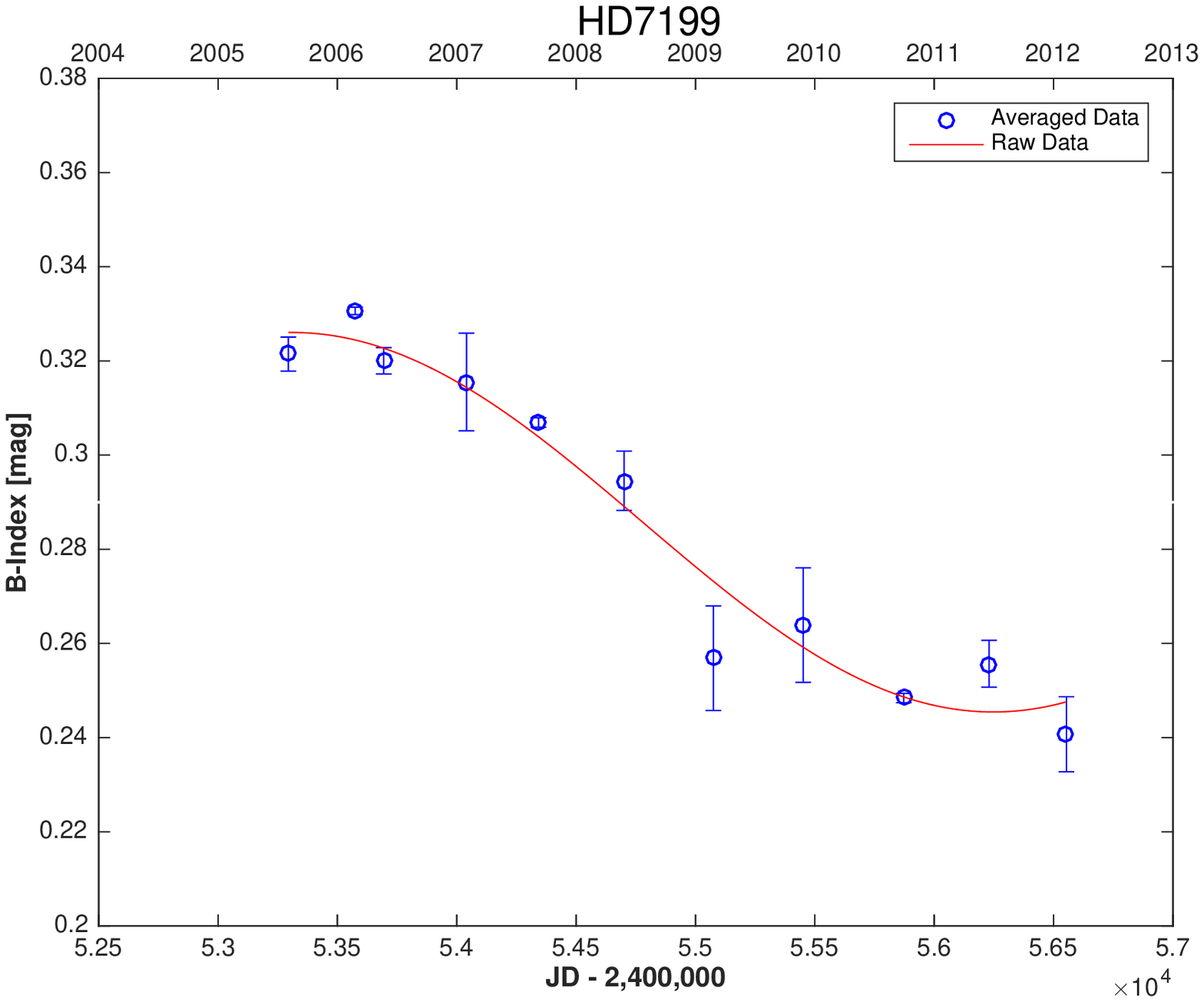}
        \end{subfigure}%
        \begin{subfigure}[b]{0.25\textwidth}
                \includegraphics[width=\linewidth]{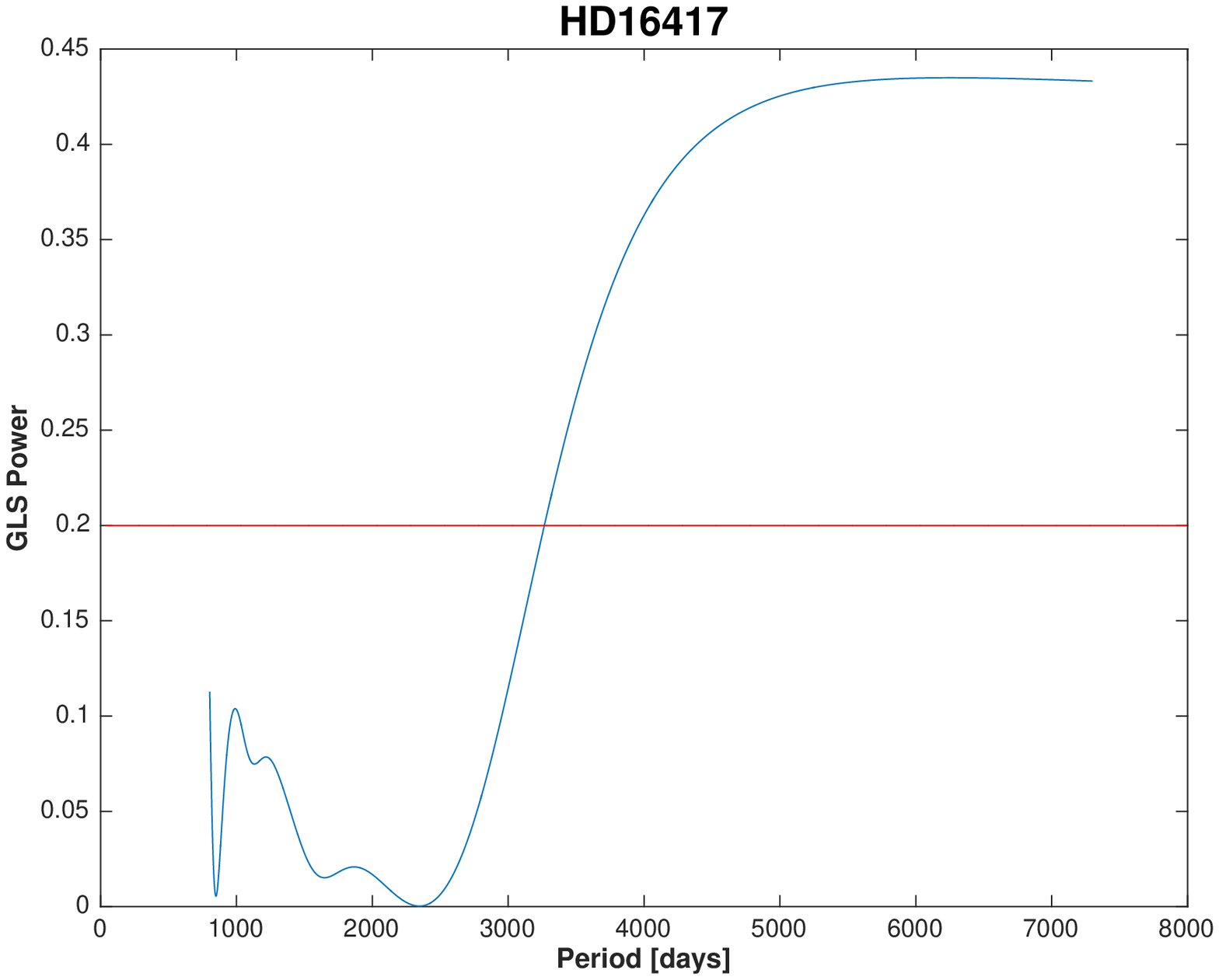}
        \end{subfigure}%
        \begin{subfigure}[b]{0.25\textwidth}
                \includegraphics[width=\linewidth]{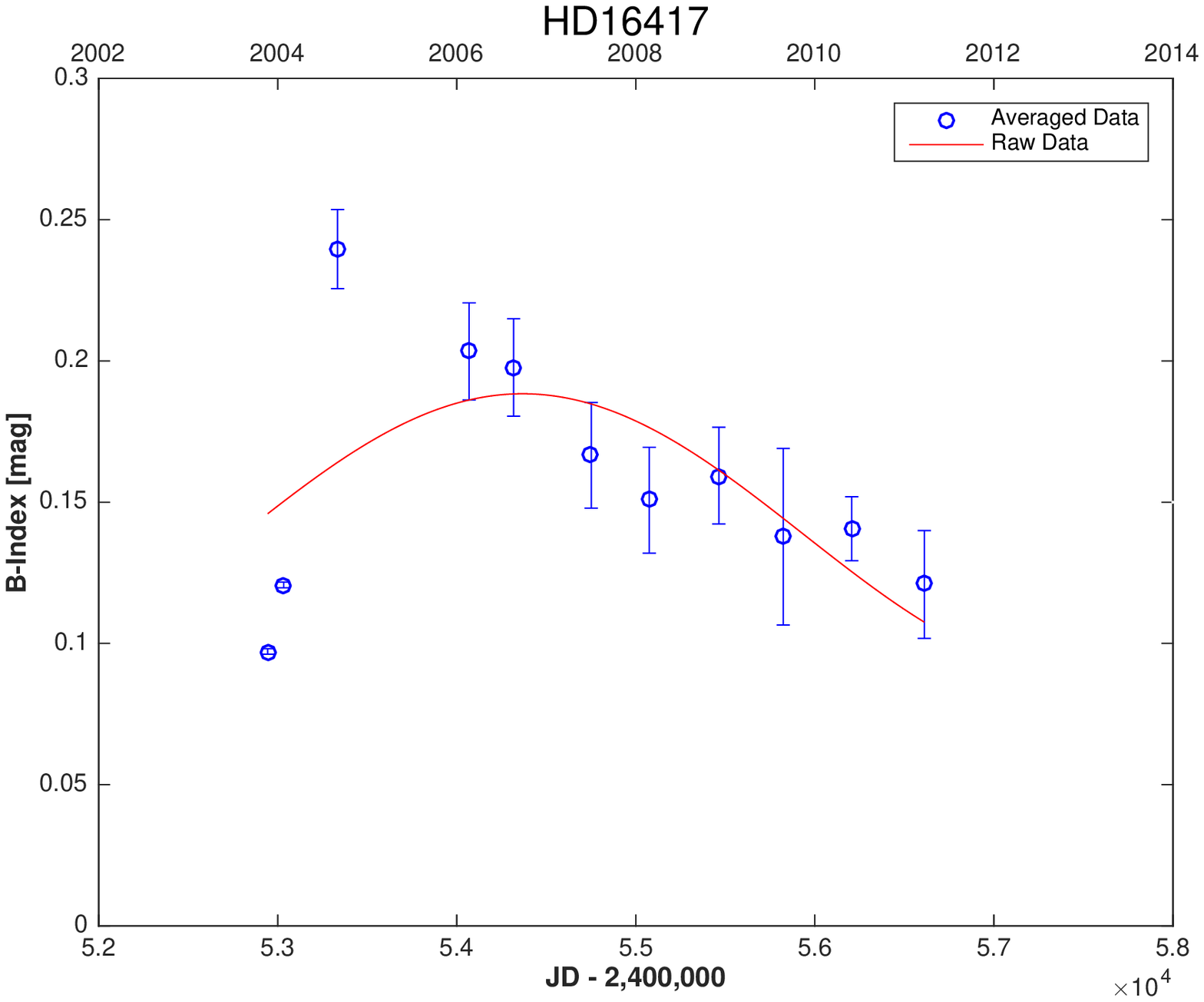}
        \end{subfigure}
         \begin{subfigure}[b]{0.25\textwidth}
                \includegraphics[width=\linewidth]{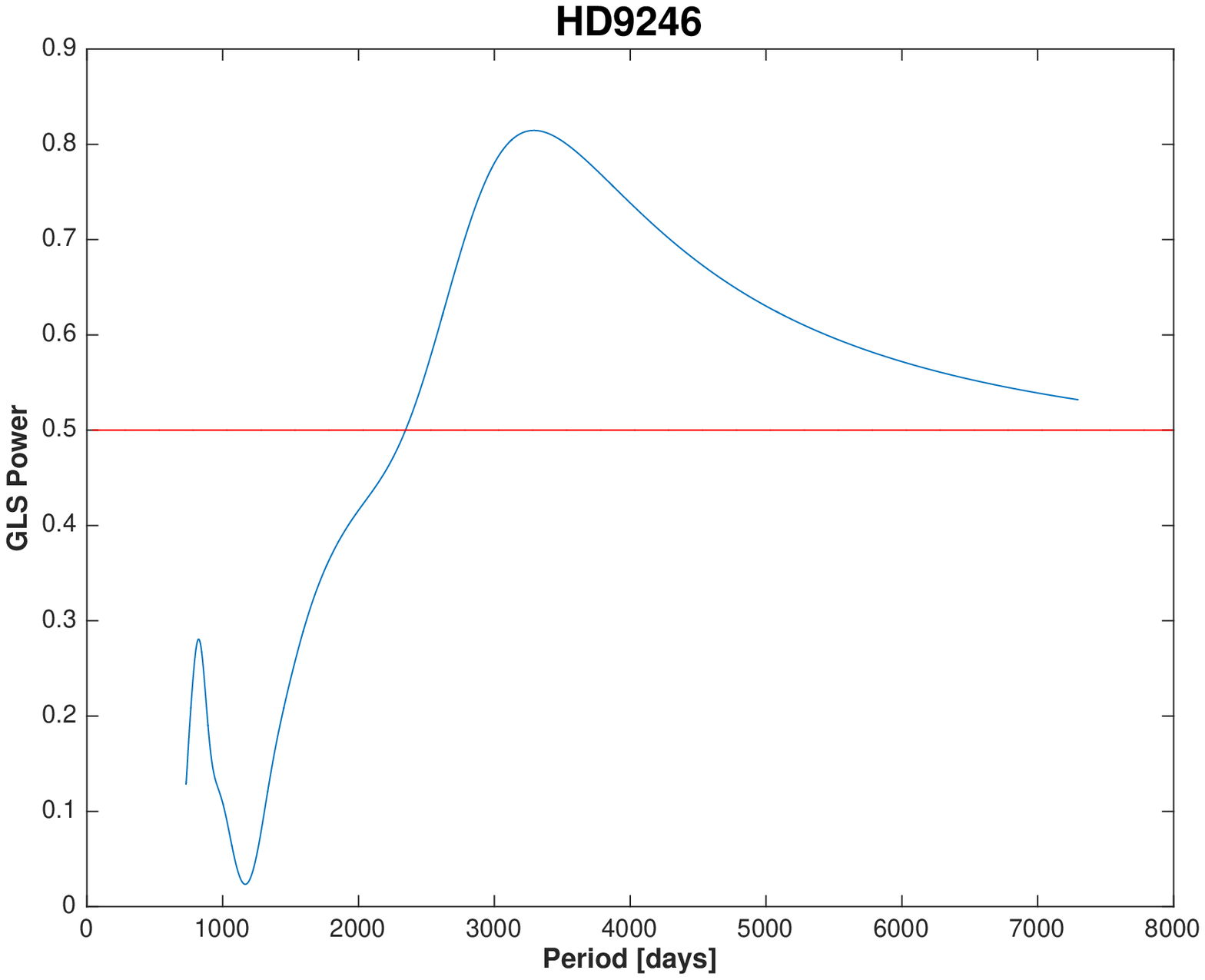}
        \end{subfigure}%
        \begin{subfigure}[b]{0.25\textwidth}
                \includegraphics[width=\linewidth]{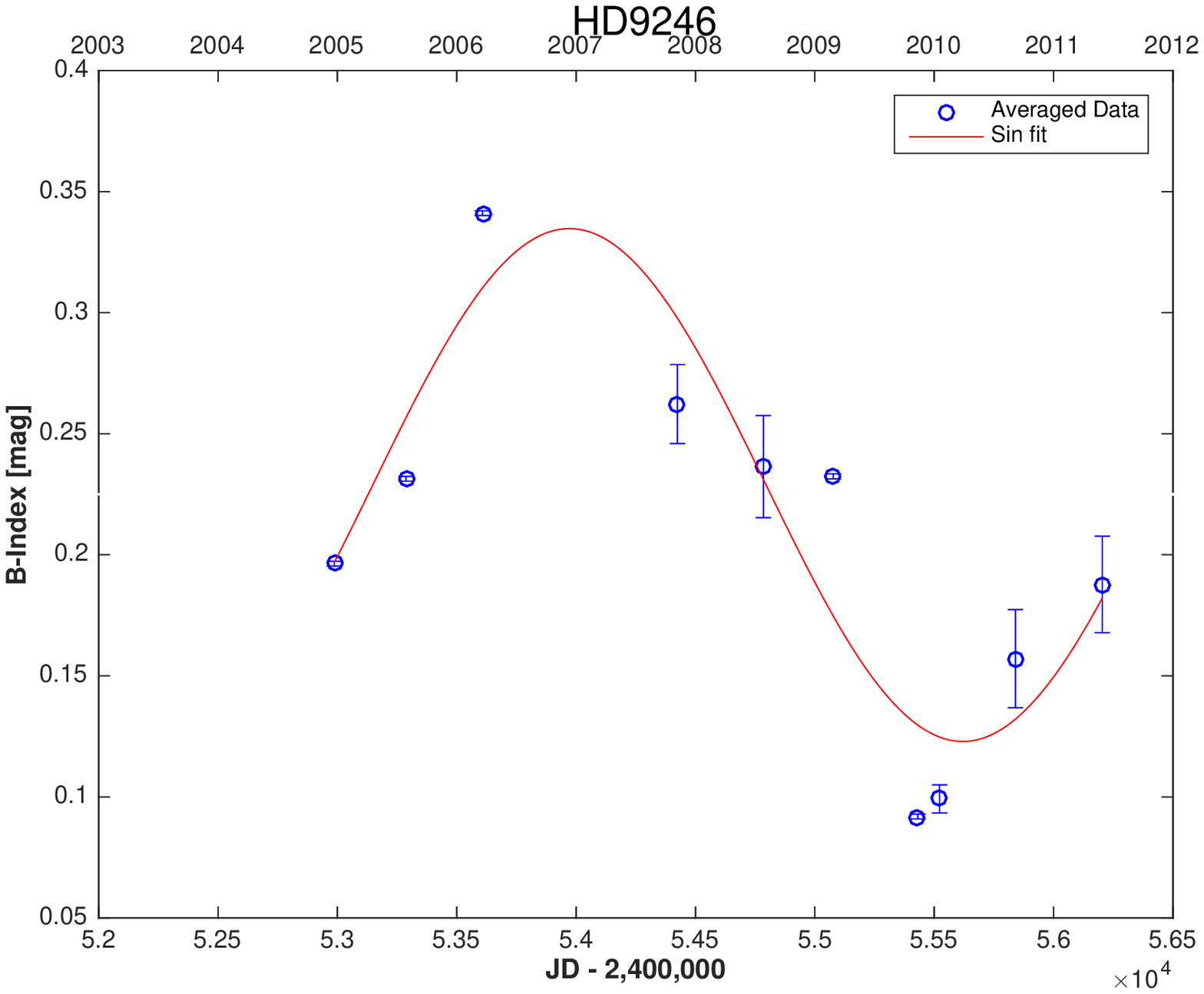}
        \end{subfigure}%
        \begin{subfigure}[b]{0.25\textwidth}
                \includegraphics[width=\linewidth]{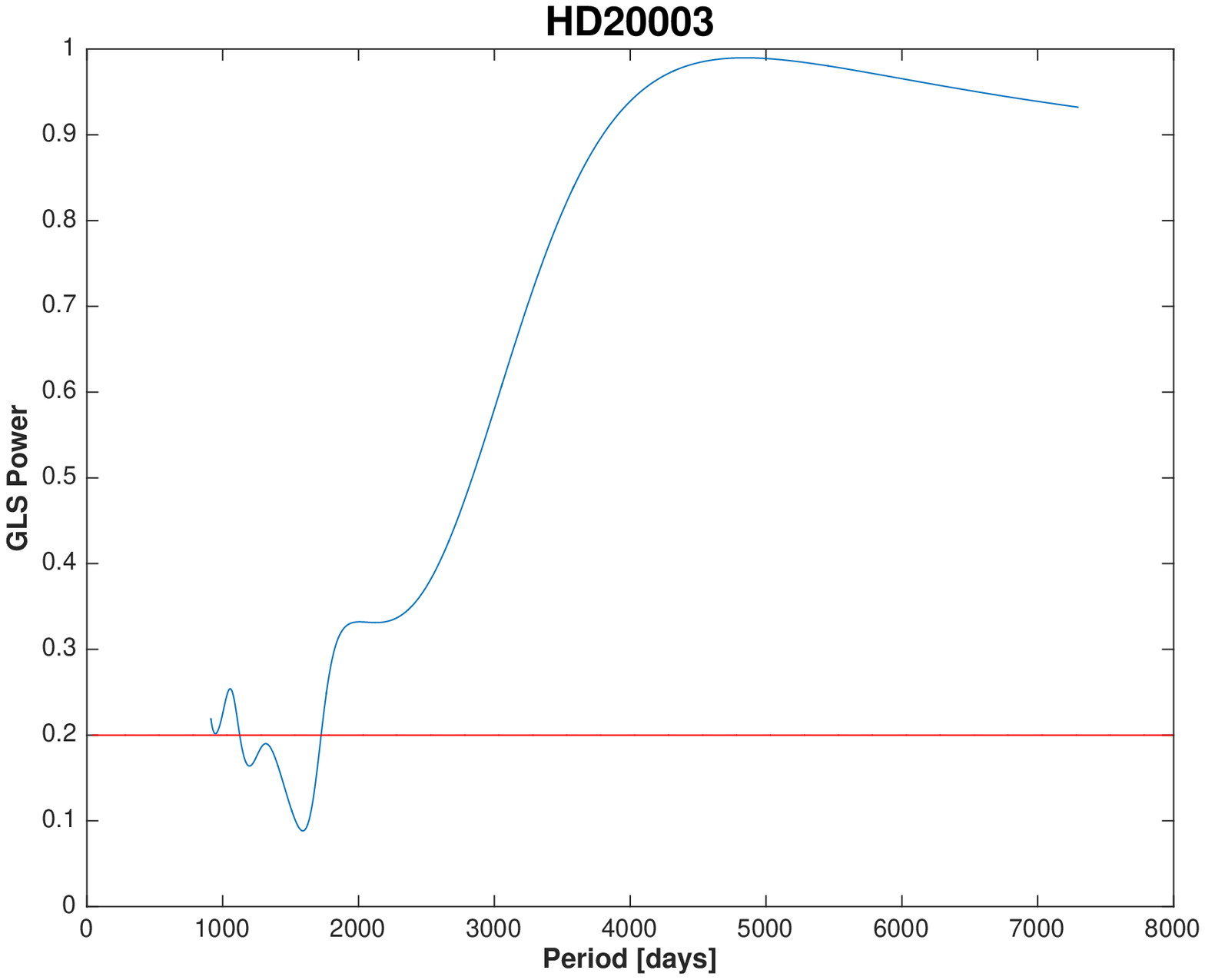}
        \end{subfigure}%
        \begin{subfigure}[b]{0.25\textwidth}
                \includegraphics[width=\linewidth]{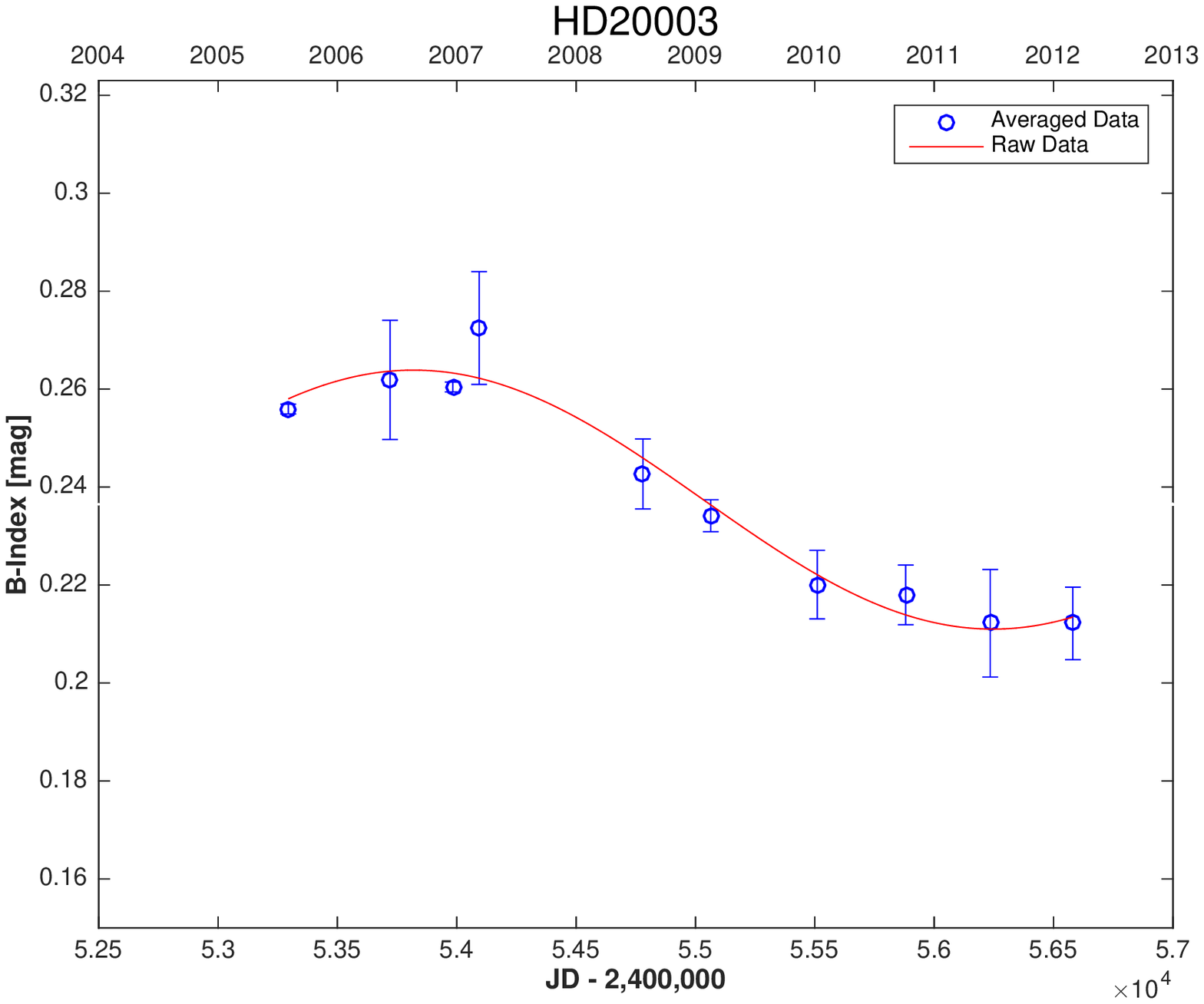}
        \end{subfigure}
        \begin{subfigure}[b]{0.25\textwidth}
                \includegraphics[width=\linewidth]{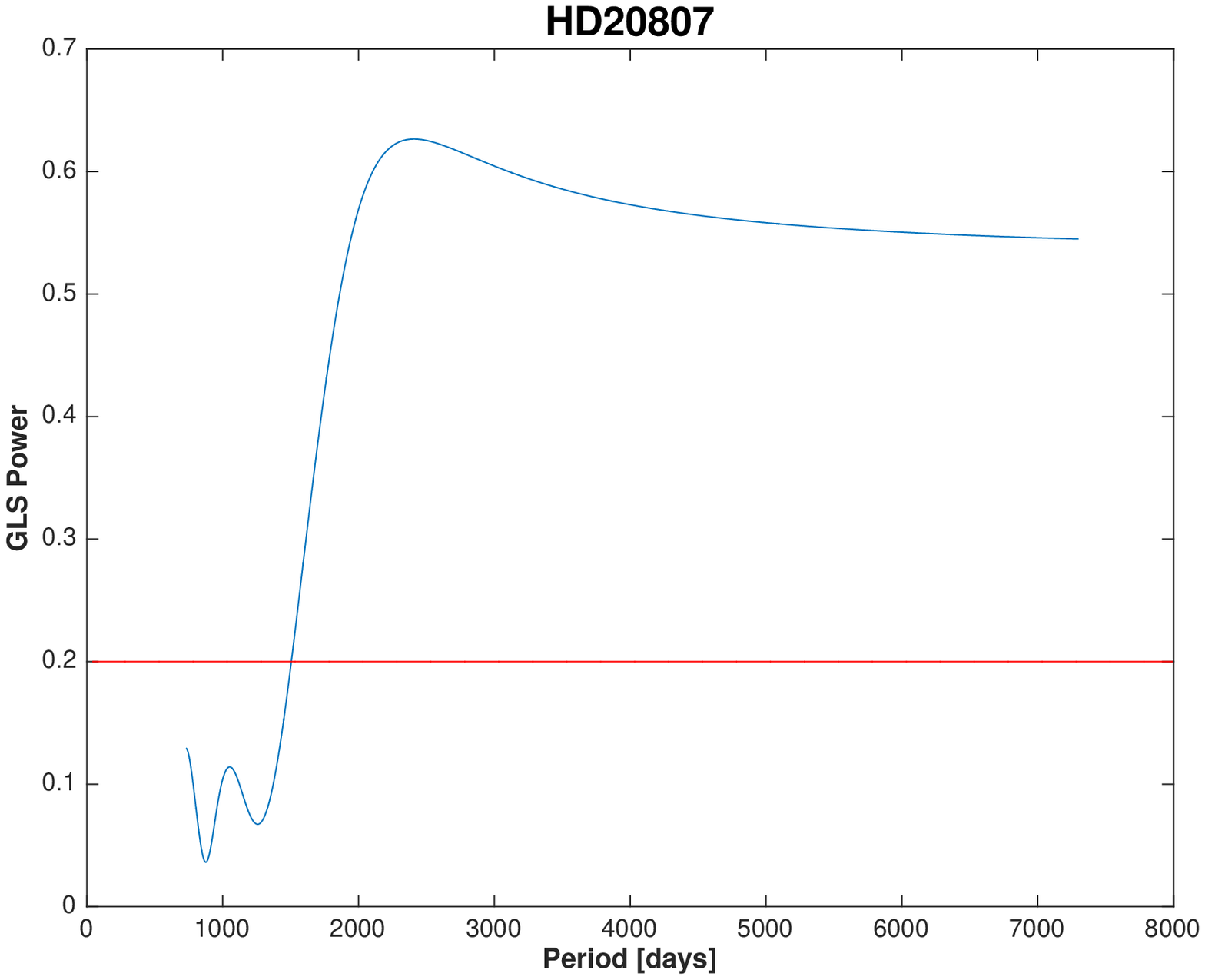}
        \end{subfigure}%
        \begin{subfigure}[b]{0.25\textwidth}
                \includegraphics[width=\linewidth]{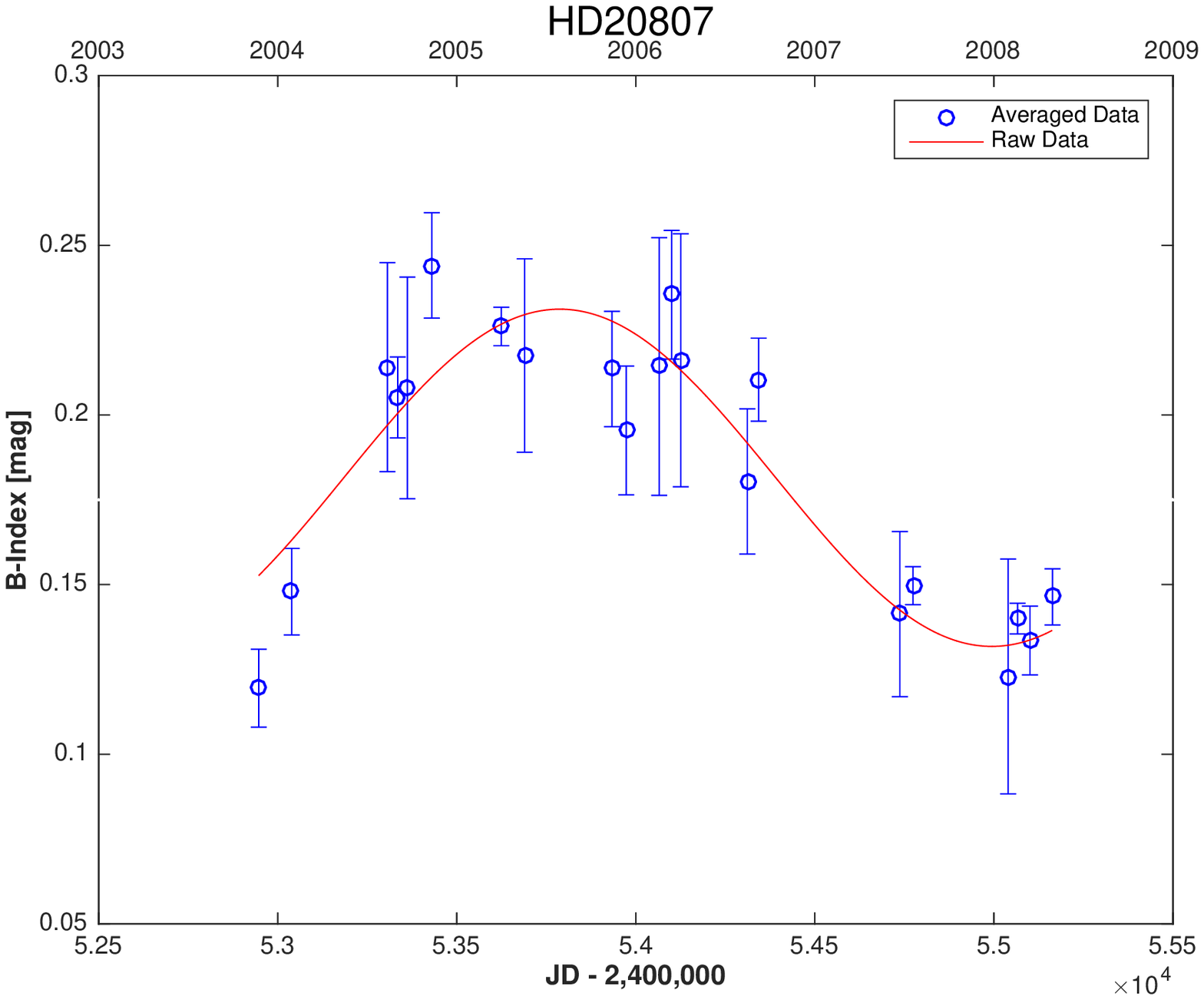}
        \end{subfigure}%
        \begin{subfigure}[b]{0.25\textwidth}
                \includegraphics[width=\linewidth]{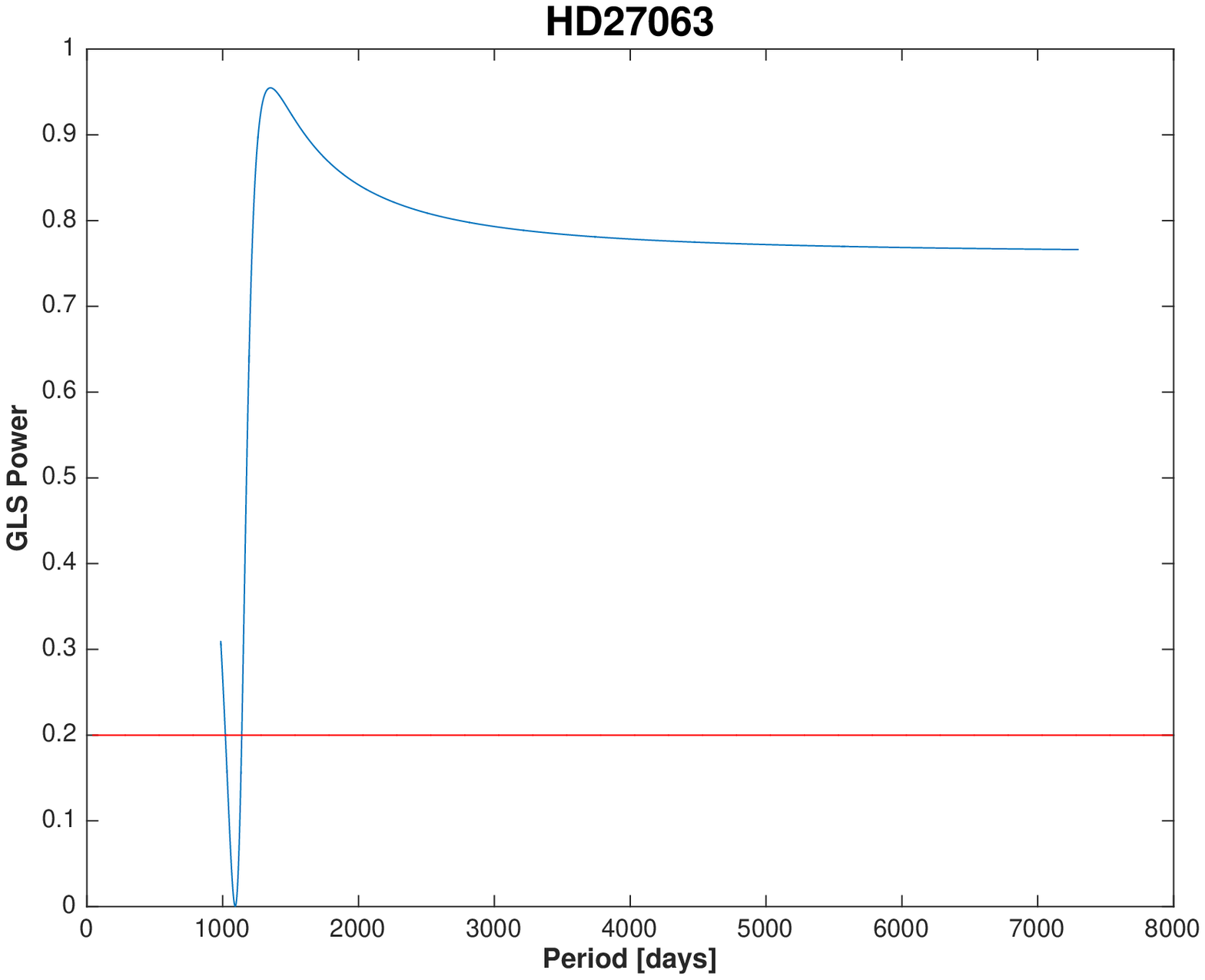}
        \end{subfigure}%
        \begin{subfigure}[b]{0.25\textwidth}
                \includegraphics[width=\linewidth]{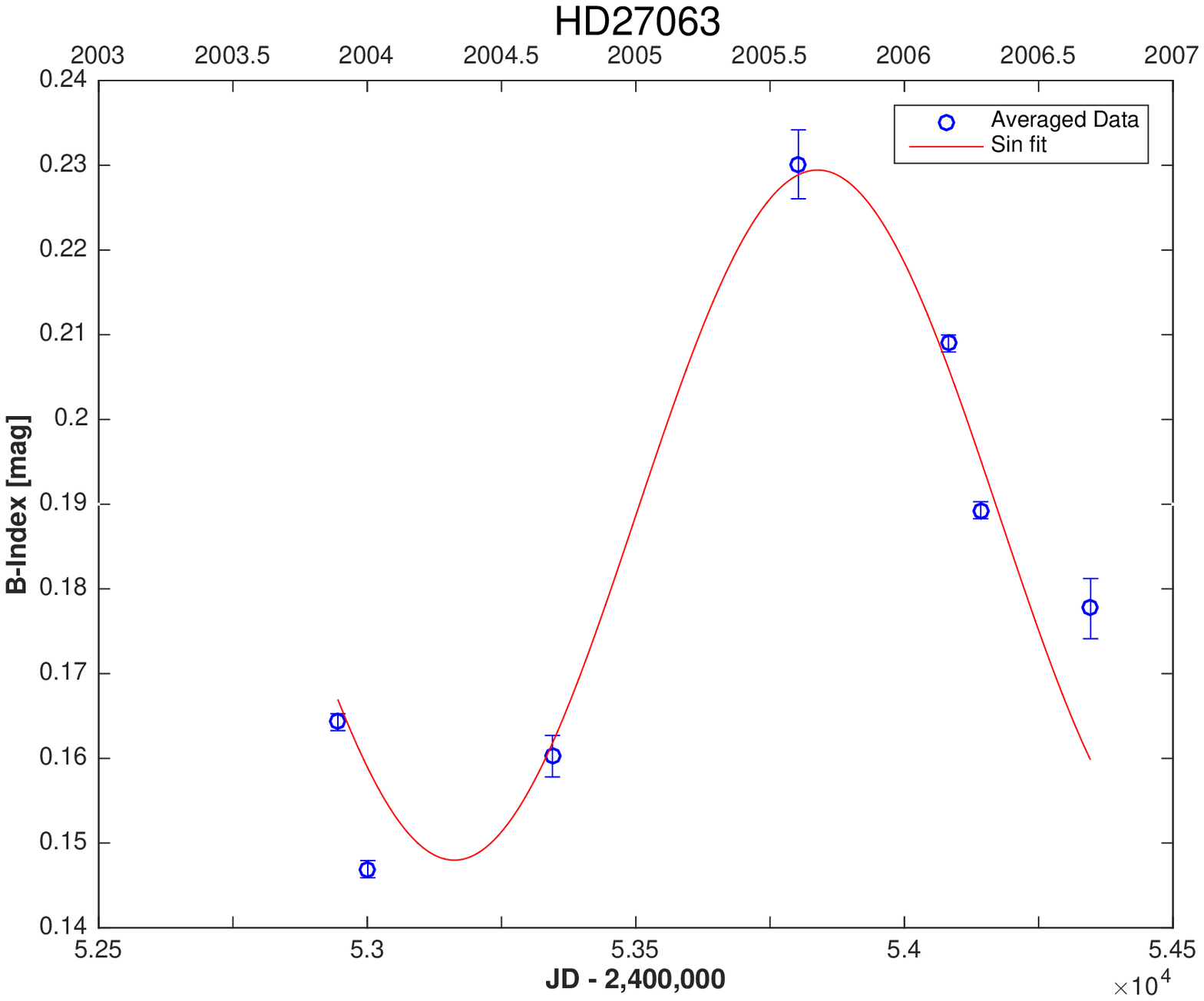}
        \end{subfigure}
 \begin{subfigure}[b]{0.25\textwidth}
                \includegraphics[width=\linewidth]{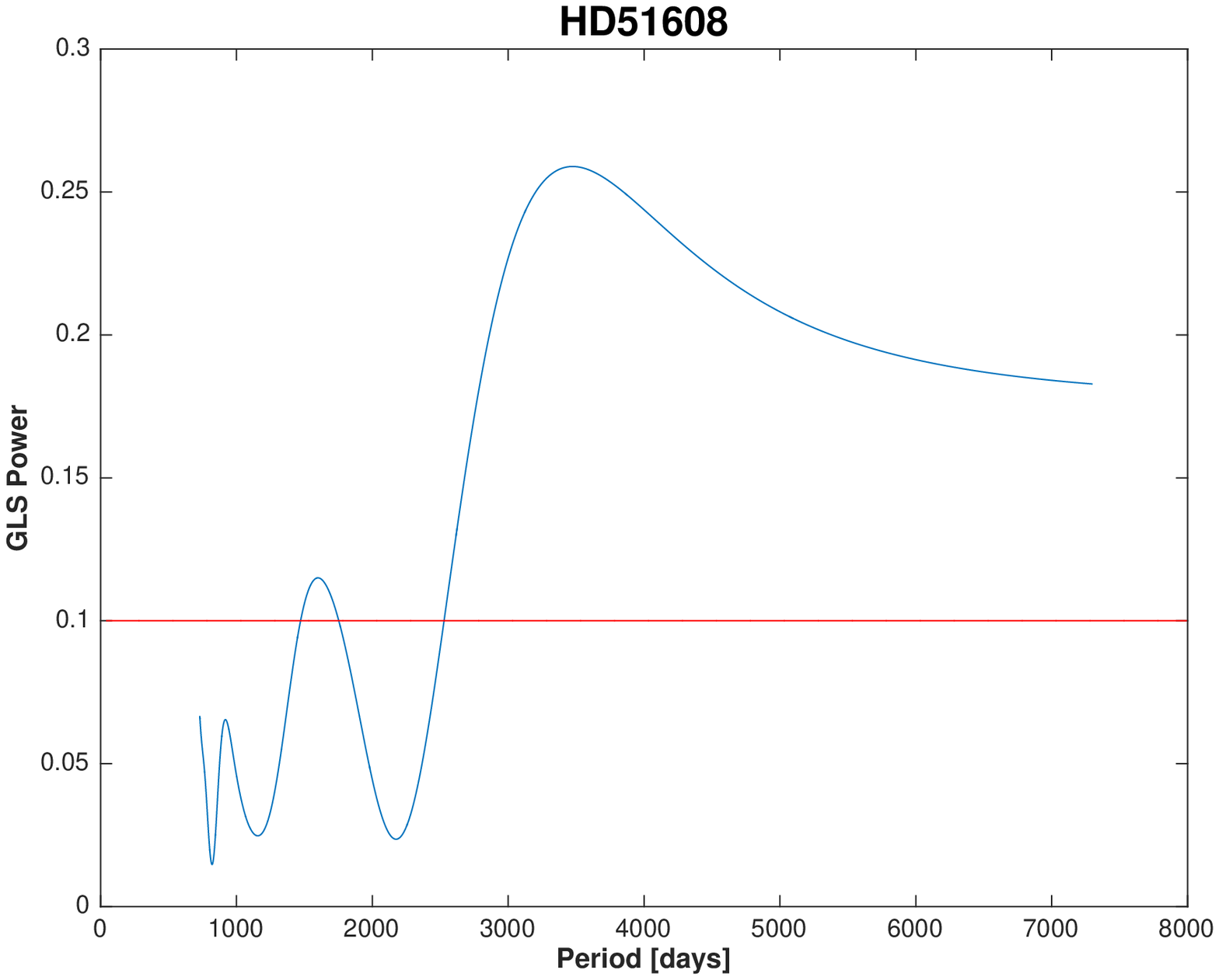}
        \end{subfigure}%
        \begin{subfigure}[b]{0.25\textwidth}
                \includegraphics[width=\linewidth]{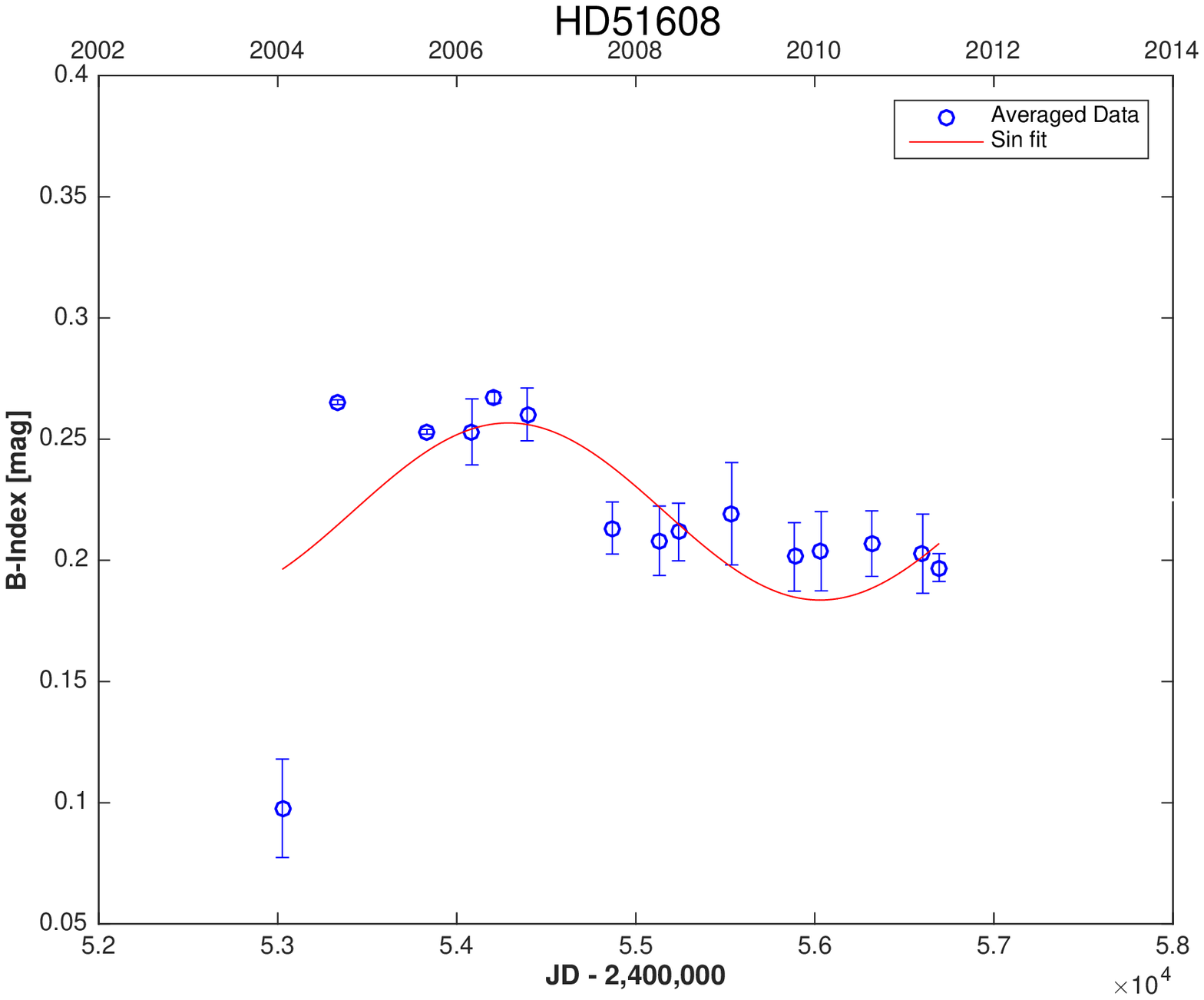}
        \end{subfigure}
 \begin{subfigure}[b]{0.25\textwidth}
                \includegraphics[width=\linewidth]{78747p.eps}
        \end{subfigure}%
        \begin{subfigure}[b]{0.25\textwidth}
                \includegraphics[width=\linewidth]{78747.eps}
        \end{subfigure}
                         
\end{figure}

\begin{figure}

 \begin{subfigure}[b]{0.25\textwidth}
                \includegraphics[width=\linewidth]{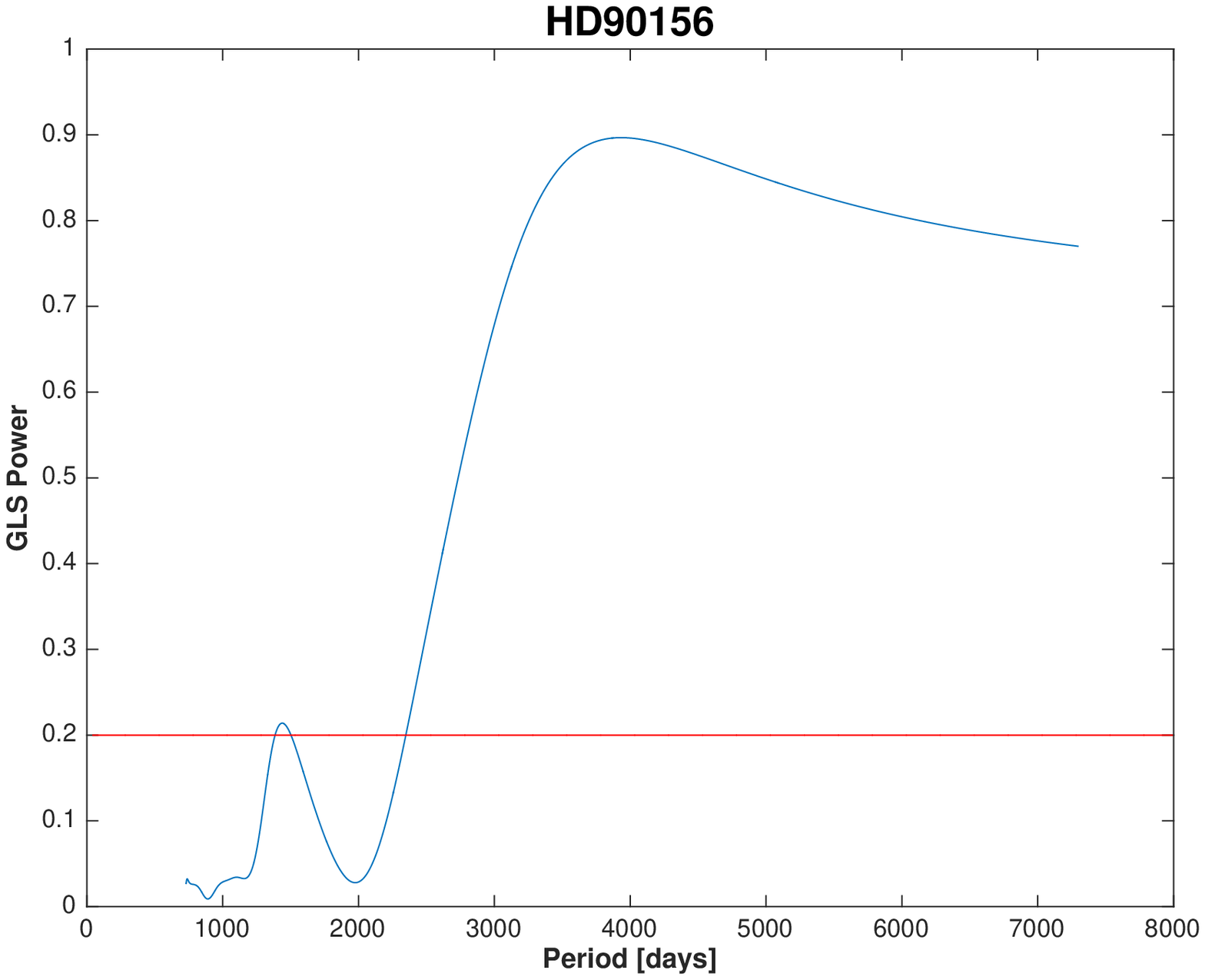}
        \end{subfigure}%
        \begin{subfigure}[b]{0.25\textwidth}
                \includegraphics[width=\linewidth]{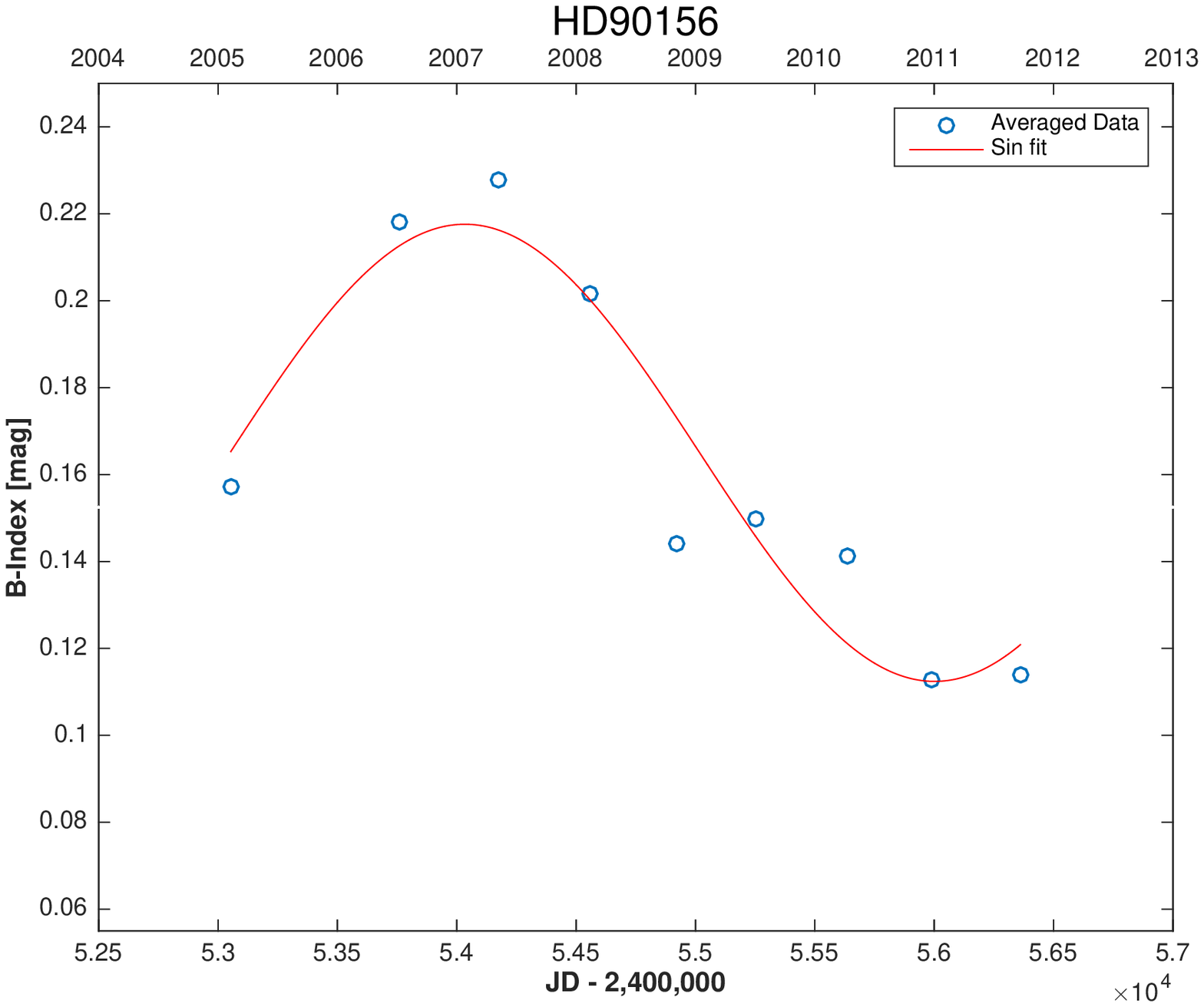}
        \end{subfigure}
         \begin{subfigure}[b]{0.25\textwidth}
                \includegraphics[width=\linewidth]{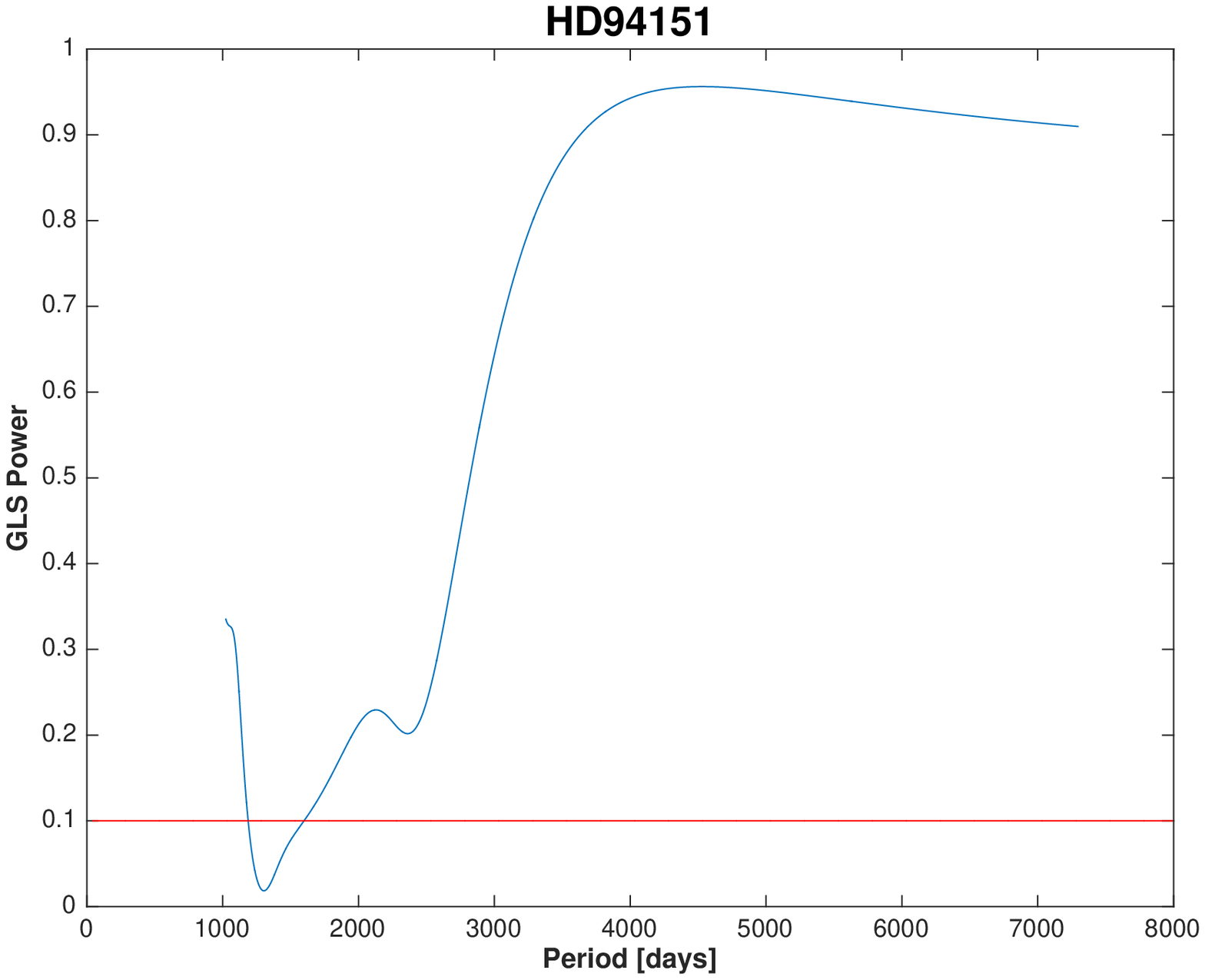}
        \end{subfigure}%
        \begin{subfigure}[b]{0.25\textwidth}
                \includegraphics[width=\linewidth]{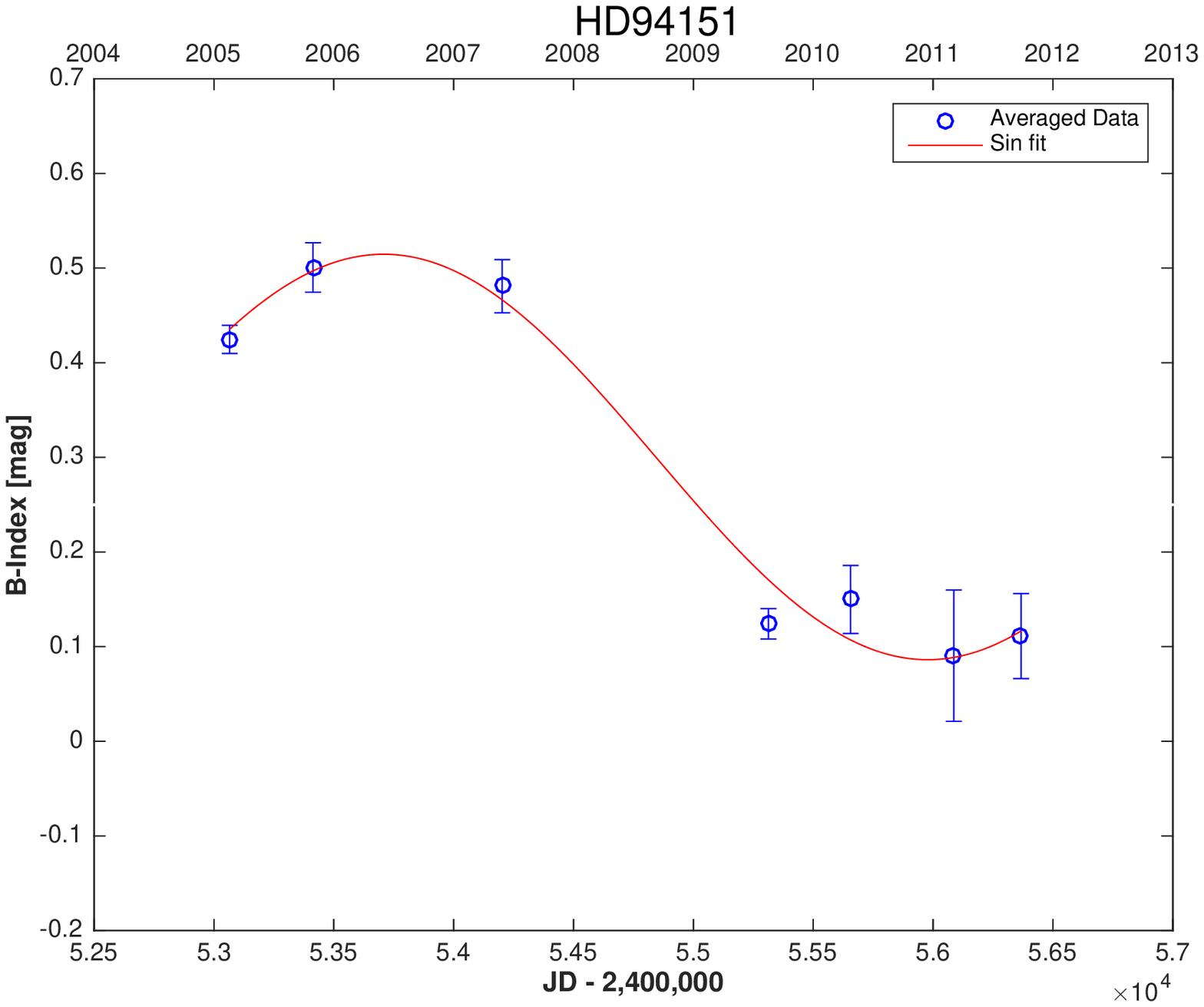}
        \end{subfigure}
 \begin{subfigure}[b]{0.25\textwidth}
                \includegraphics[width=\linewidth]{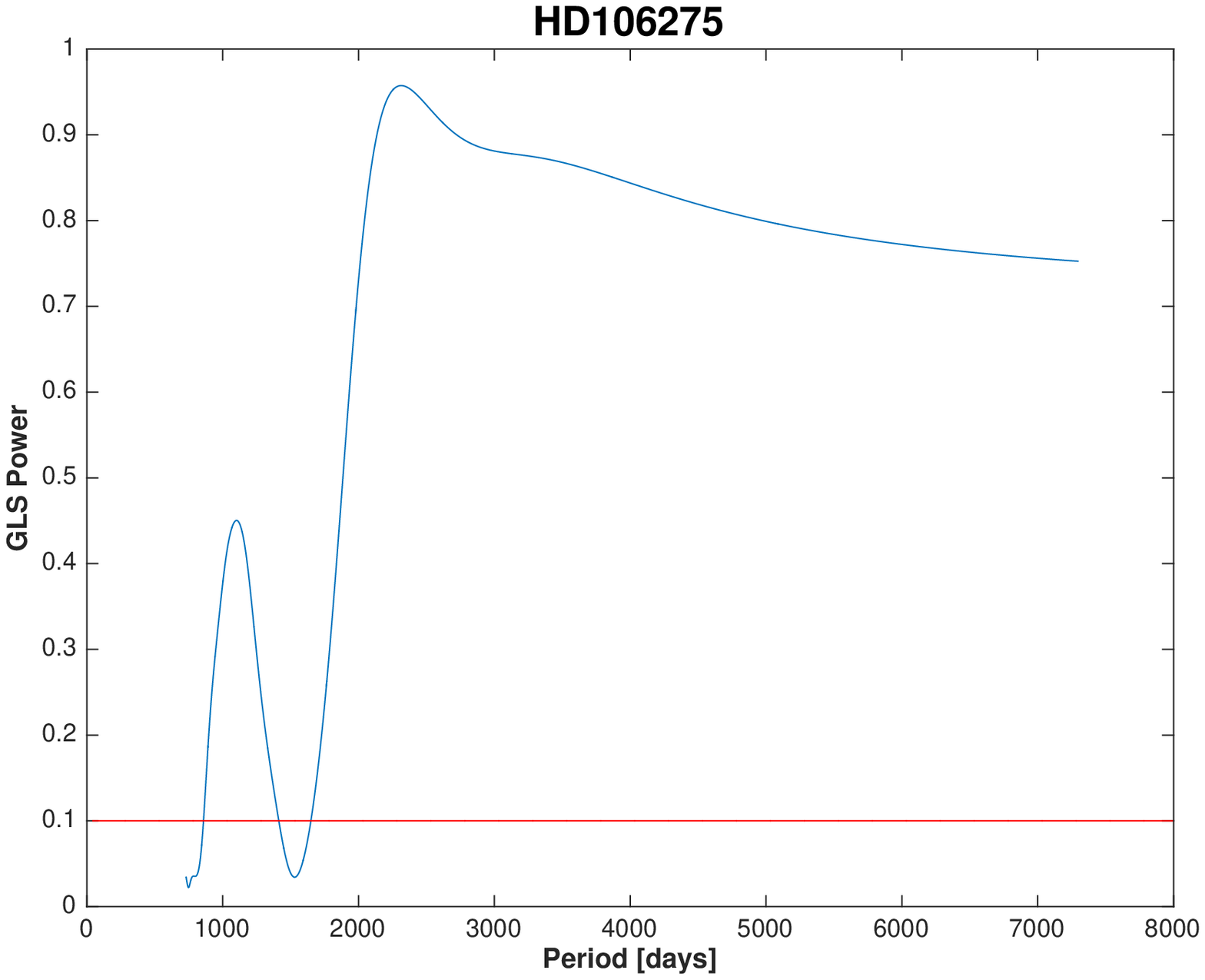}
        \end{subfigure}%
        \begin{subfigure}[b]{0.25\textwidth}
                \includegraphics[width=\linewidth]{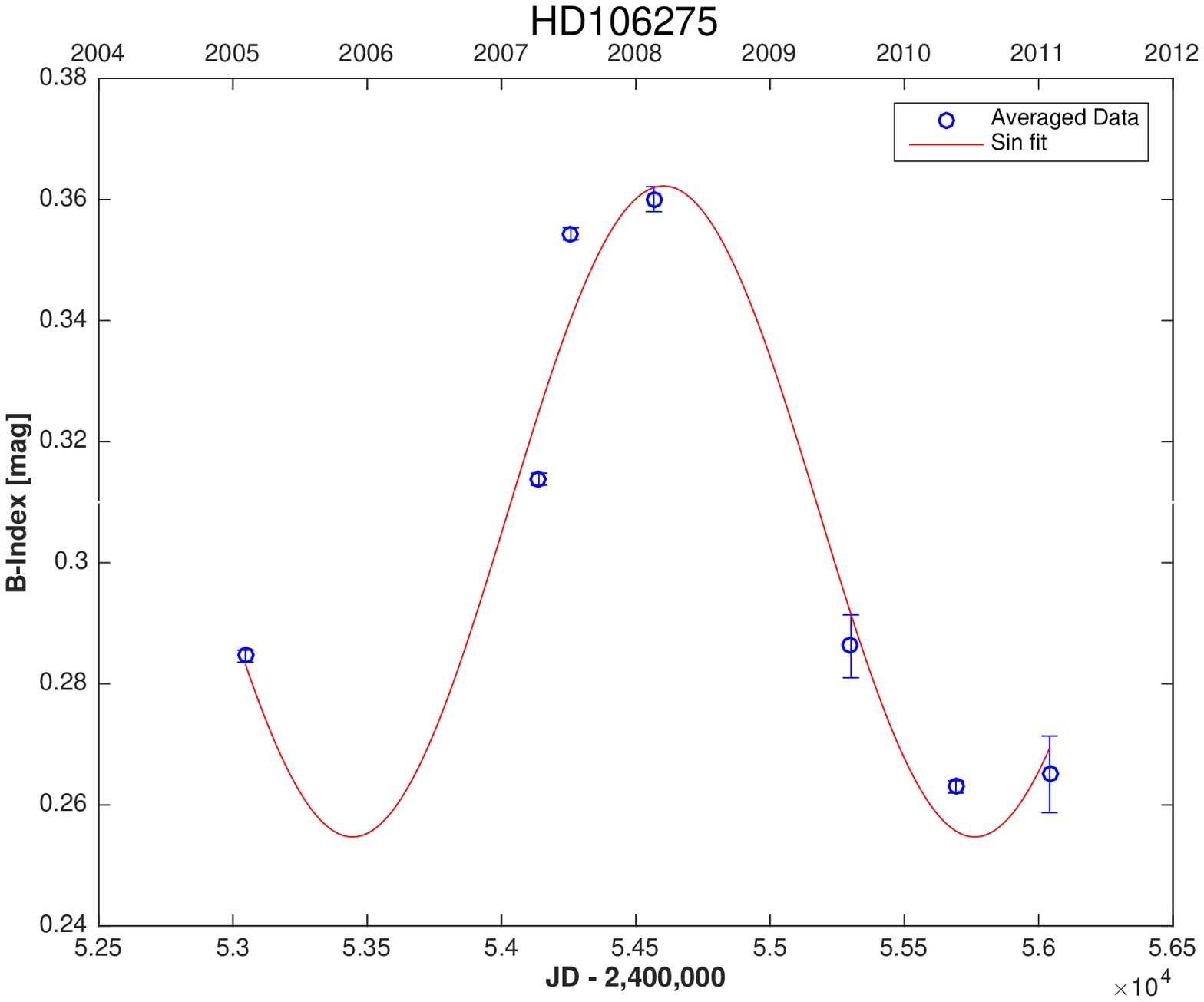}
        \end{subfigure}
        \begin{subfigure}[b]{0.25\textwidth}
                \includegraphics[width=\linewidth]{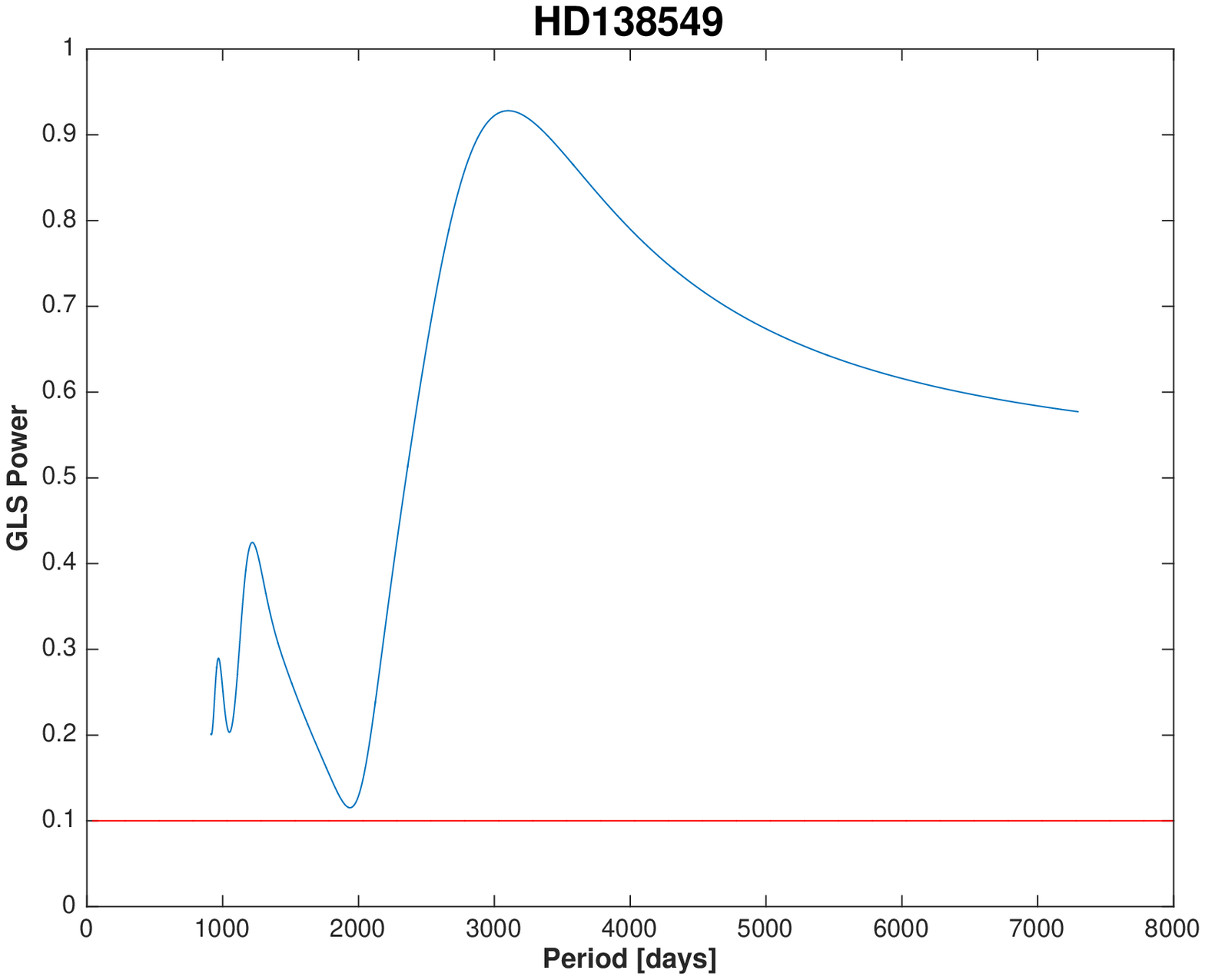}
        \end{subfigure}%
        \begin{subfigure}[b]{0.25\textwidth}
                \includegraphics[width=\linewidth]{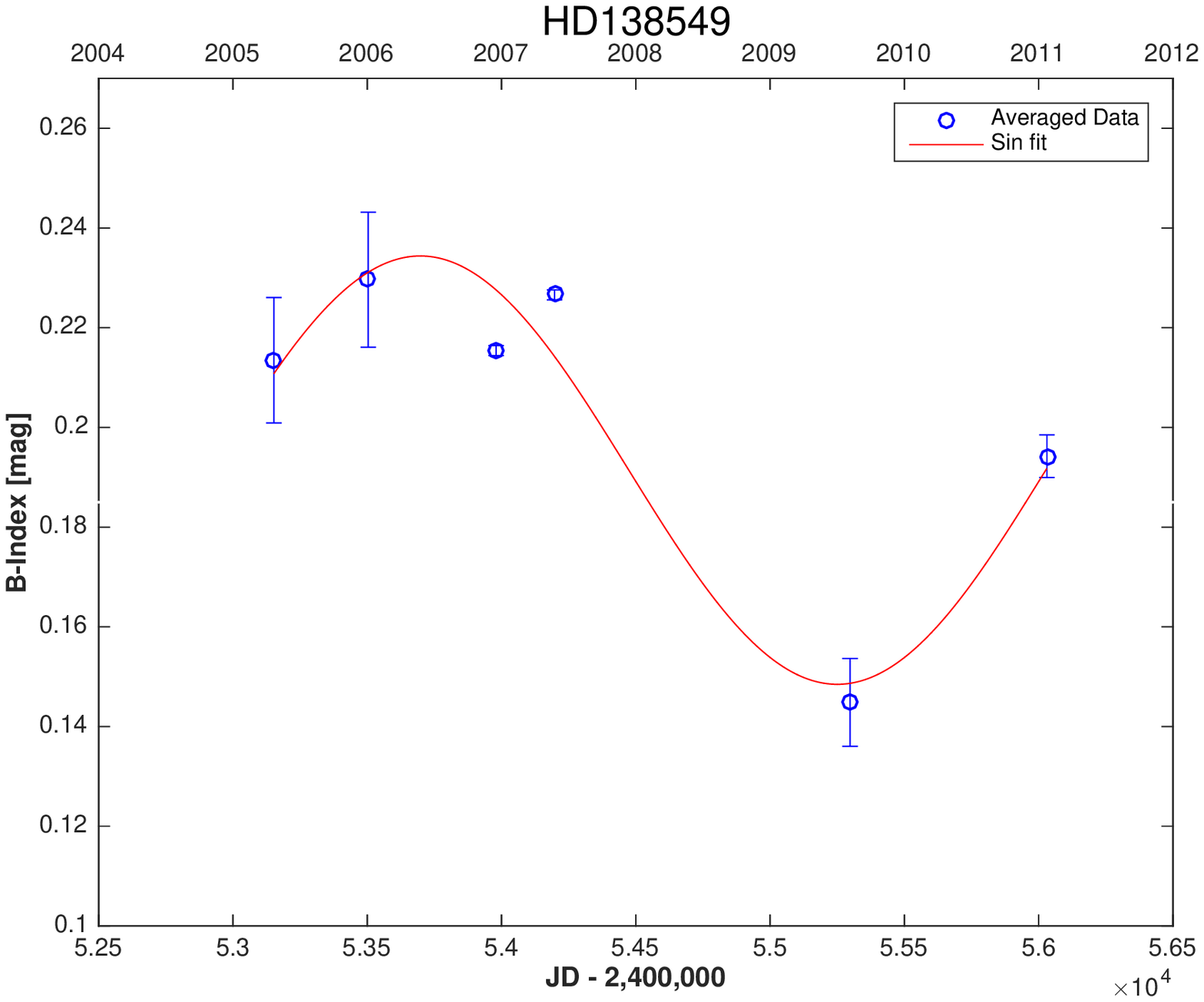}
        \end{subfigure}
        \begin{subfigure}[b]{0.25\textwidth}
                \includegraphics[width=\linewidth]{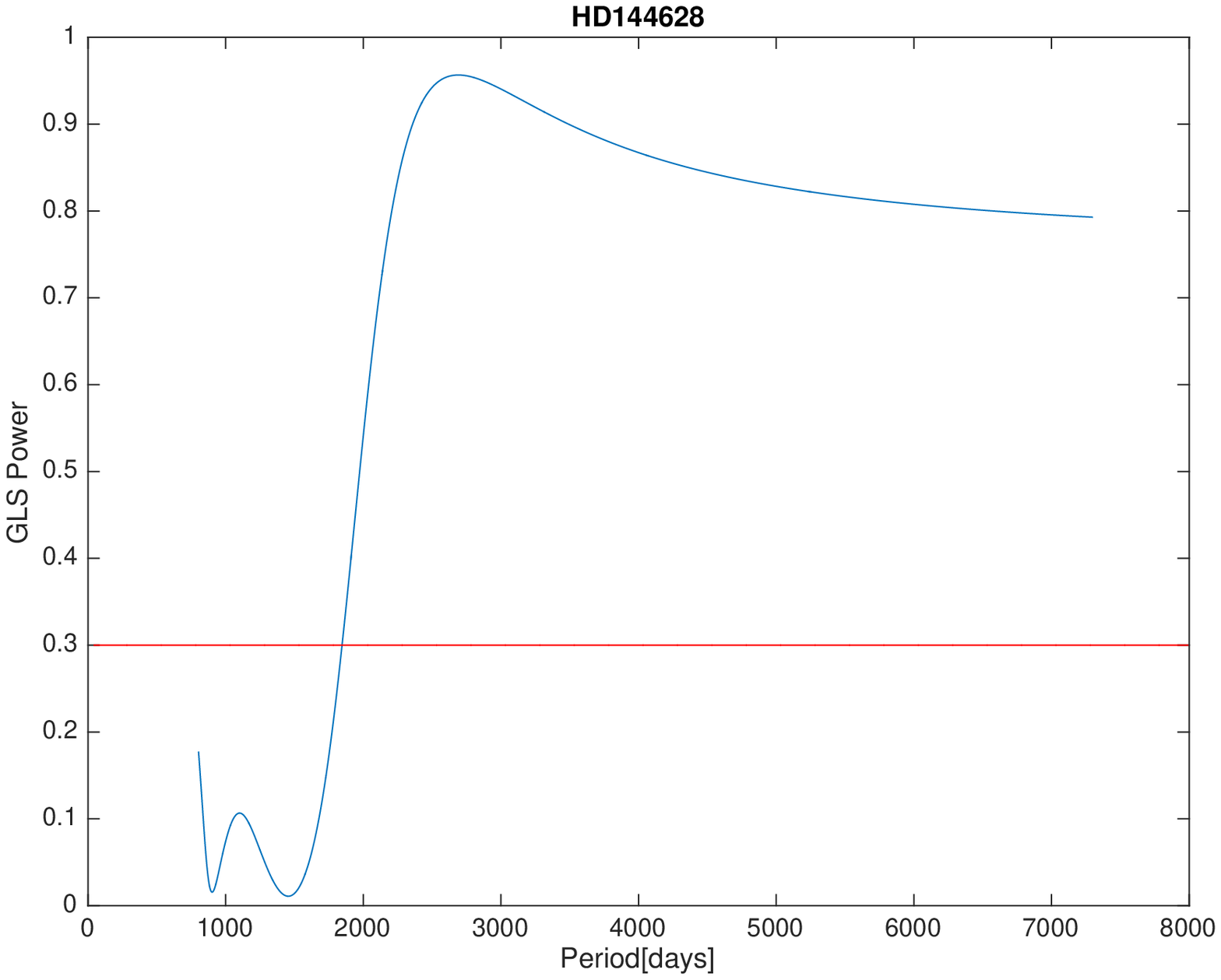}
        \end{subfigure}%
        \begin{subfigure}[b]{0.25\textwidth}
                \includegraphics[width=\linewidth]{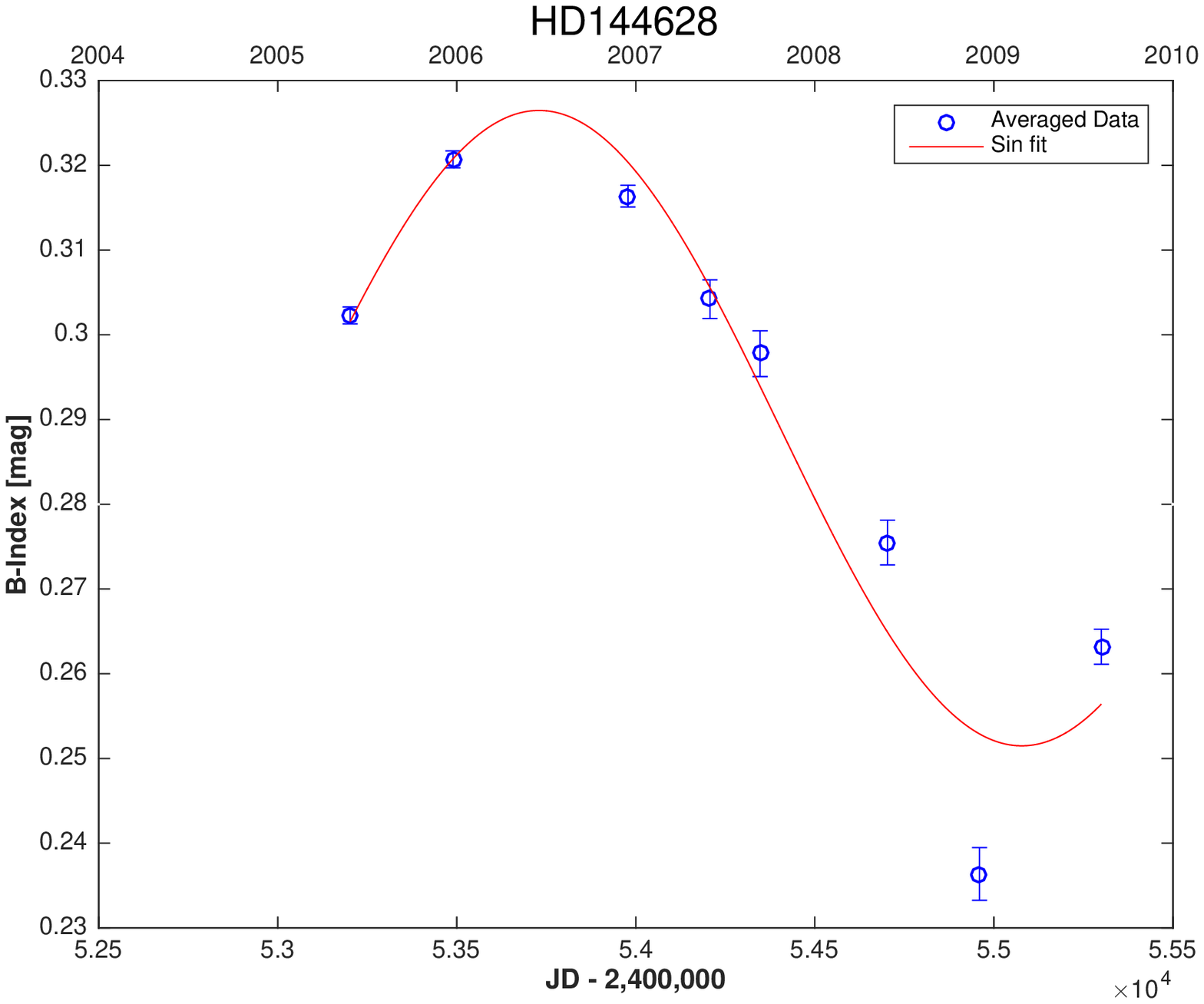}
        \end{subfigure}
        \begin{subfigure}[b]{0.25\textwidth}
                \includegraphics[width=\linewidth]{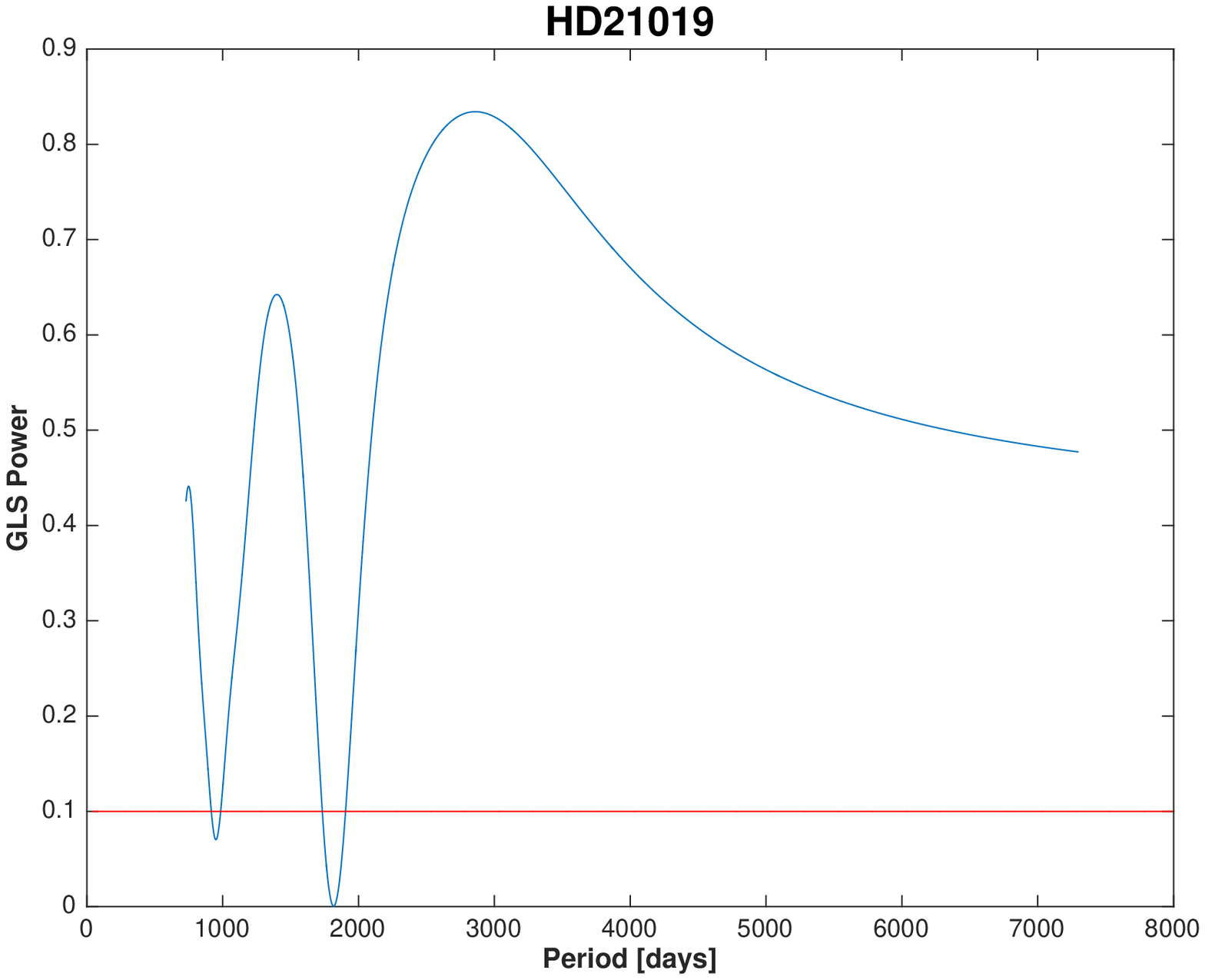}
        \end{subfigure}%
        \begin{subfigure}[b]{0.25\textwidth}
                \includegraphics[width=\linewidth]{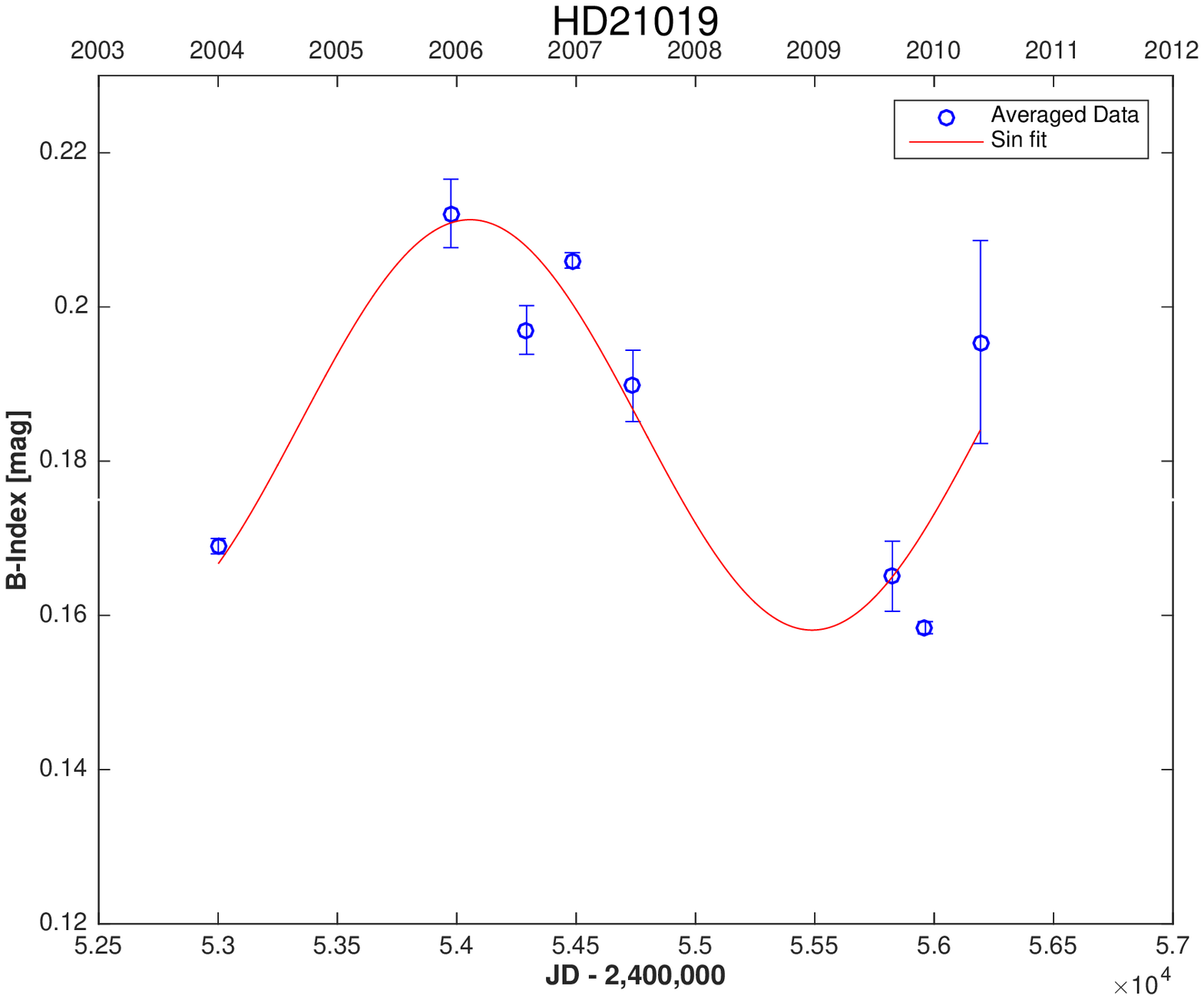}
        \end{subfigure} 
           \begin{subfigure}[b]{0.25\textwidth}
                \includegraphics[width=\linewidth]{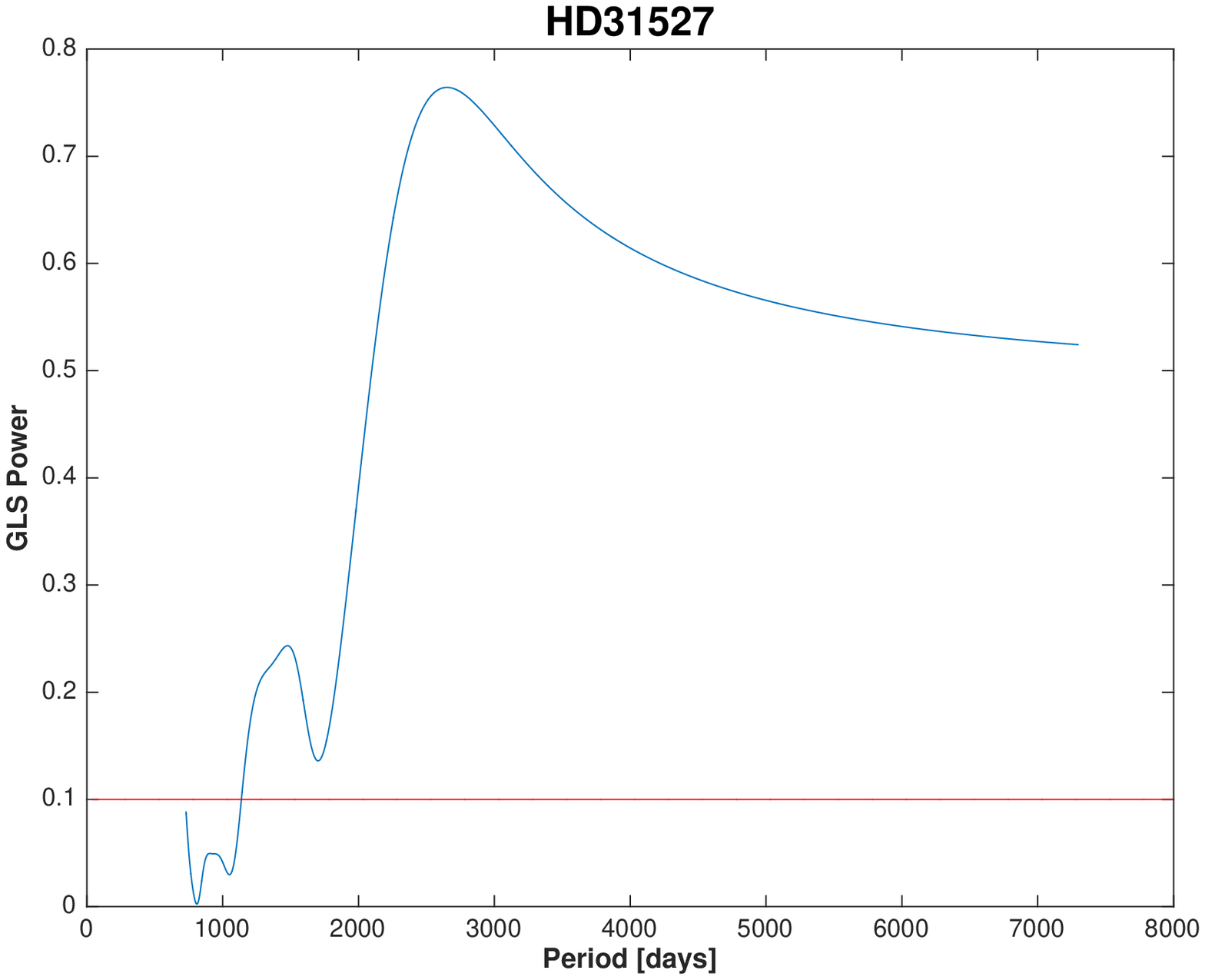}
        \end{subfigure}%
        \begin{subfigure}[b]{0.25\textwidth}
                \includegraphics[width=\linewidth]{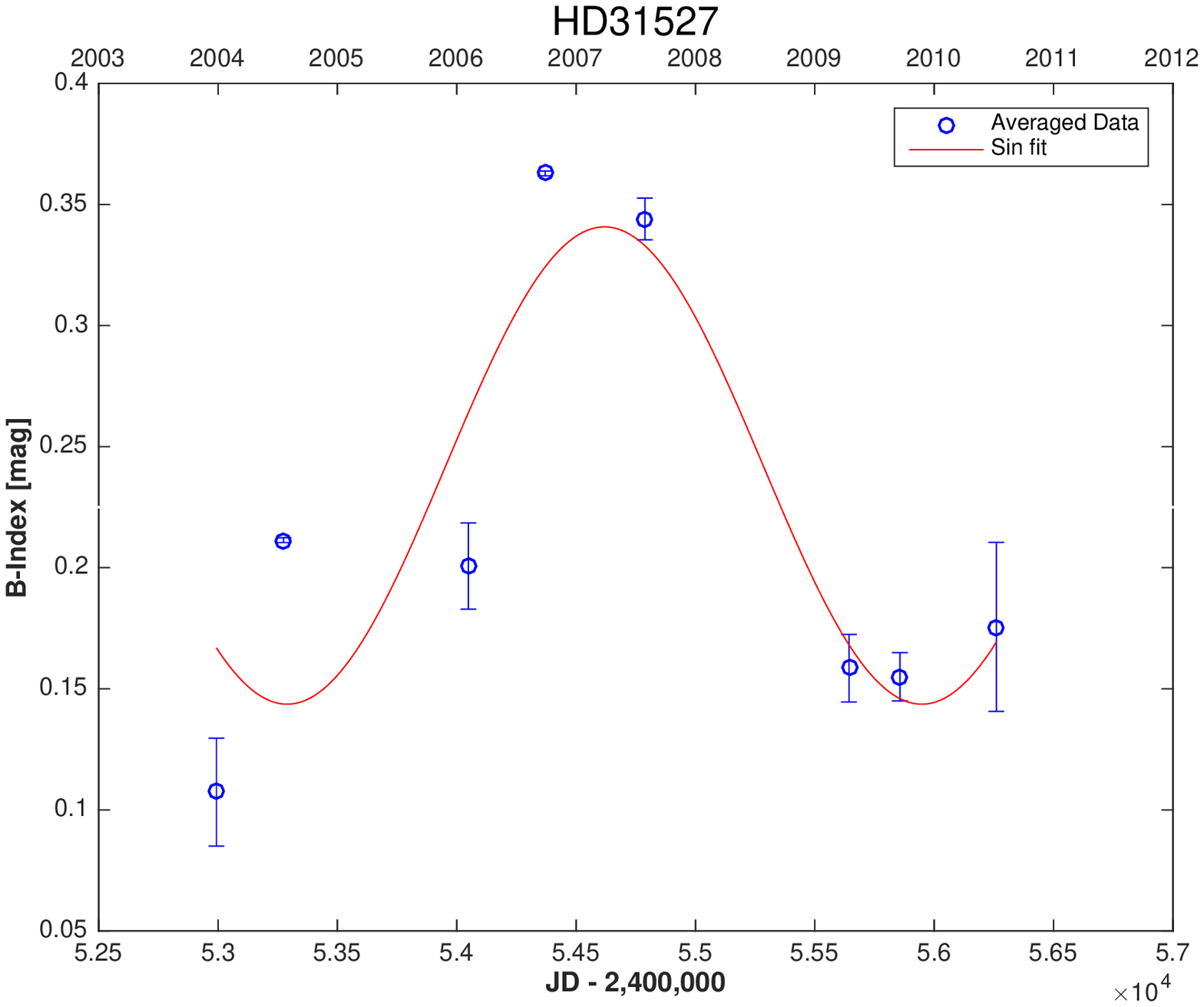}
        \end{subfigure}
        \begin{subfigure}[b]{0.25\textwidth}
                \includegraphics[width=\linewidth]{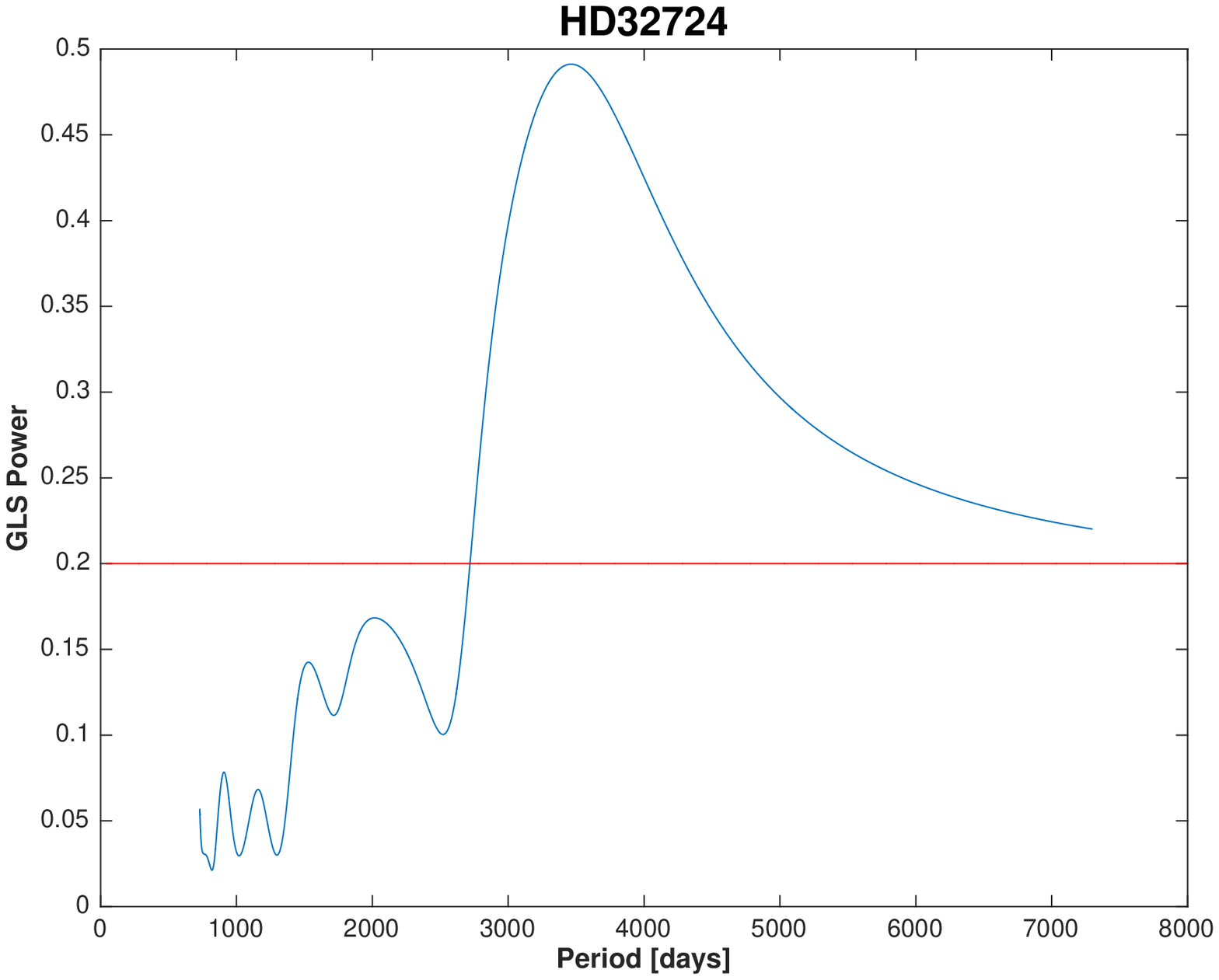}
        \end{subfigure}%
        \begin{subfigure}[b]{0.25\textwidth}
                \includegraphics[width=\linewidth]{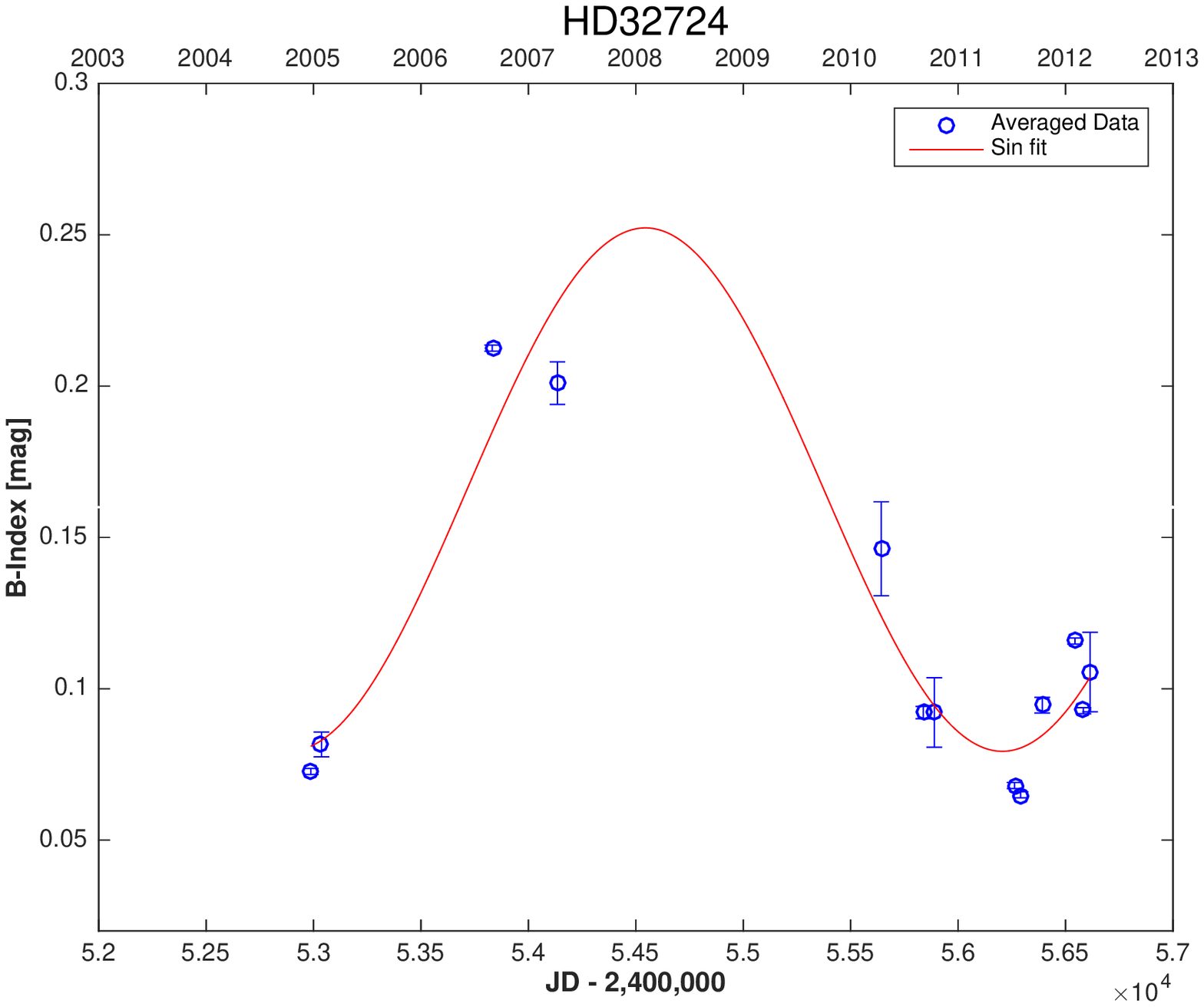}
        \end{subfigure} 
   \begin{subfigure}[b]{0.25\textwidth}
                \includegraphics[width=\linewidth]{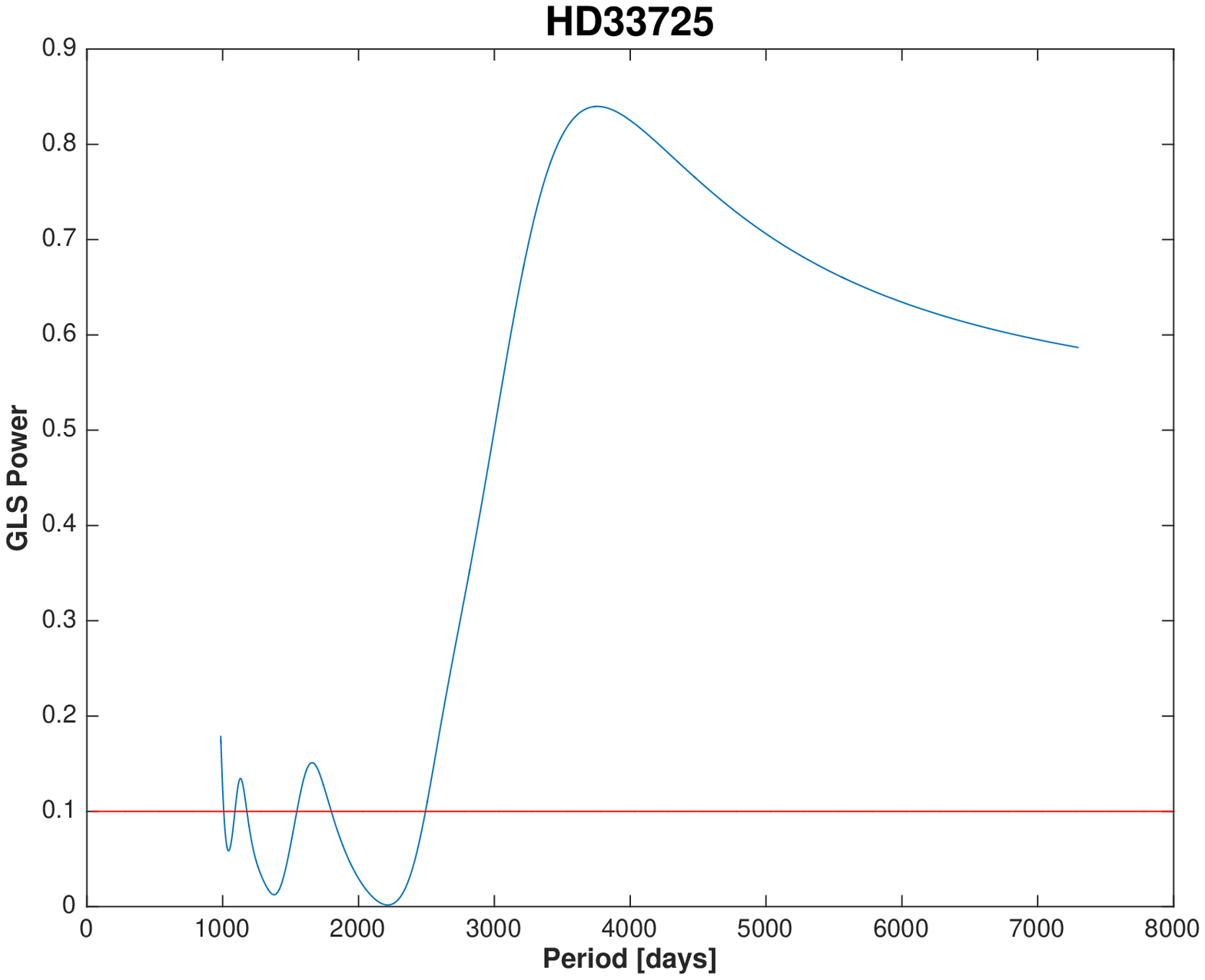}
        \end{subfigure}%
        \begin{subfigure}[b]{0.25\textwidth}
                \includegraphics[width=\linewidth]{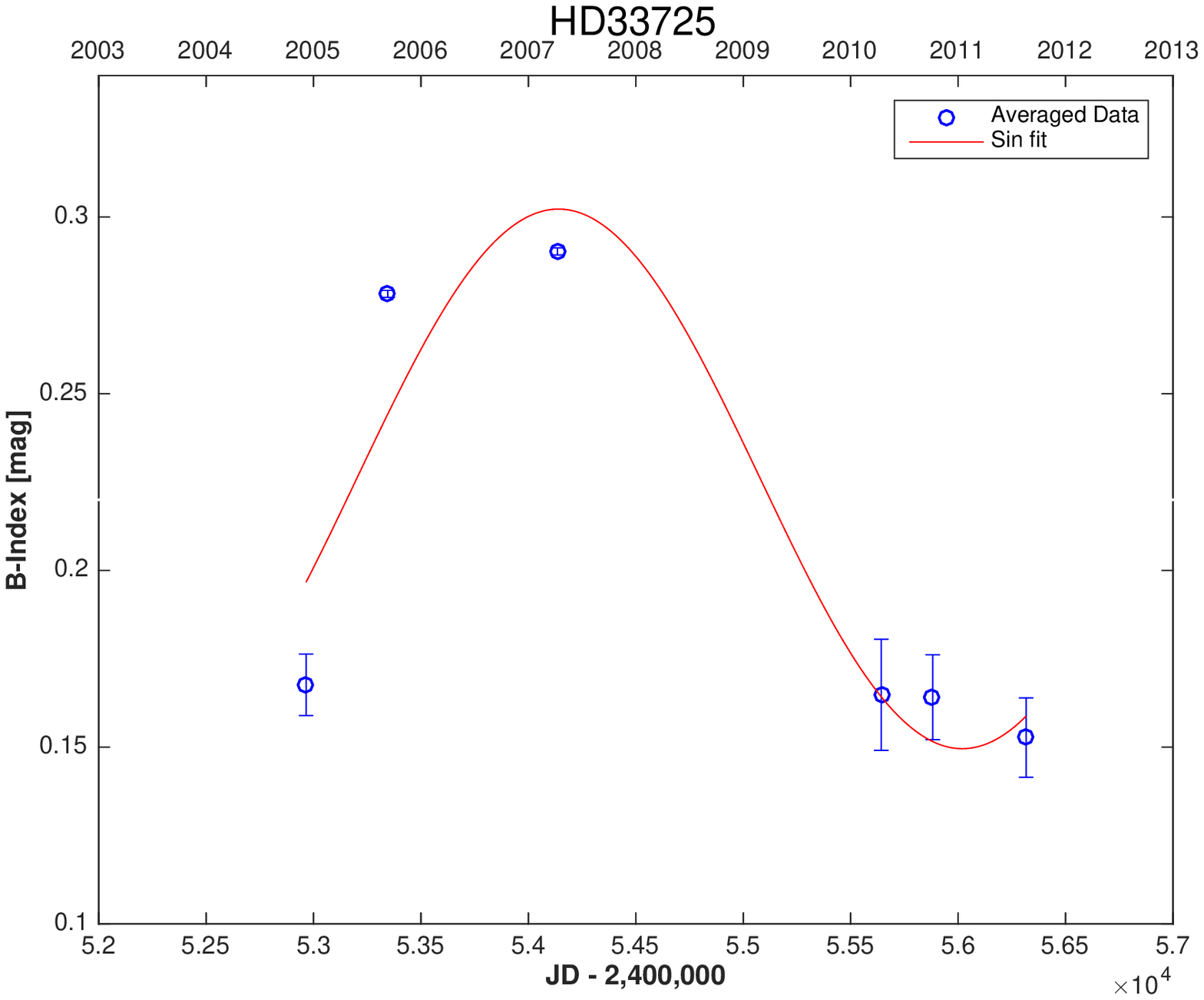}
        \end{subfigure}
        \begin{subfigure}[b]{0.25\textwidth}
                \includegraphics[width=\linewidth]{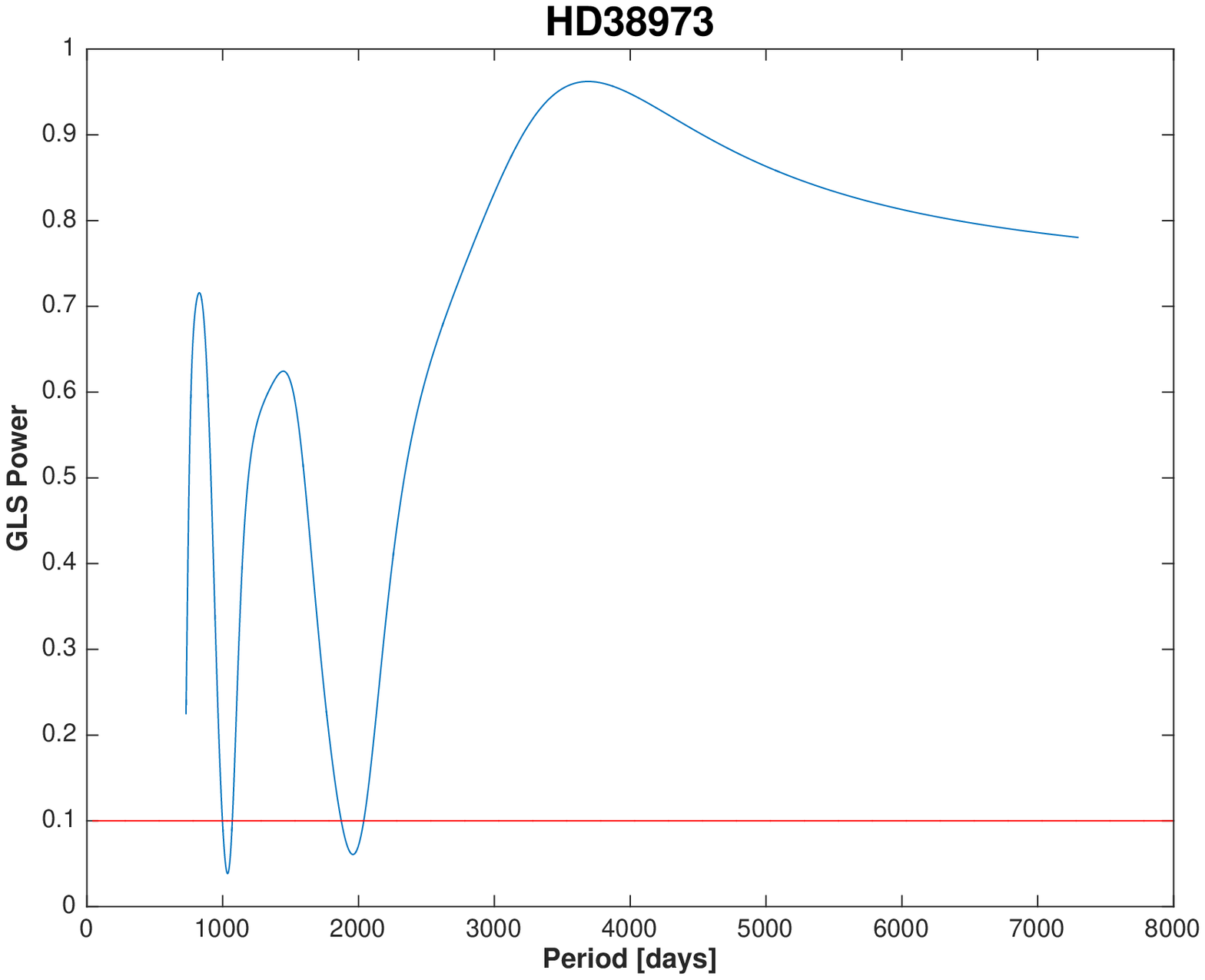}
        \end{subfigure}%
        \begin{subfigure}[b]{0.25\textwidth}
                \includegraphics[width=\linewidth]{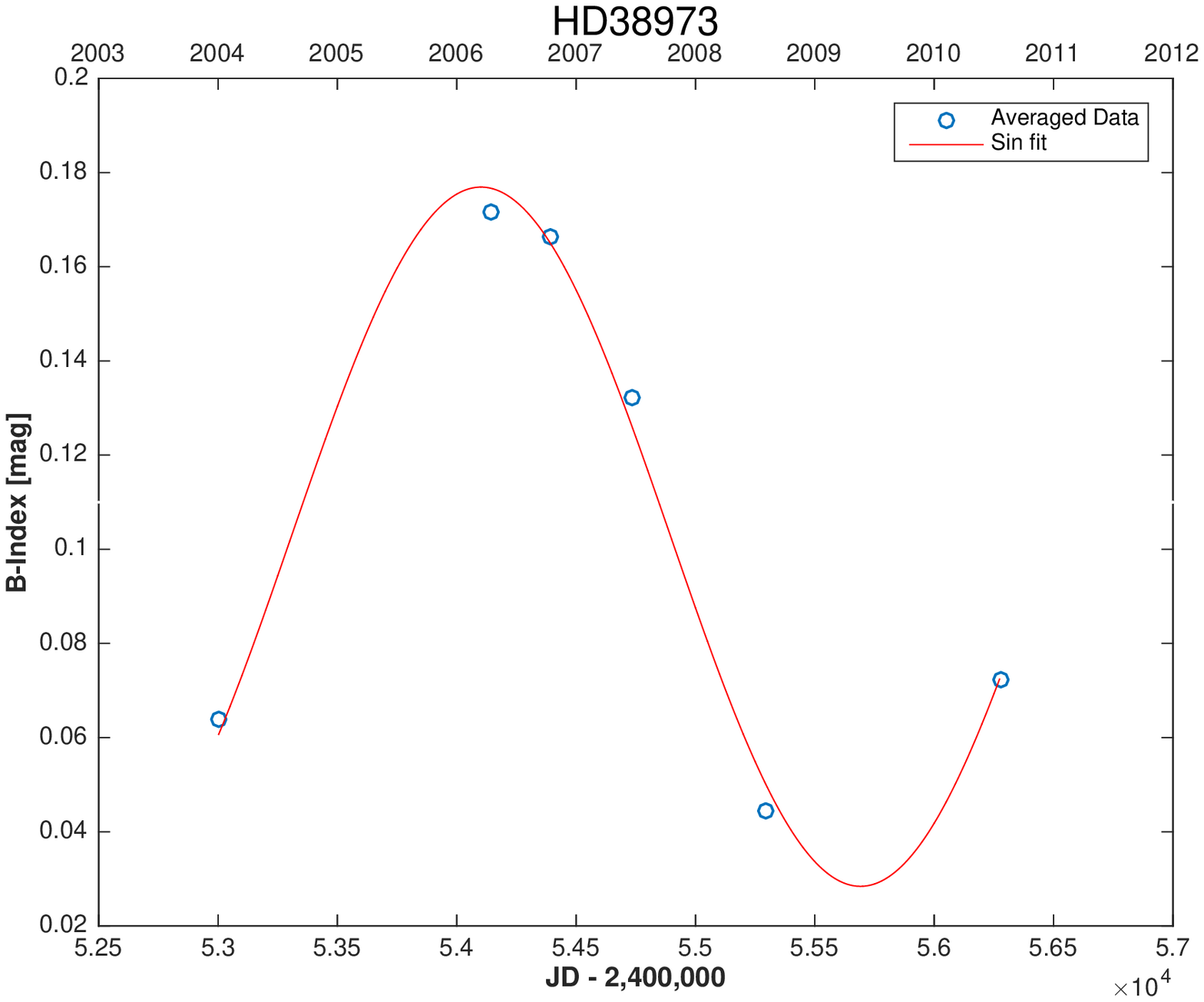}
        \end{subfigure} 
         \begin{subfigure}[b]{0.25\textwidth}
                 \includegraphics[width=\linewidth]{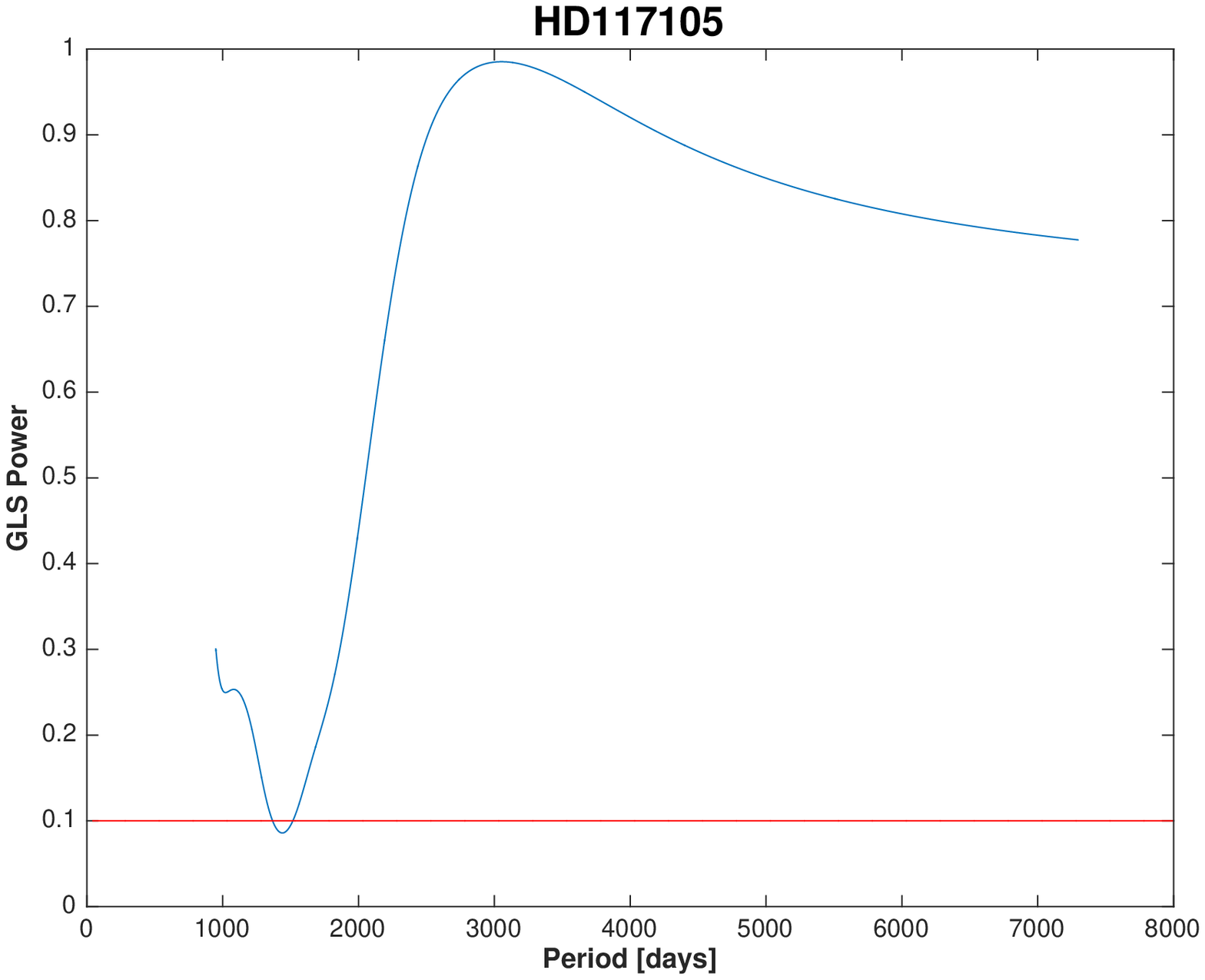}
        \end{subfigure}%
        \begin{subfigure}[b]{0.25\textwidth}
                \includegraphics[width=\linewidth]{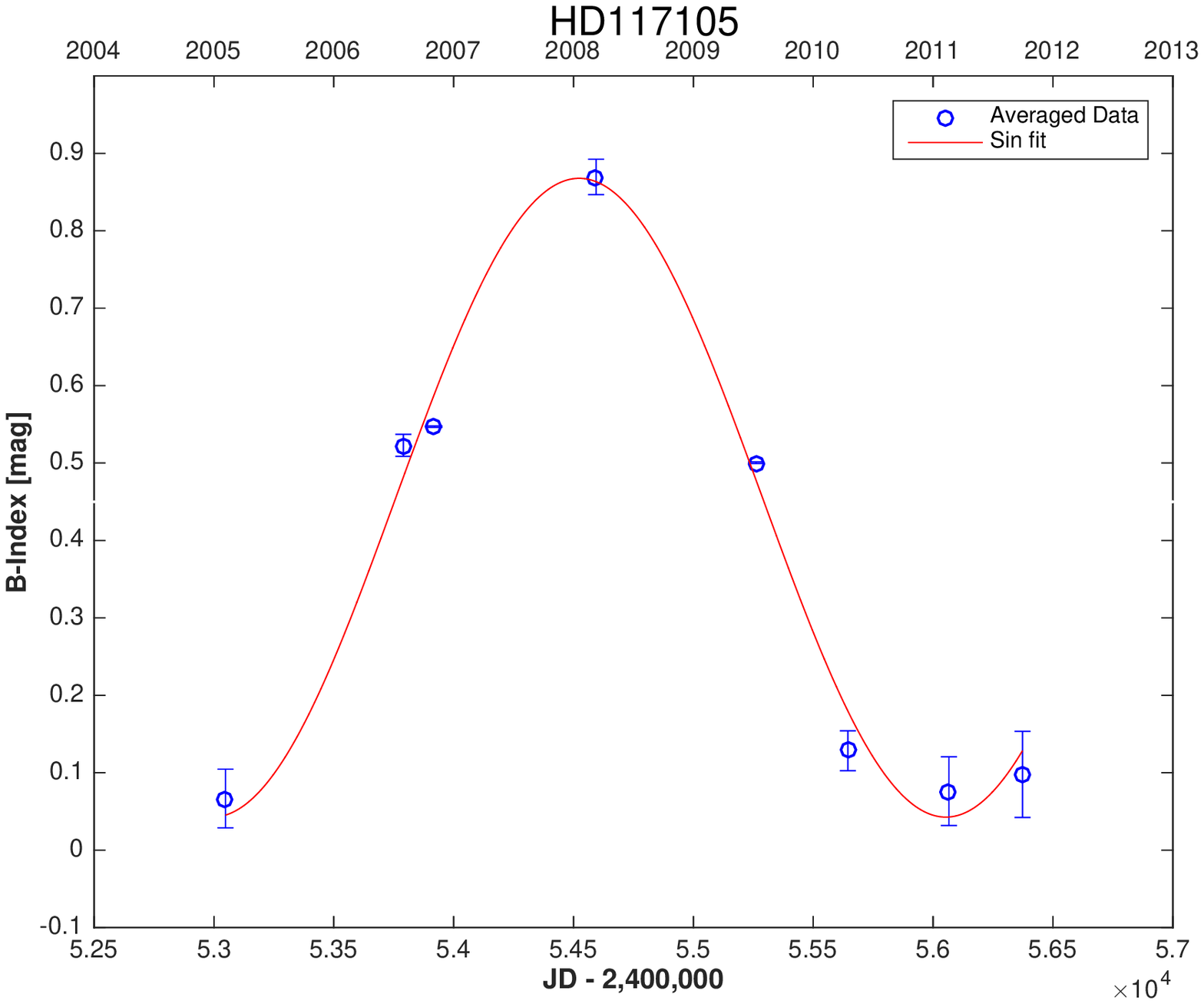}
        \end{subfigure}
        \begin{subfigure}[b]{0.25\textwidth}
                \includegraphics[width=\linewidth]{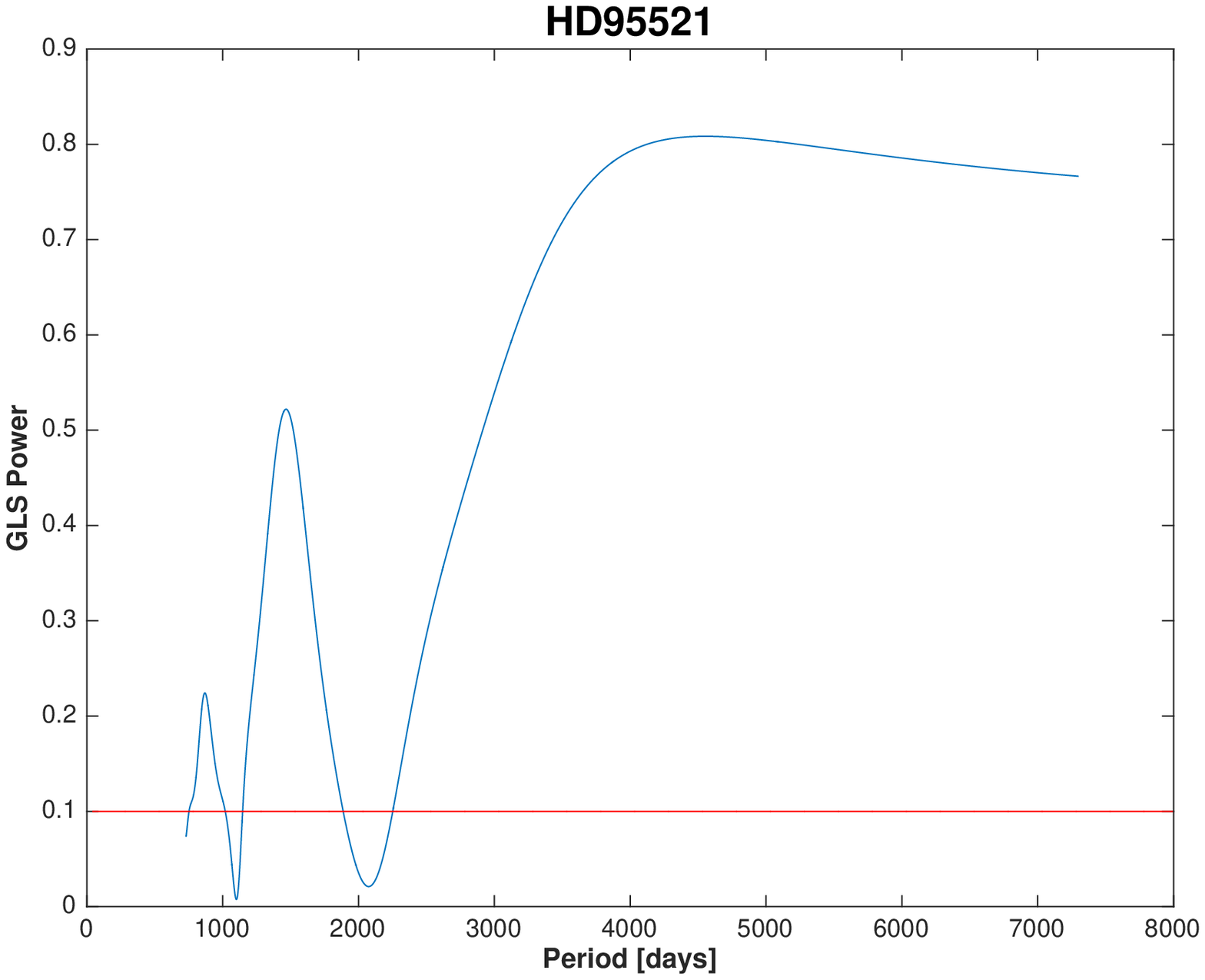}
        \end{subfigure}%
        \begin{subfigure}[b]{0.25\textwidth}
                \includegraphics[width=\linewidth]{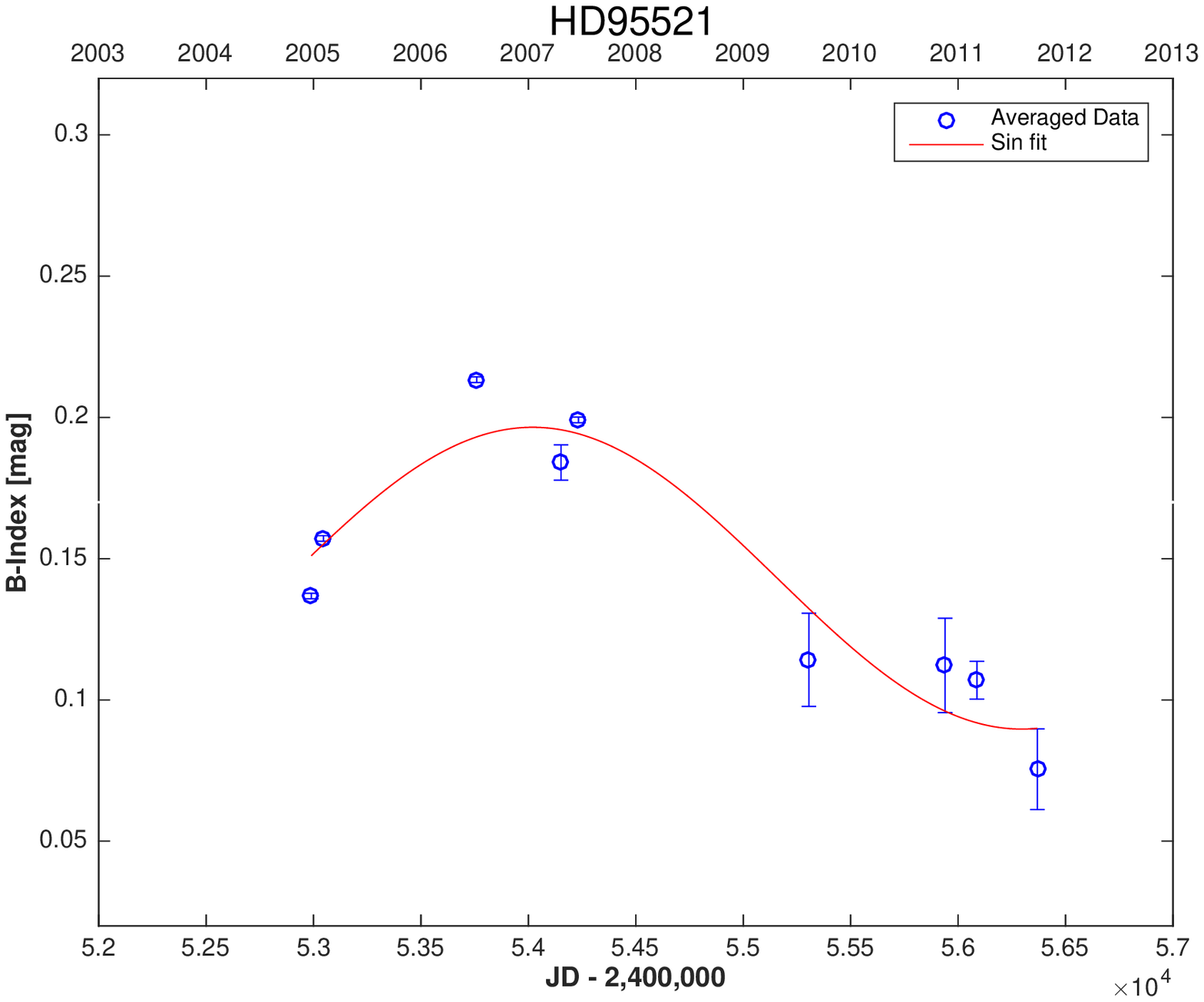}
        \end{subfigure} 
 \begin{subfigure}[b]{0.25\textwidth}
                \includegraphics[width=\linewidth]{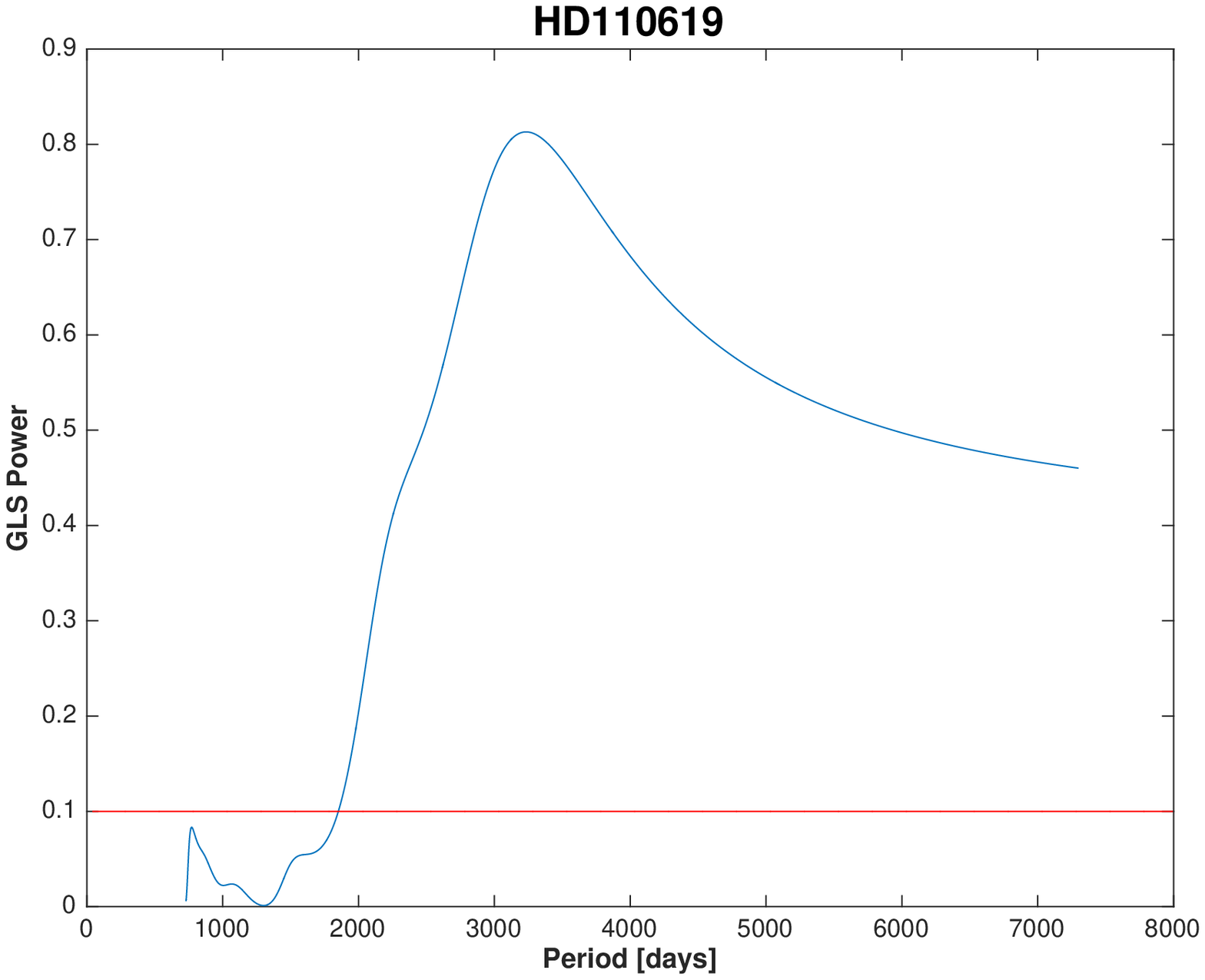}
        \end{subfigure}%
        \begin{subfigure}[b]{0.25\textwidth}
                \includegraphics[width=\linewidth]{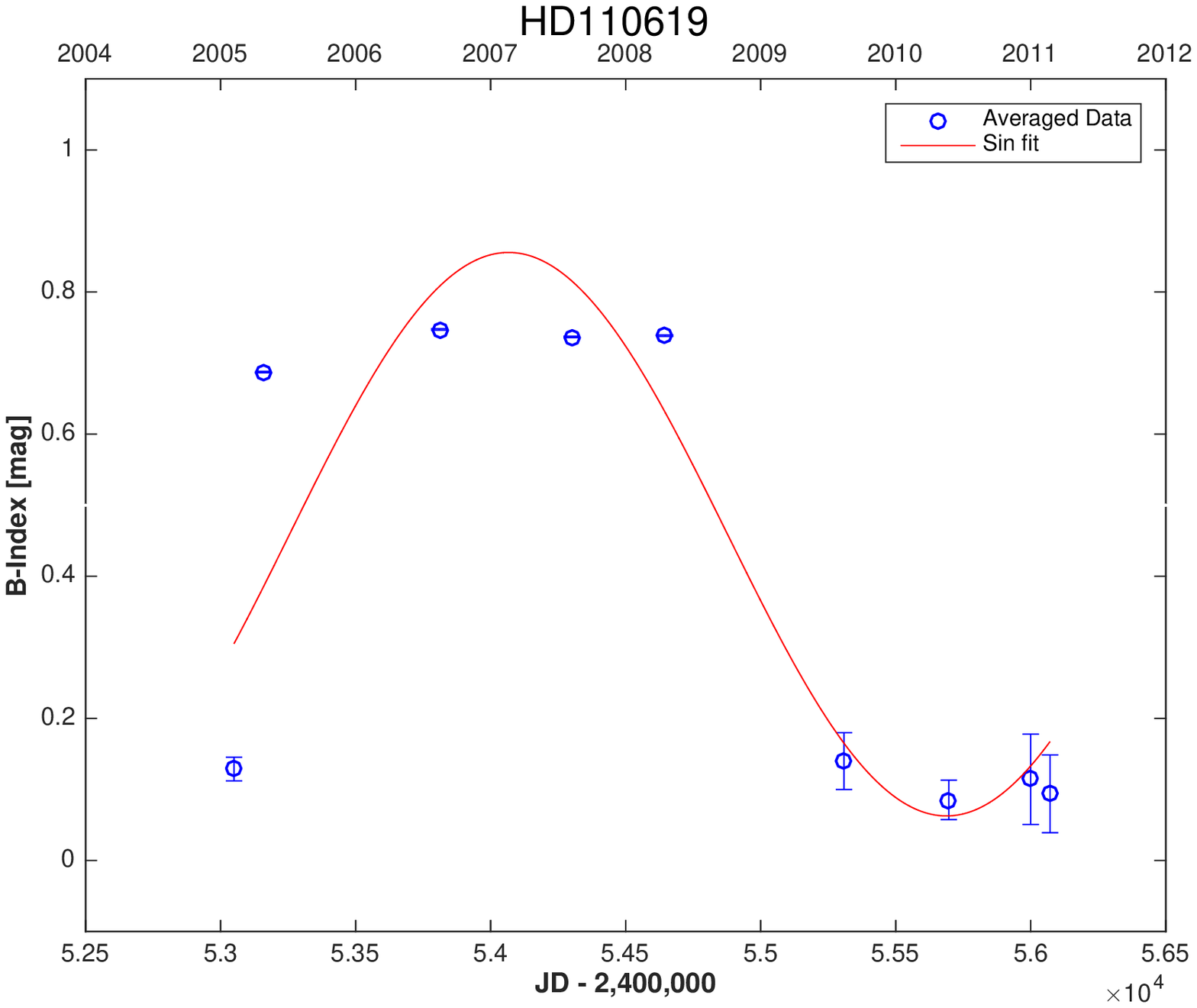}
        \end{subfigure}
        \begin{subfigure}[b]{0.25\textwidth}
                \includegraphics[width=\linewidth]{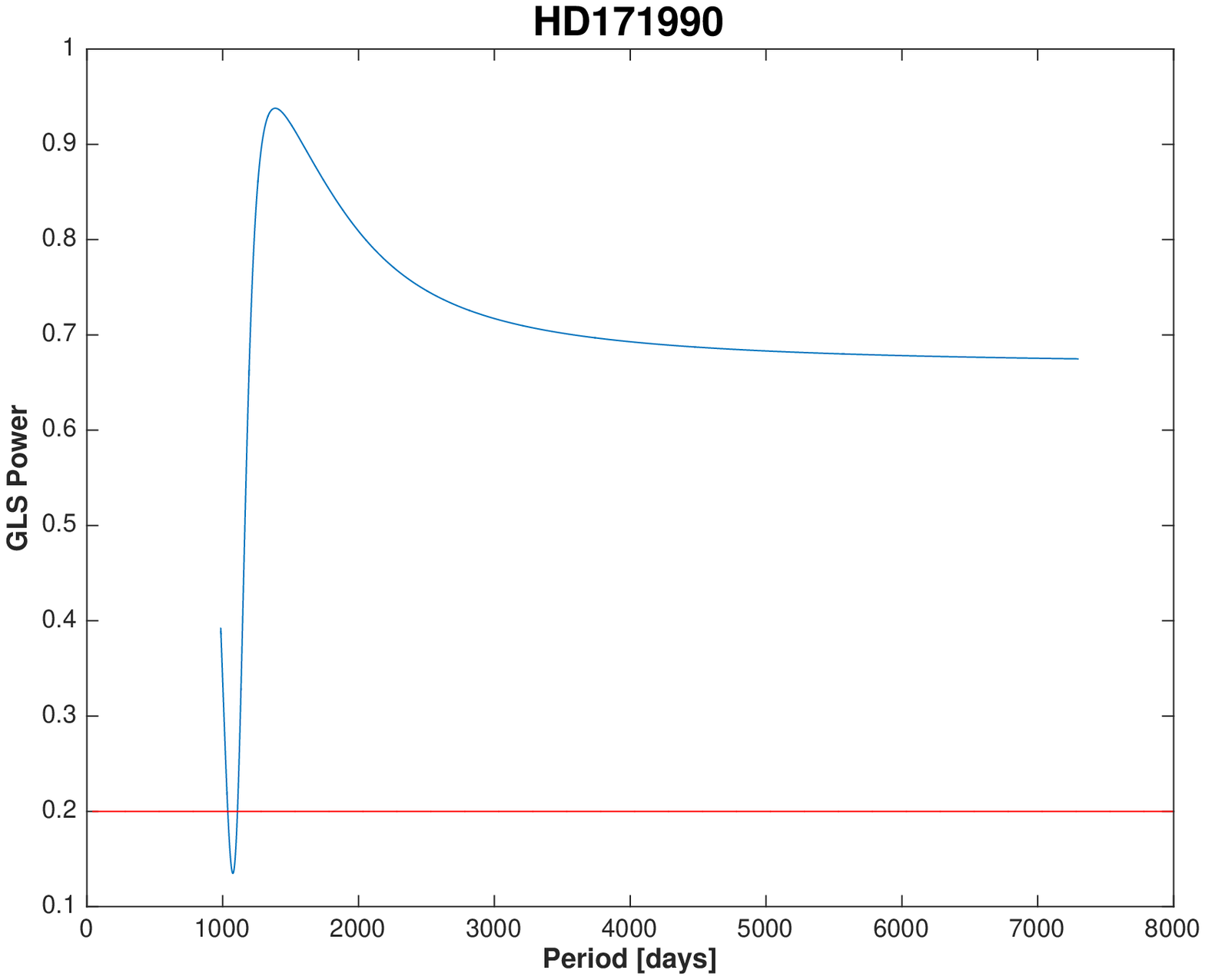}
        \end{subfigure}%
        \begin{subfigure}[b]{0.25\textwidth}
                \includegraphics[width=\linewidth]{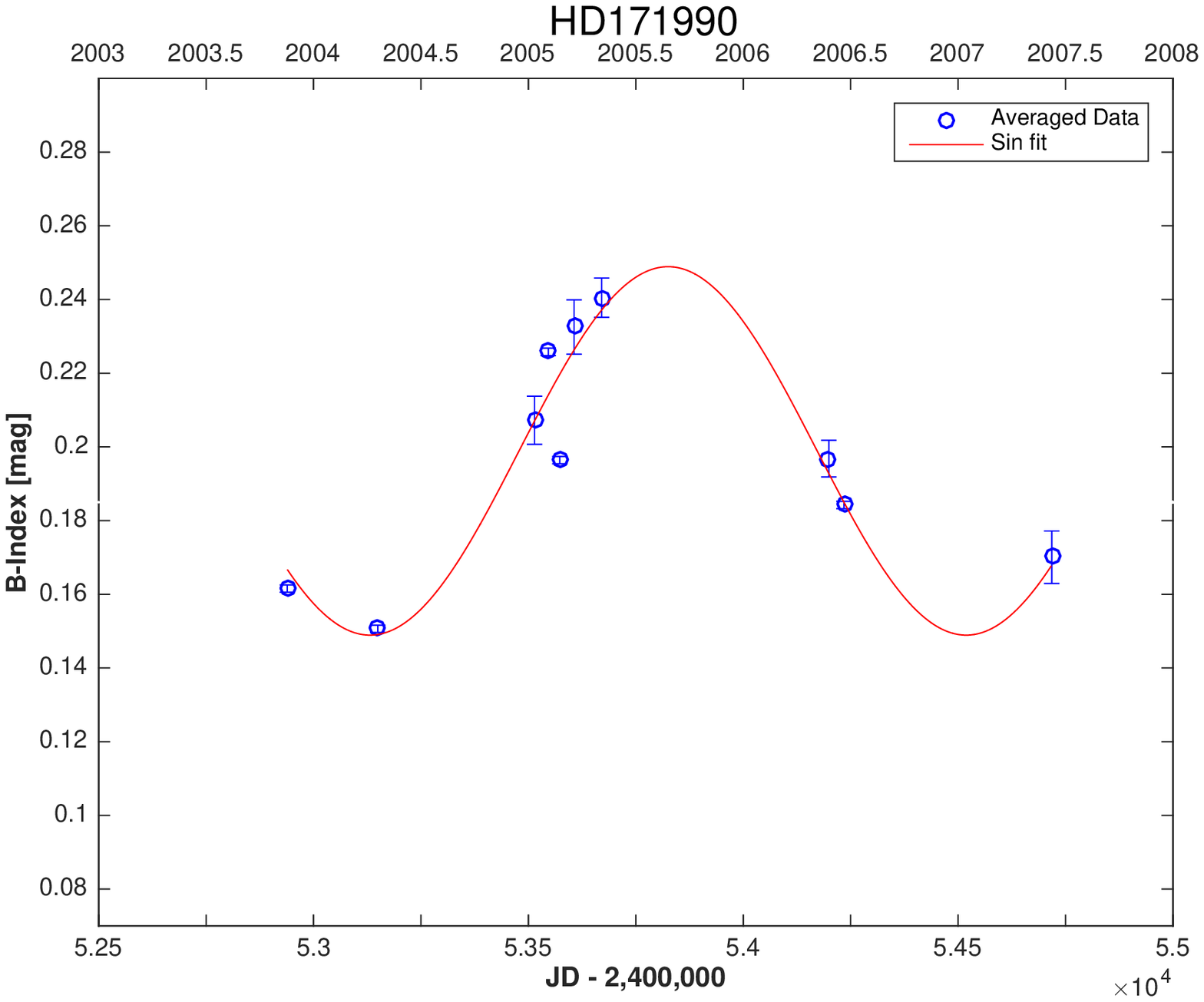}
        \end{subfigure} 
        
\end{figure}

\begin{figure}
 \begin{subfigure}[b]{0.25\textwidth}
                \includegraphics[width=\linewidth]{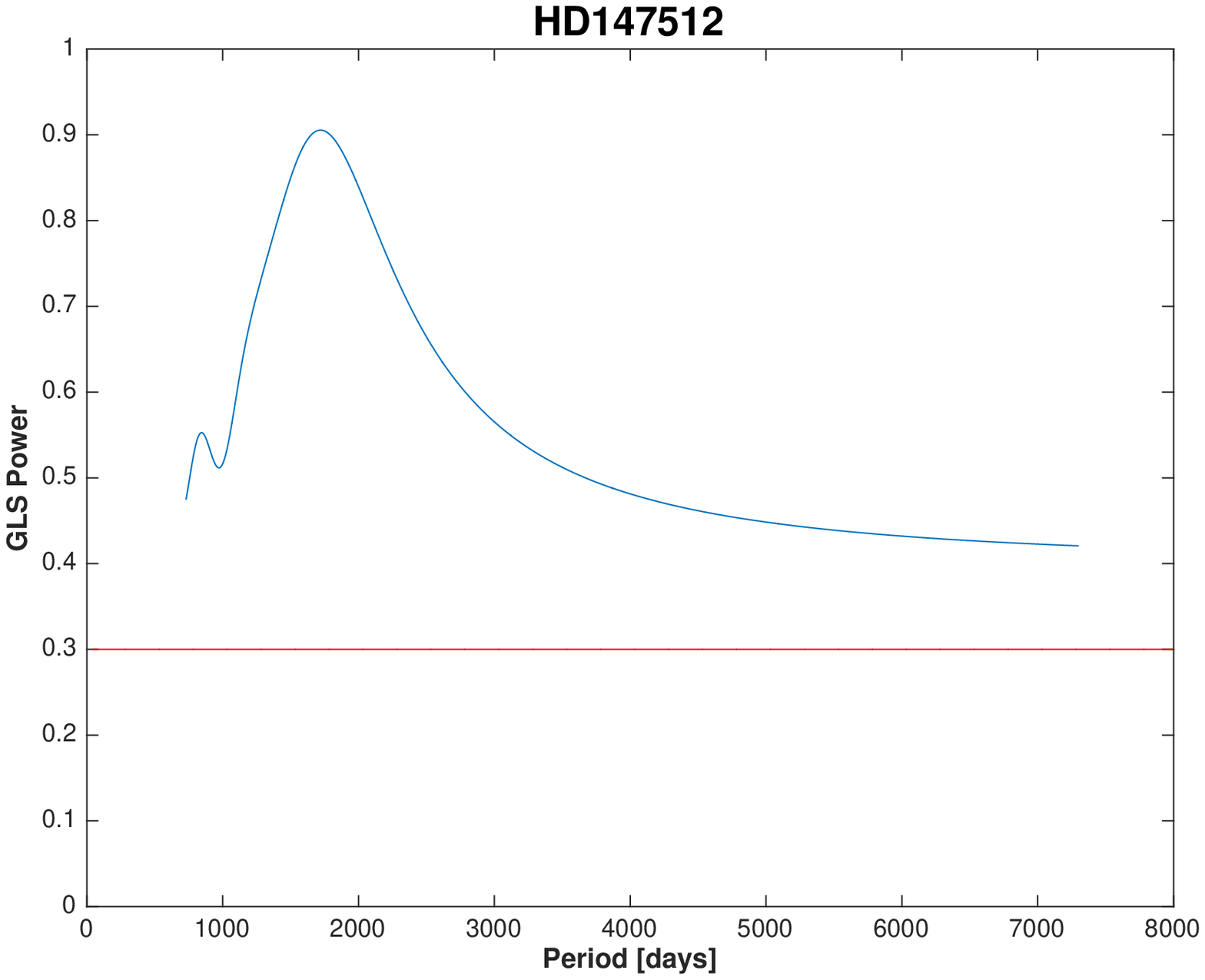}
        \end{subfigure}%
        \begin{subfigure}[b]{0.25\textwidth}
                \includegraphics[width=\linewidth]{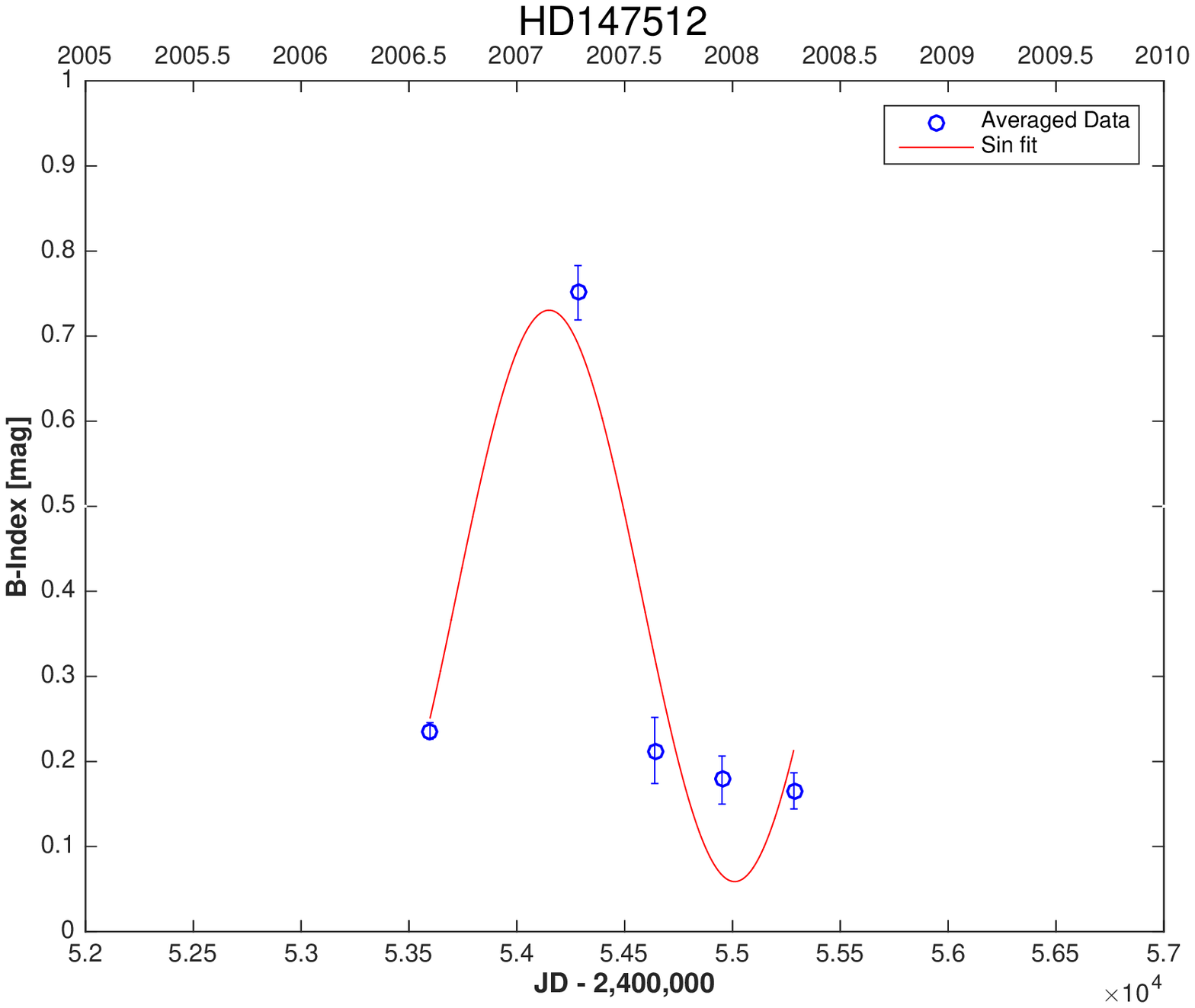}
        \end{subfigure} 
 \begin{subfigure}[b]{0.25\textwidth}
                \includegraphics[width=\linewidth]{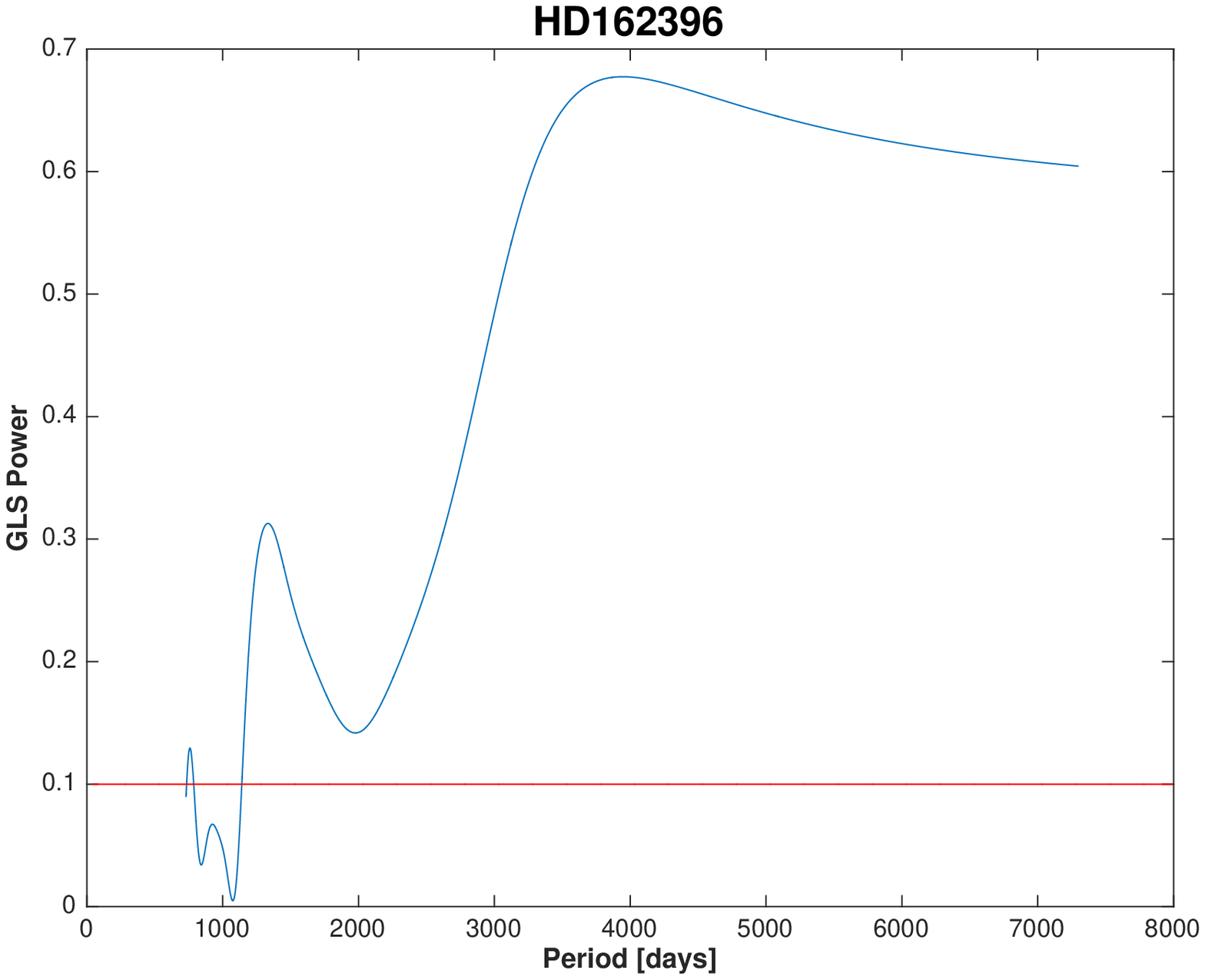}
        \end{subfigure}%
        \begin{subfigure}[b]{0.25\textwidth}
                \includegraphics[width=\linewidth]{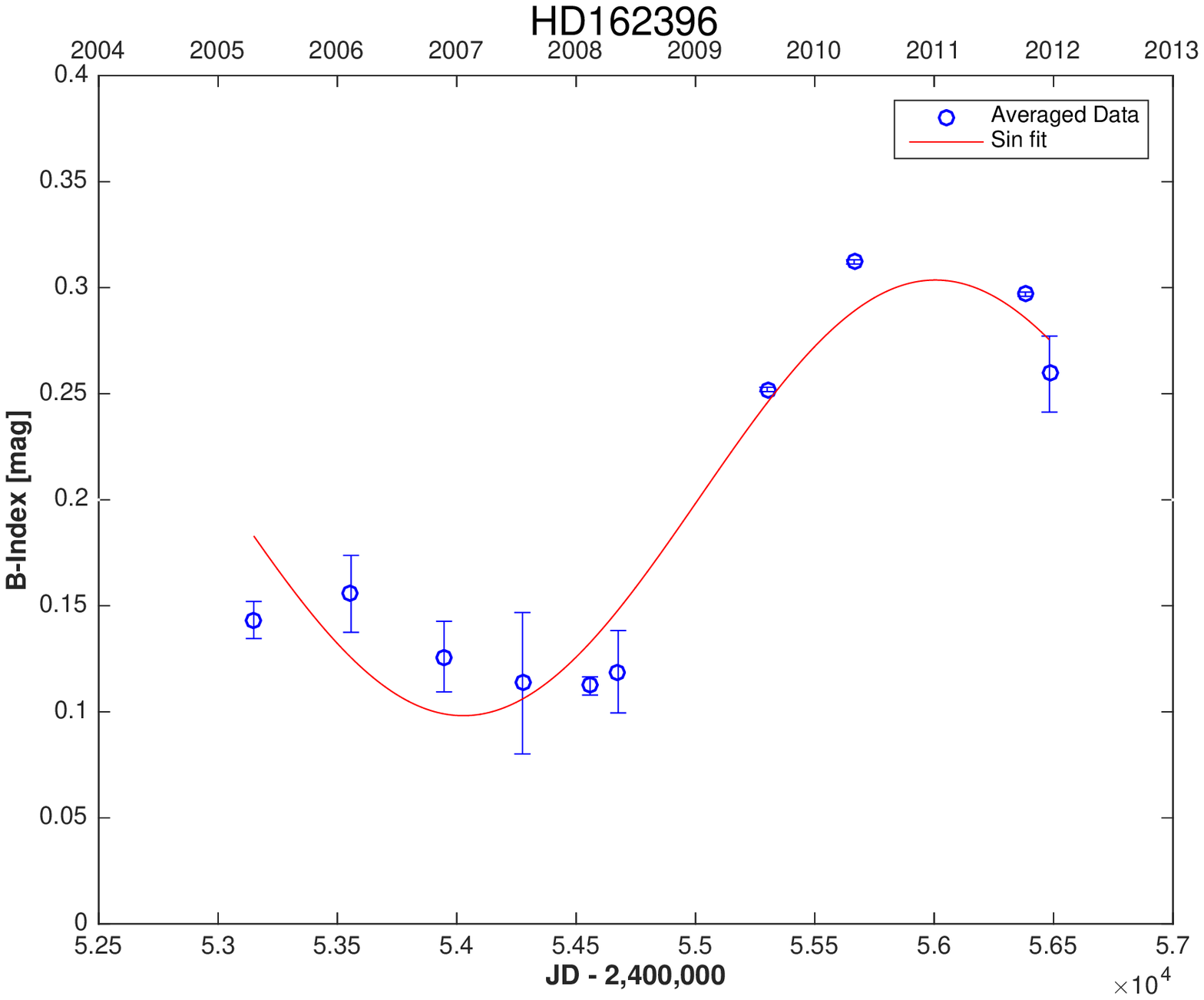}
        \end{subfigure} 
\begin{subfigure}[b]{0.25\textwidth}
                \includegraphics[width=\linewidth]{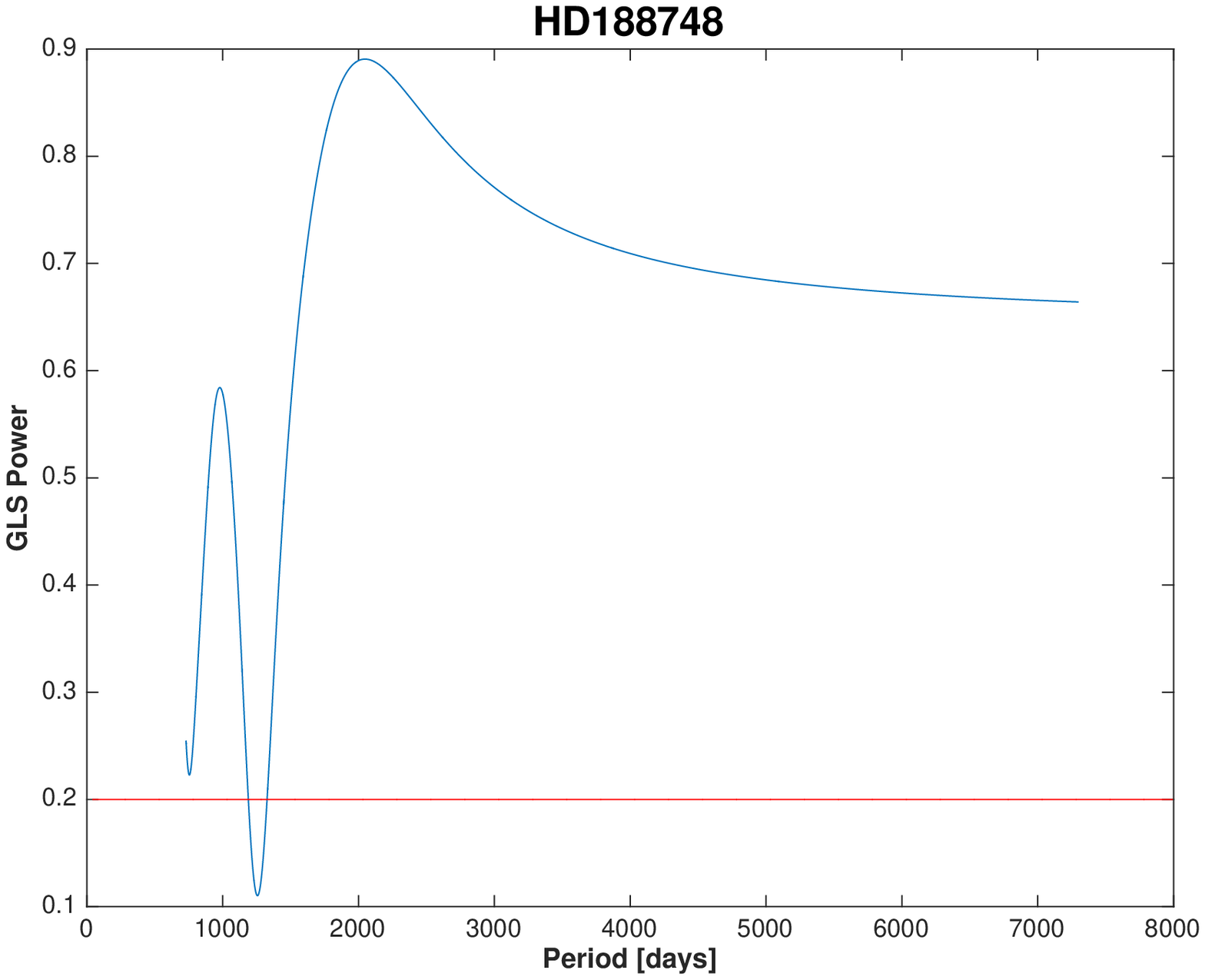}
        \end{subfigure}%
        \begin{subfigure}[b]{0.25\textwidth}
                \includegraphics[width=\linewidth]{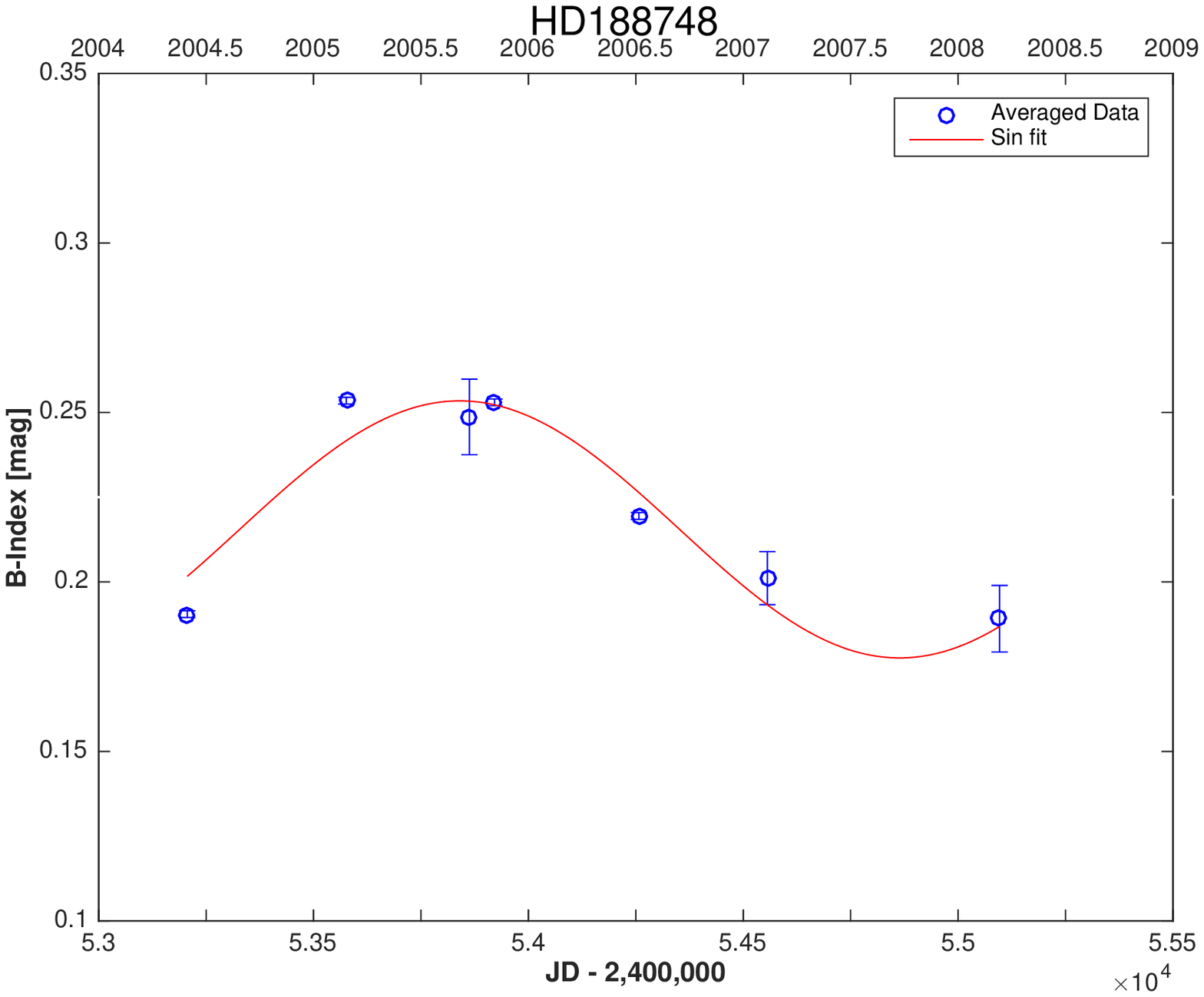}
        \end{subfigure} 
 \begin{subfigure}[b]{0.25\textwidth}
                \includegraphics[width=\linewidth]{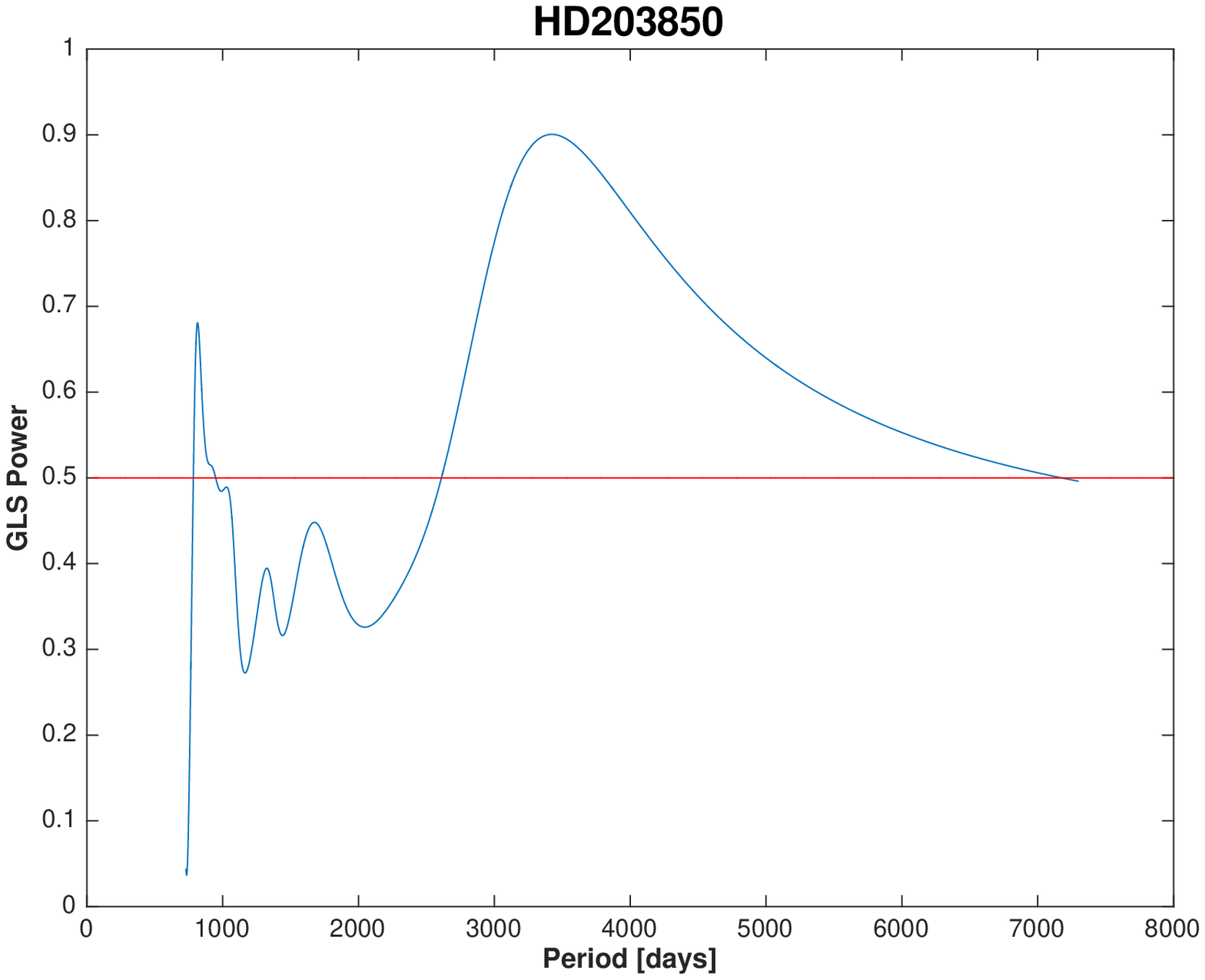}
        \end{subfigure}%
        \begin{subfigure}[b]{0.25\textwidth}
                \includegraphics[width=\linewidth]{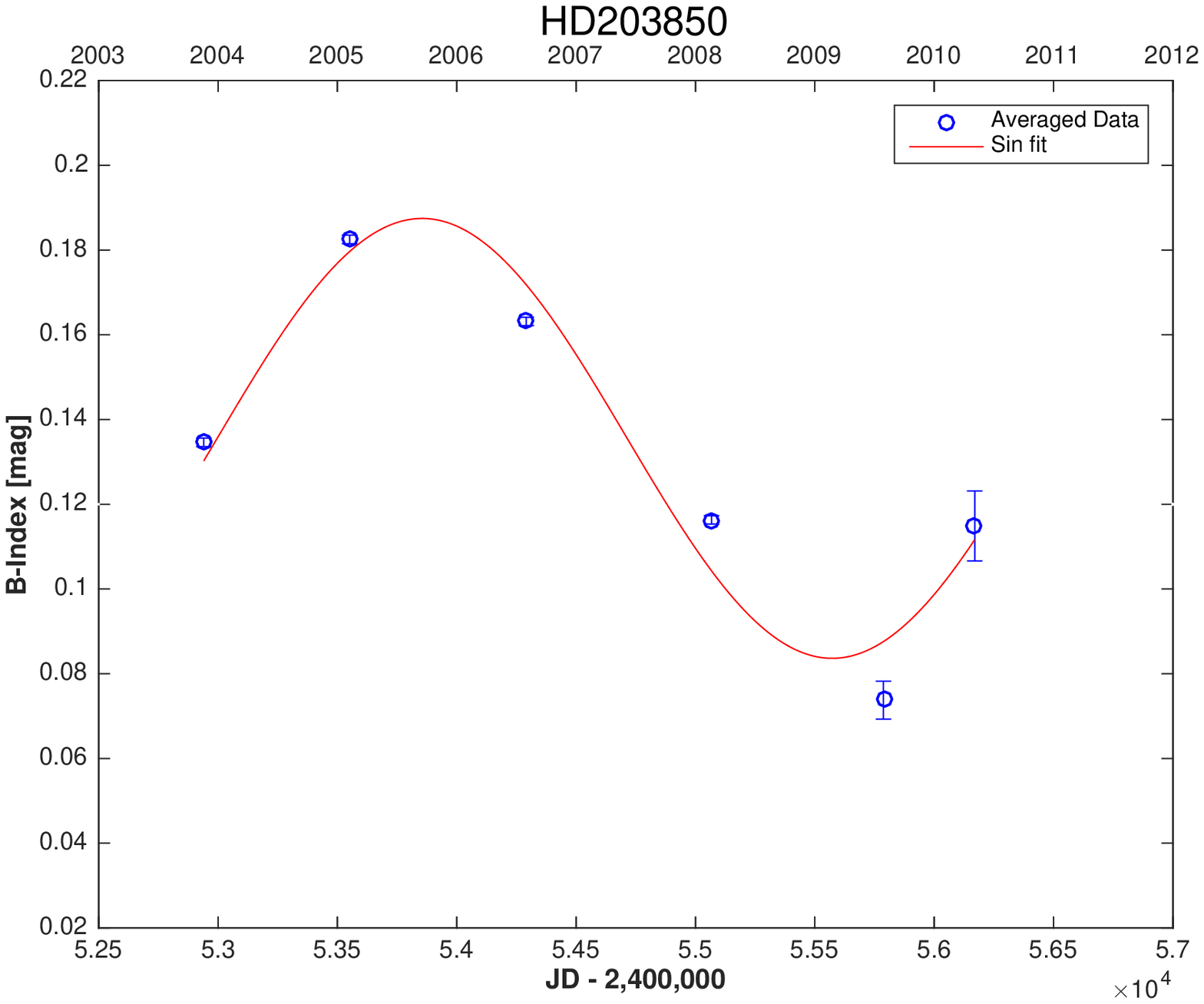}
        \end{subfigure} 
\begin{subfigure}[b]{0.25\textwidth}
                \includegraphics[width=\linewidth]{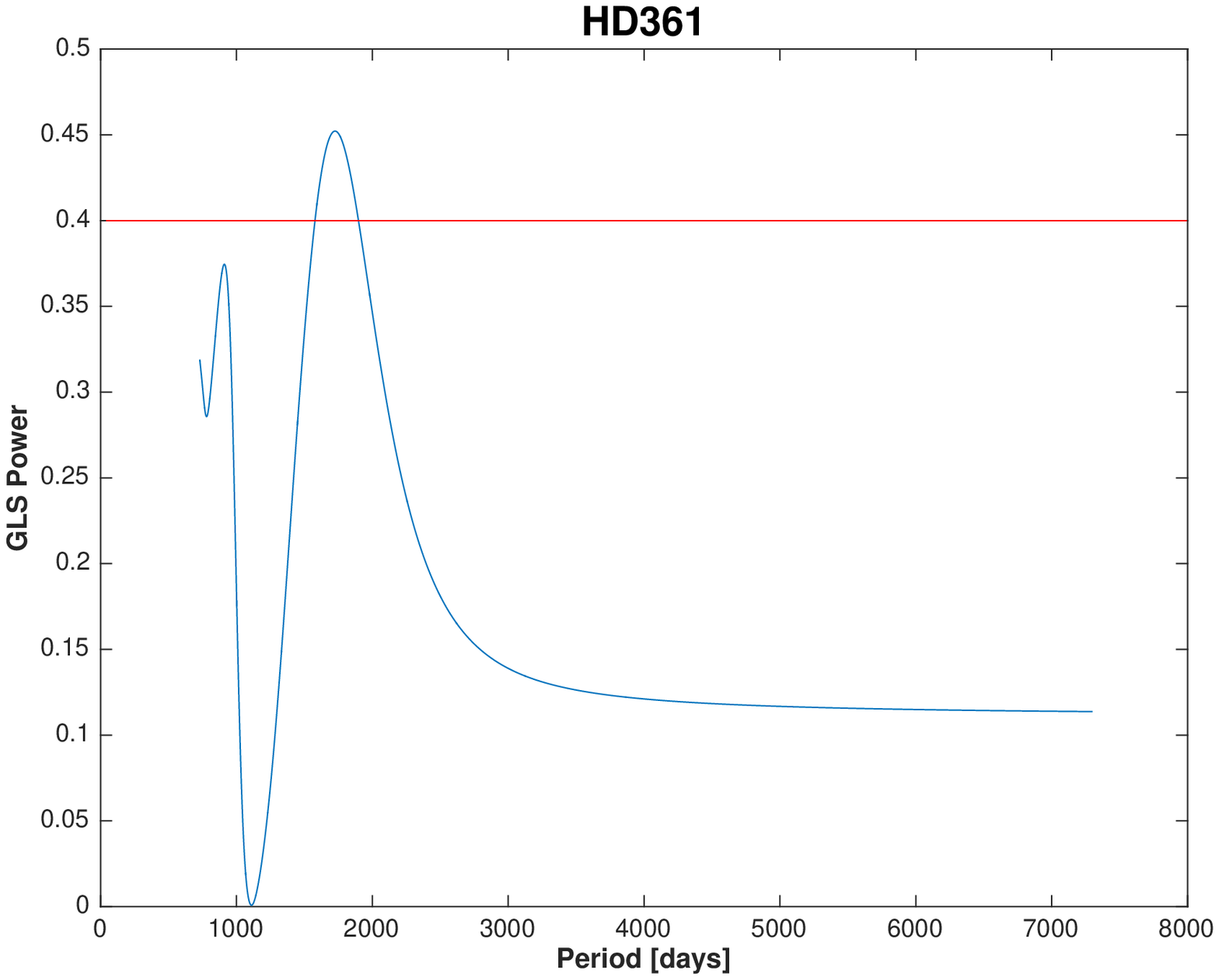}
        \end{subfigure}%
        \begin{subfigure}[b]{0.25\textwidth}
                \includegraphics[width=\linewidth]{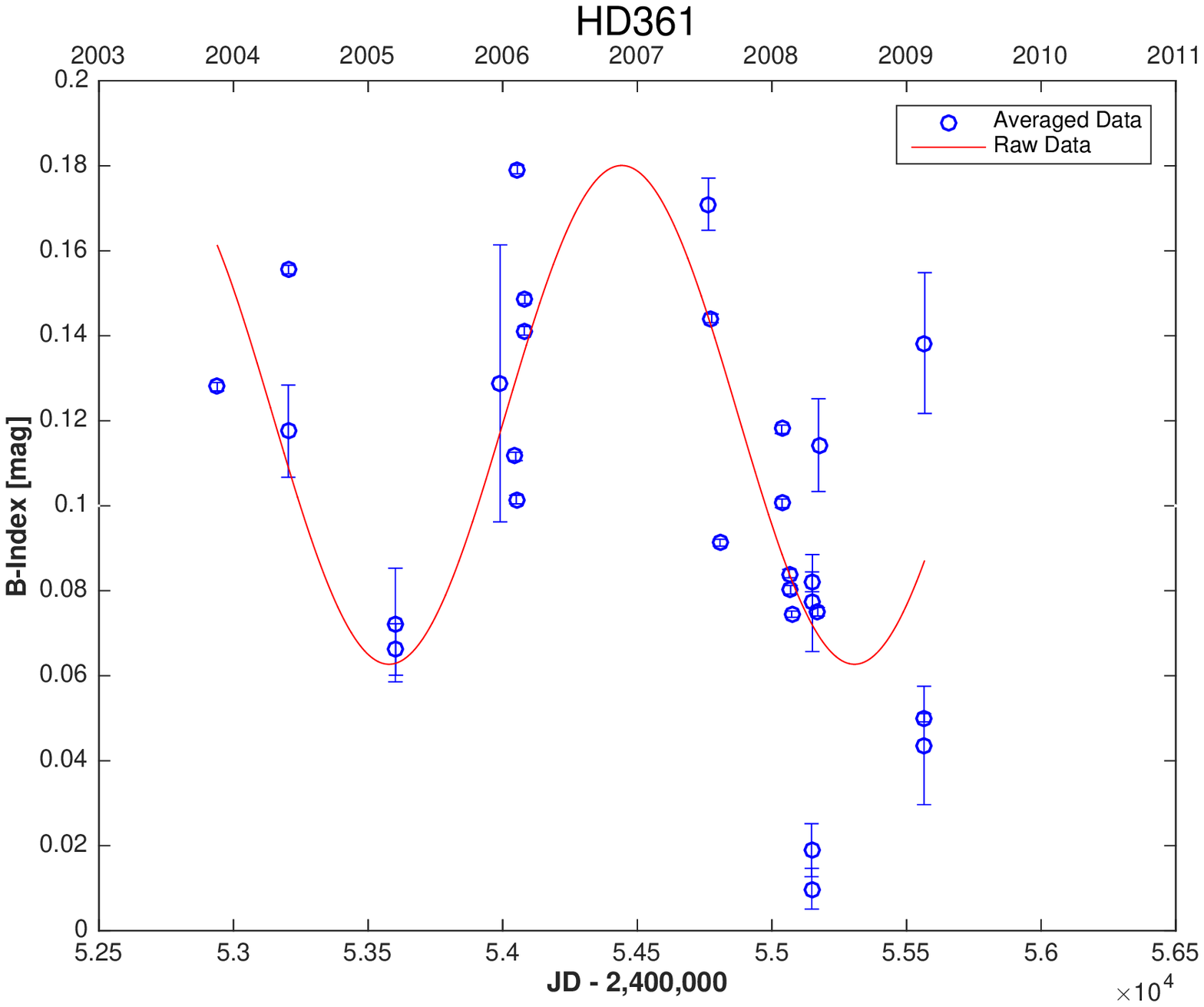}
        \end{subfigure} 
 \begin{subfigure}[b]{0.25\textwidth}
                \includegraphics[width=\linewidth]{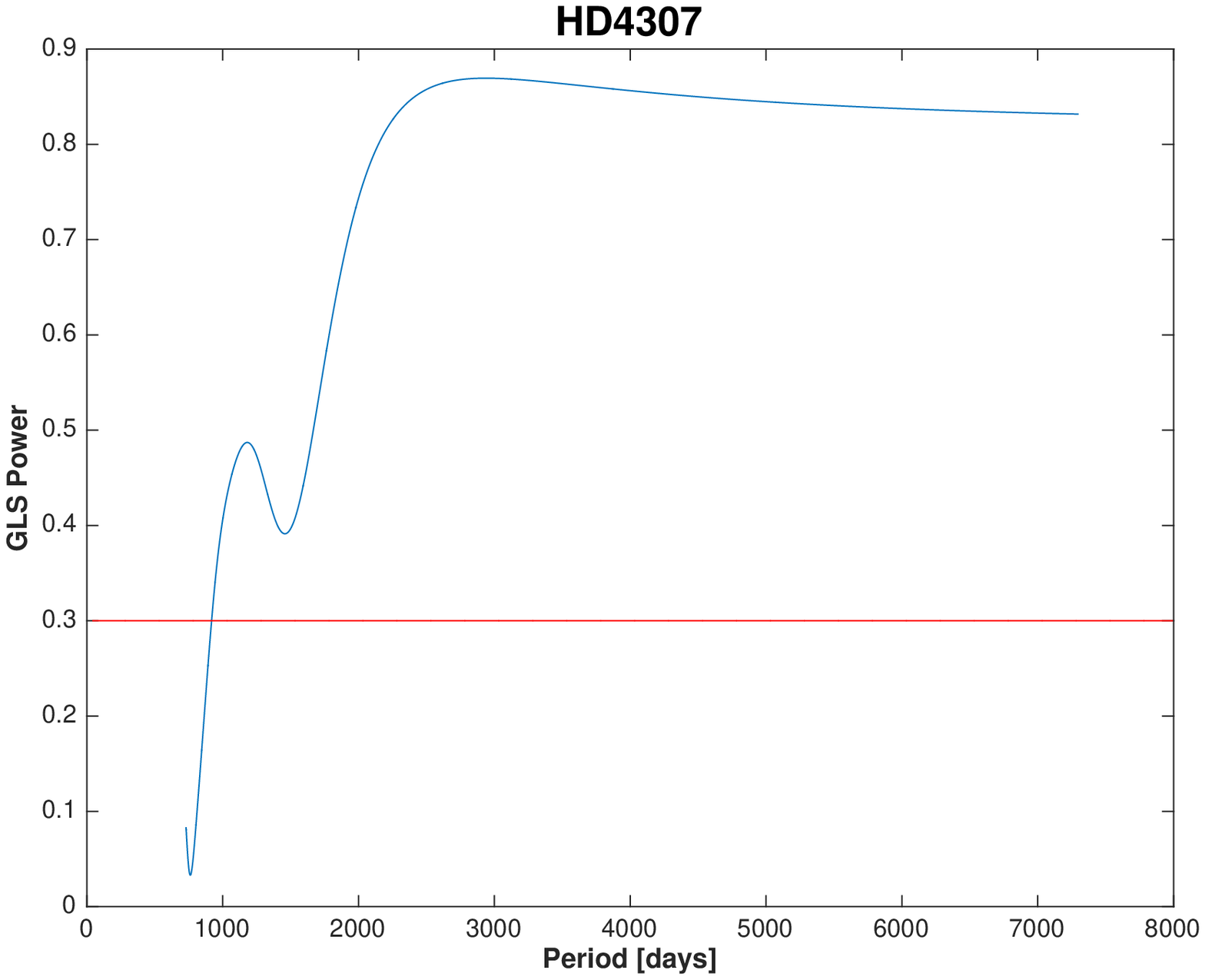}
        \end{subfigure}%
        \begin{subfigure}[b]{0.25\textwidth}
                \includegraphics[width=\linewidth]{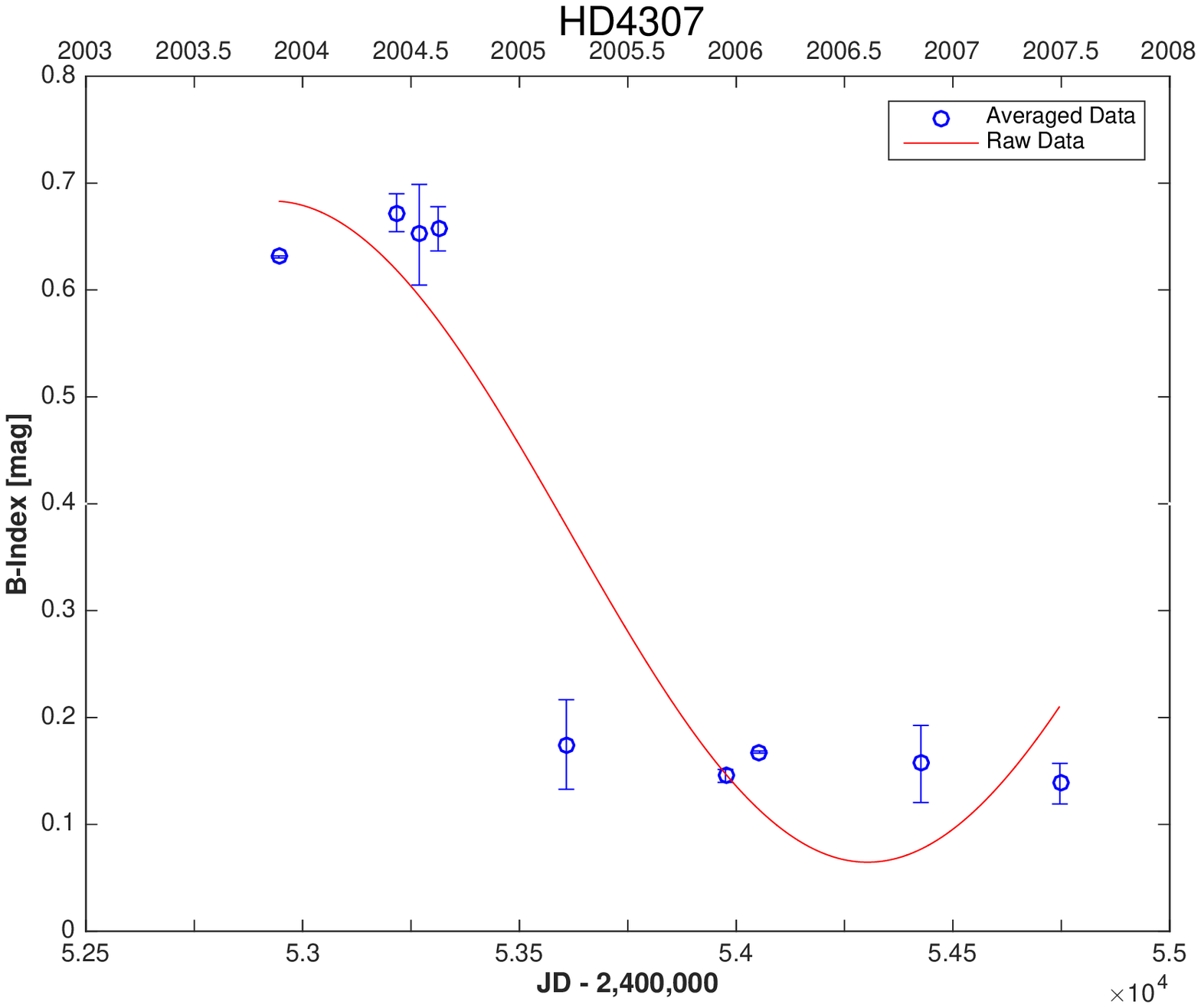}
        \end{subfigure} 
\begin{subfigure}[b]{0.25\textwidth}
                \includegraphics[width=\linewidth]{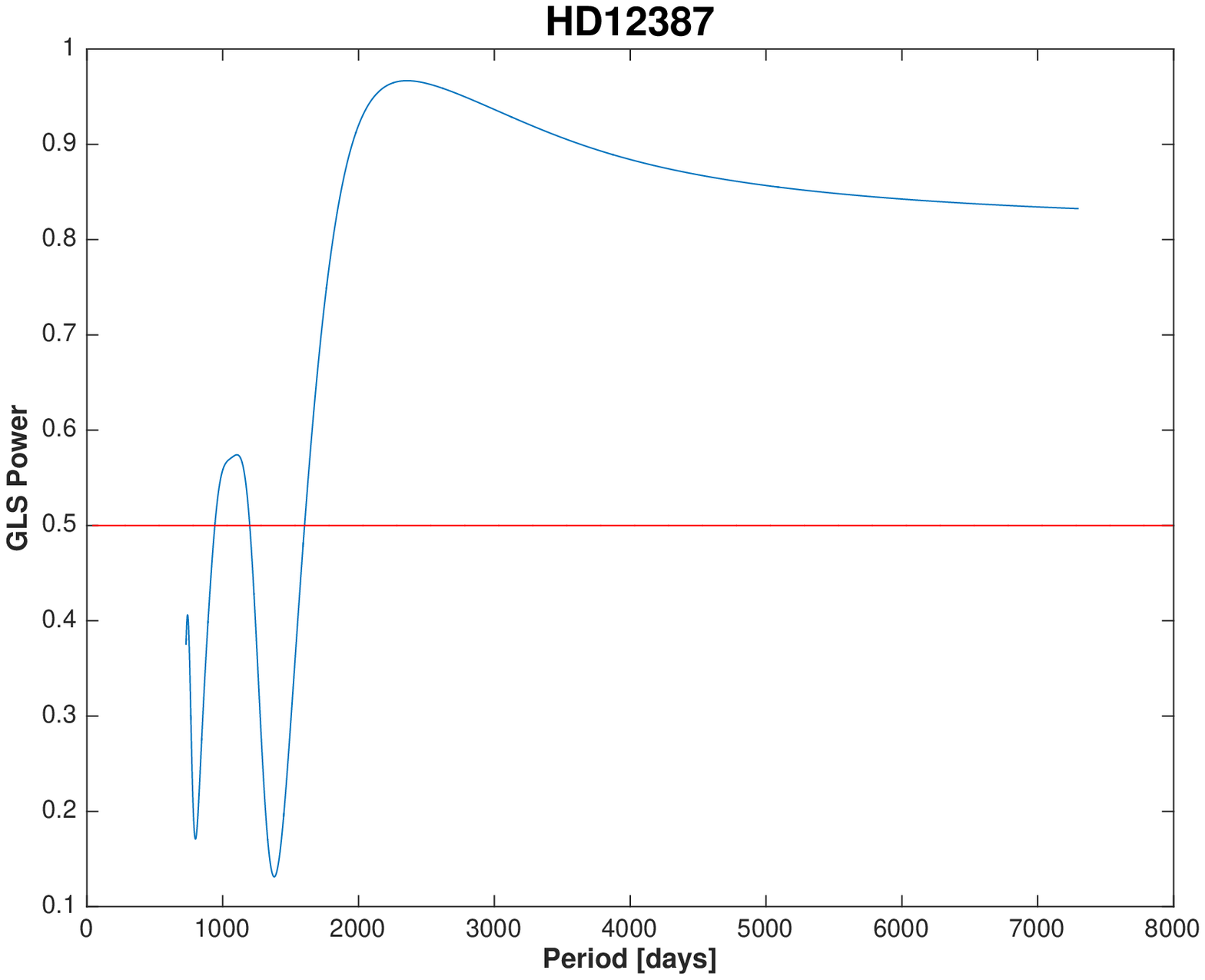}
        \end{subfigure}%
        \begin{subfigure}[b]{0.25\textwidth}
                \includegraphics[width=\linewidth]{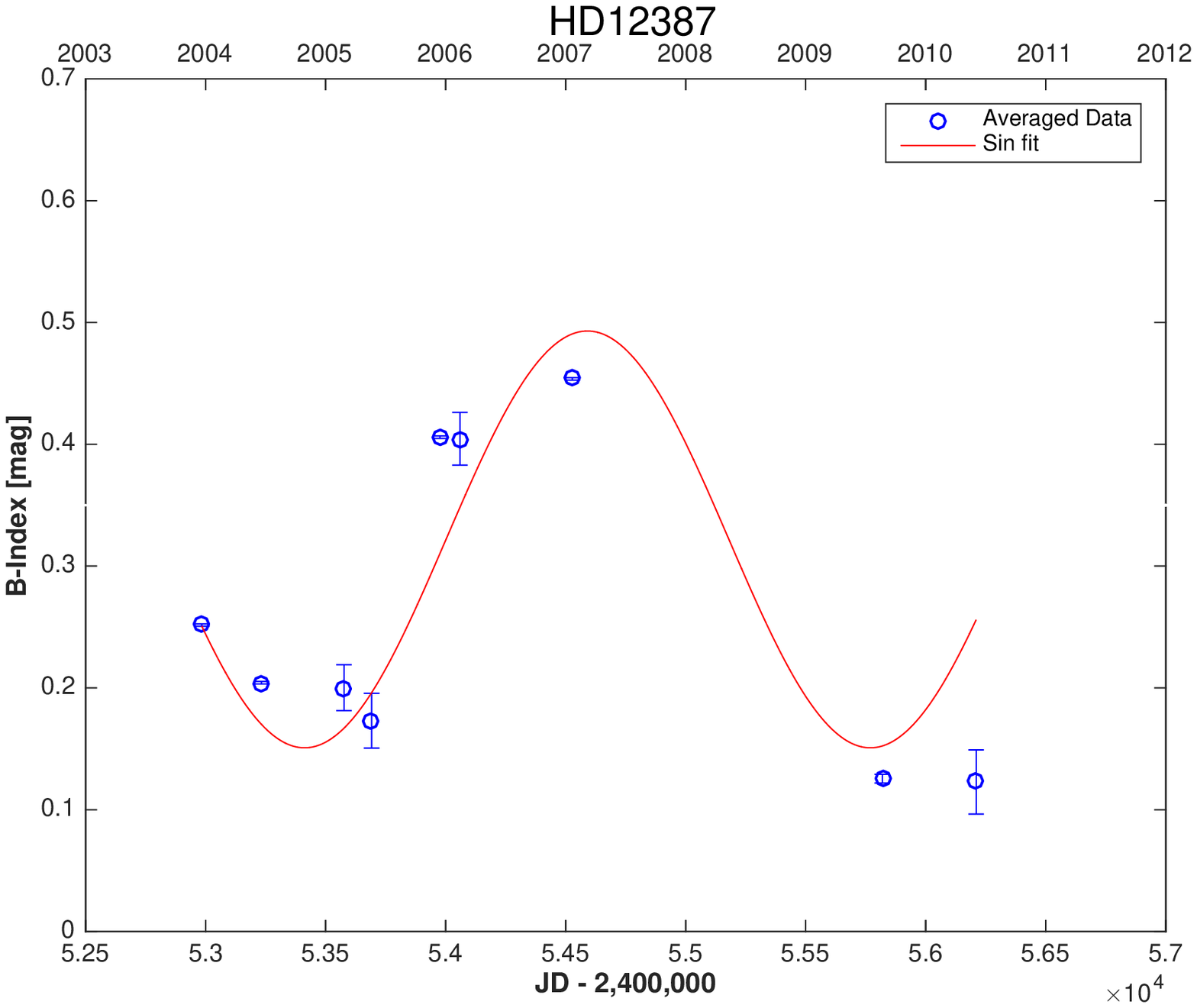}
        \end{subfigure} 
 \begin{subfigure}[b]{0.25\textwidth}
                \includegraphics[width=\linewidth]{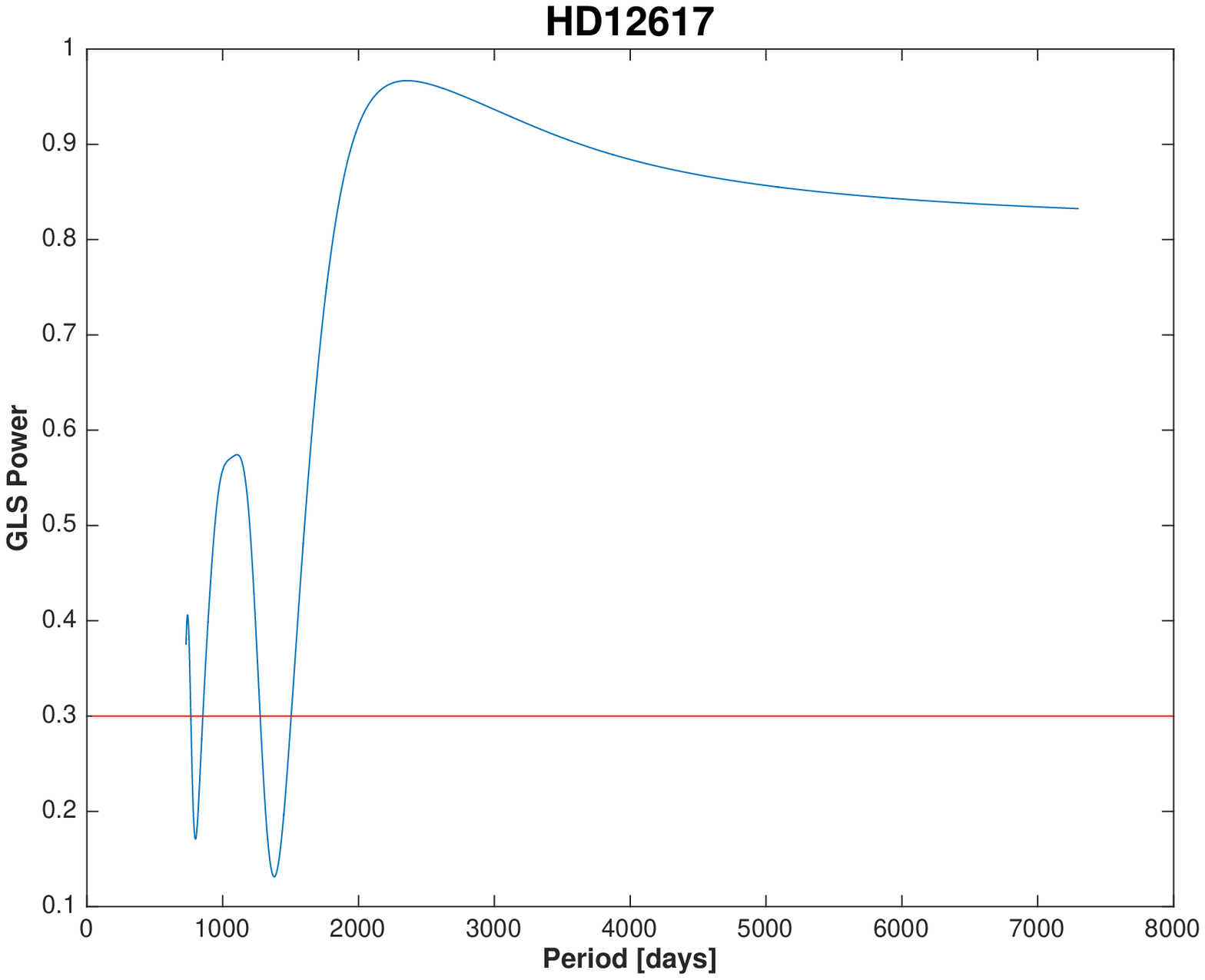}
        \end{subfigure}%
        \begin{subfigure}[b]{0.25\textwidth}
                \includegraphics[width=\linewidth]{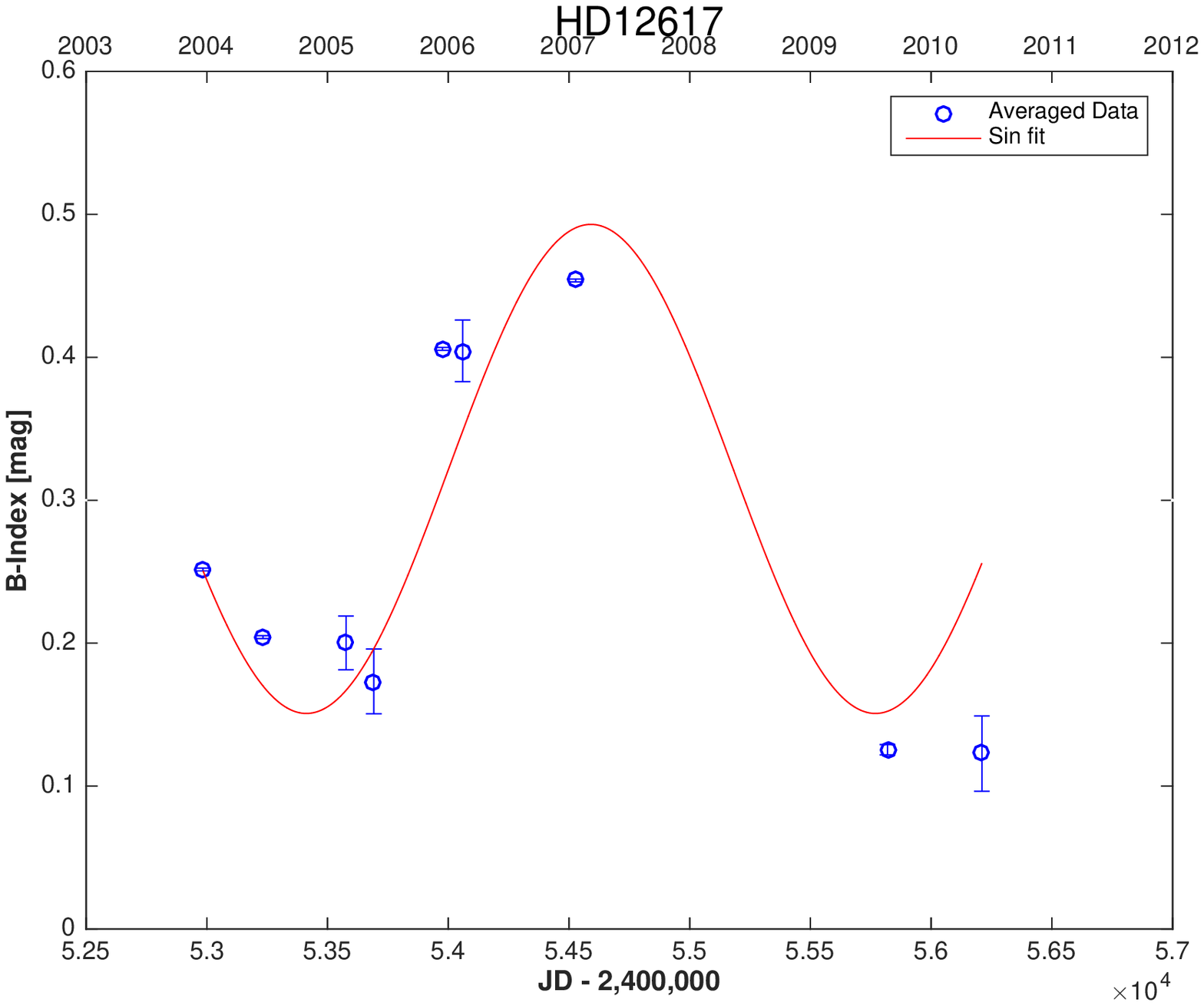}
        \end{subfigure} 
        \begin{subfigure}[b]{0.25\textwidth}
                \includegraphics[width=\linewidth]{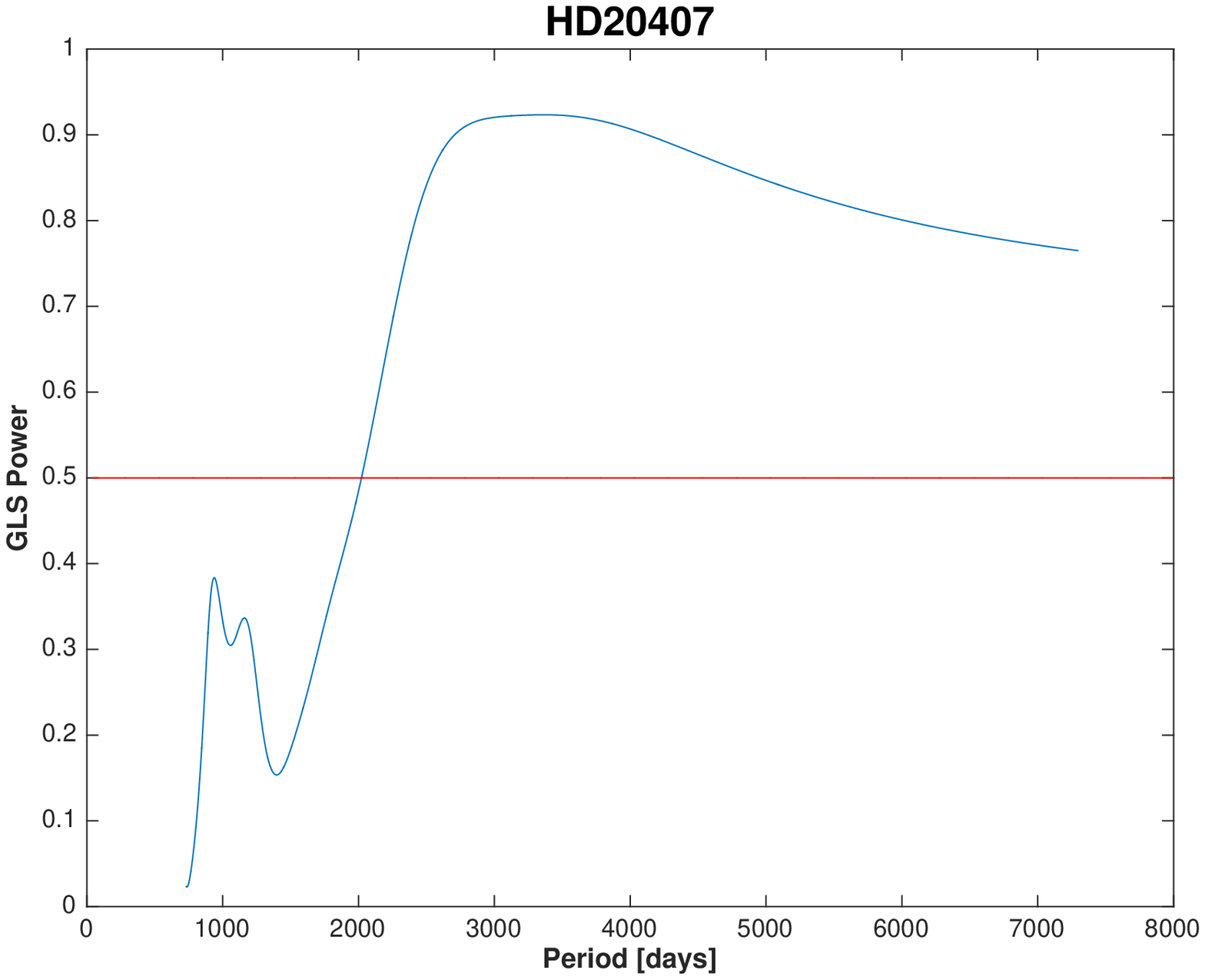}
        \end{subfigure}%
        \begin{subfigure}[b]{0.25\textwidth}
                \includegraphics[width=\linewidth]{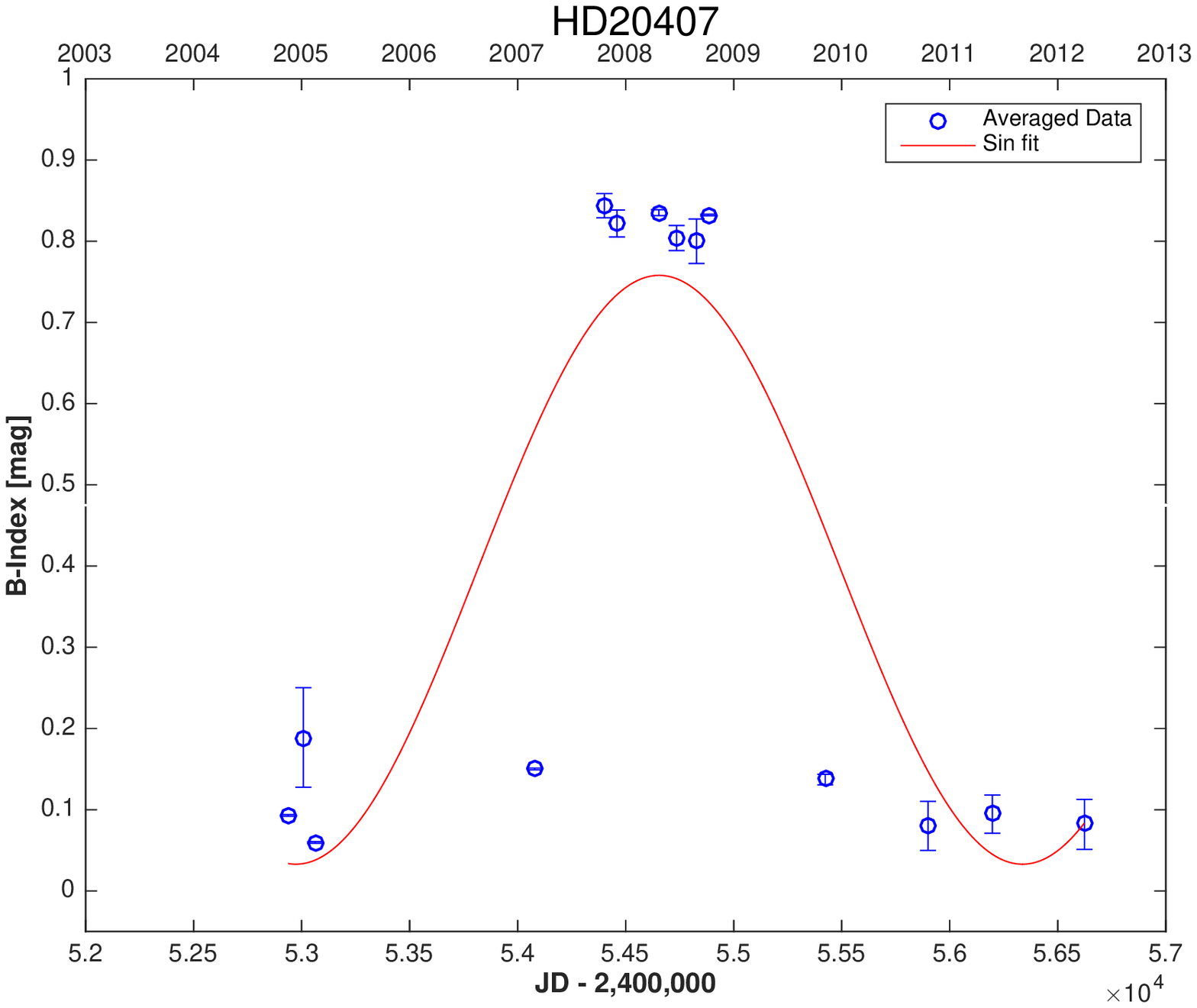}
        \end{subfigure} 
 \begin{subfigure}[b]{0.25\textwidth}
                \includegraphics[width=\linewidth]{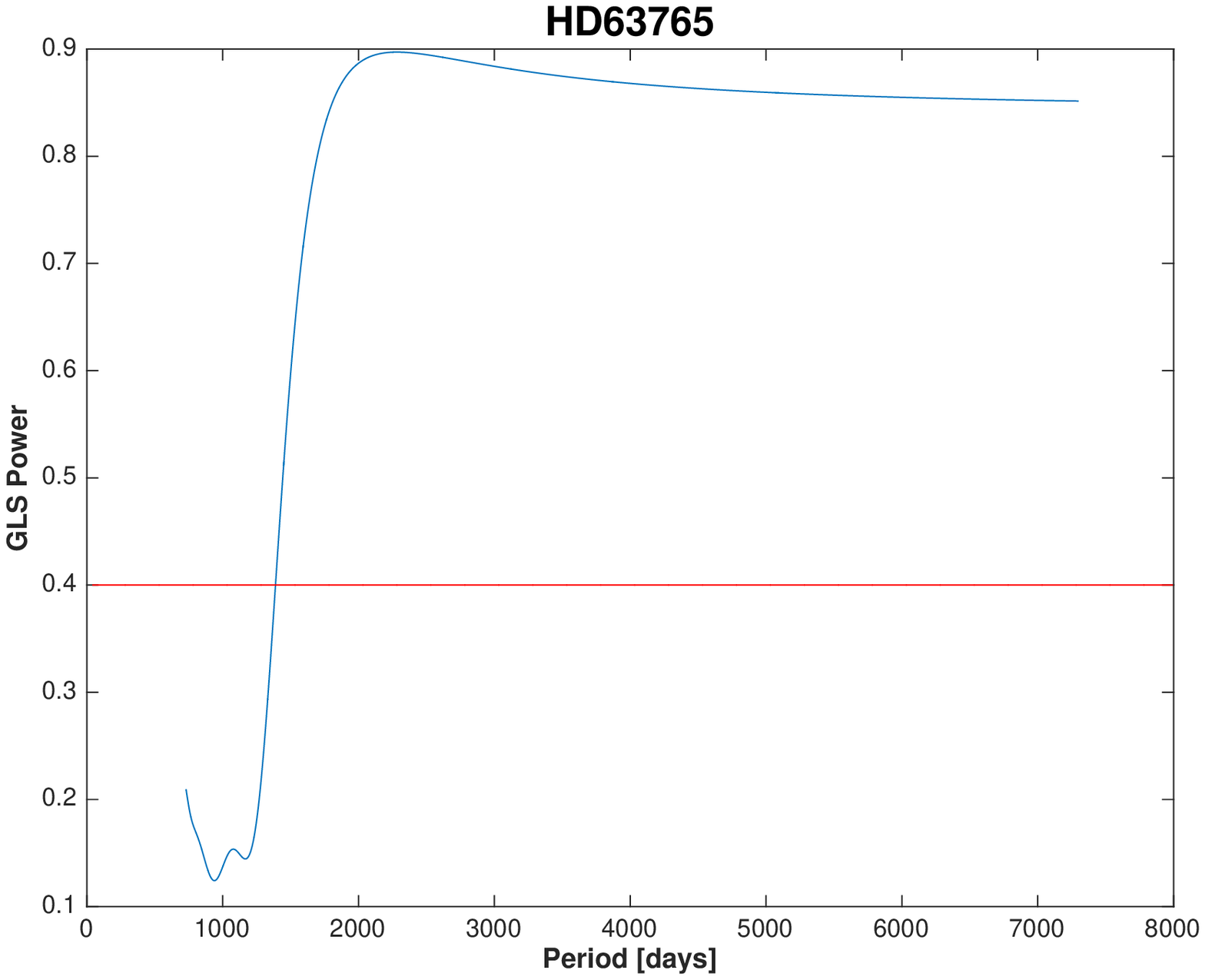}
        \end{subfigure}%
        \begin{subfigure}[b]{0.25\textwidth}
                \includegraphics[width=\linewidth]{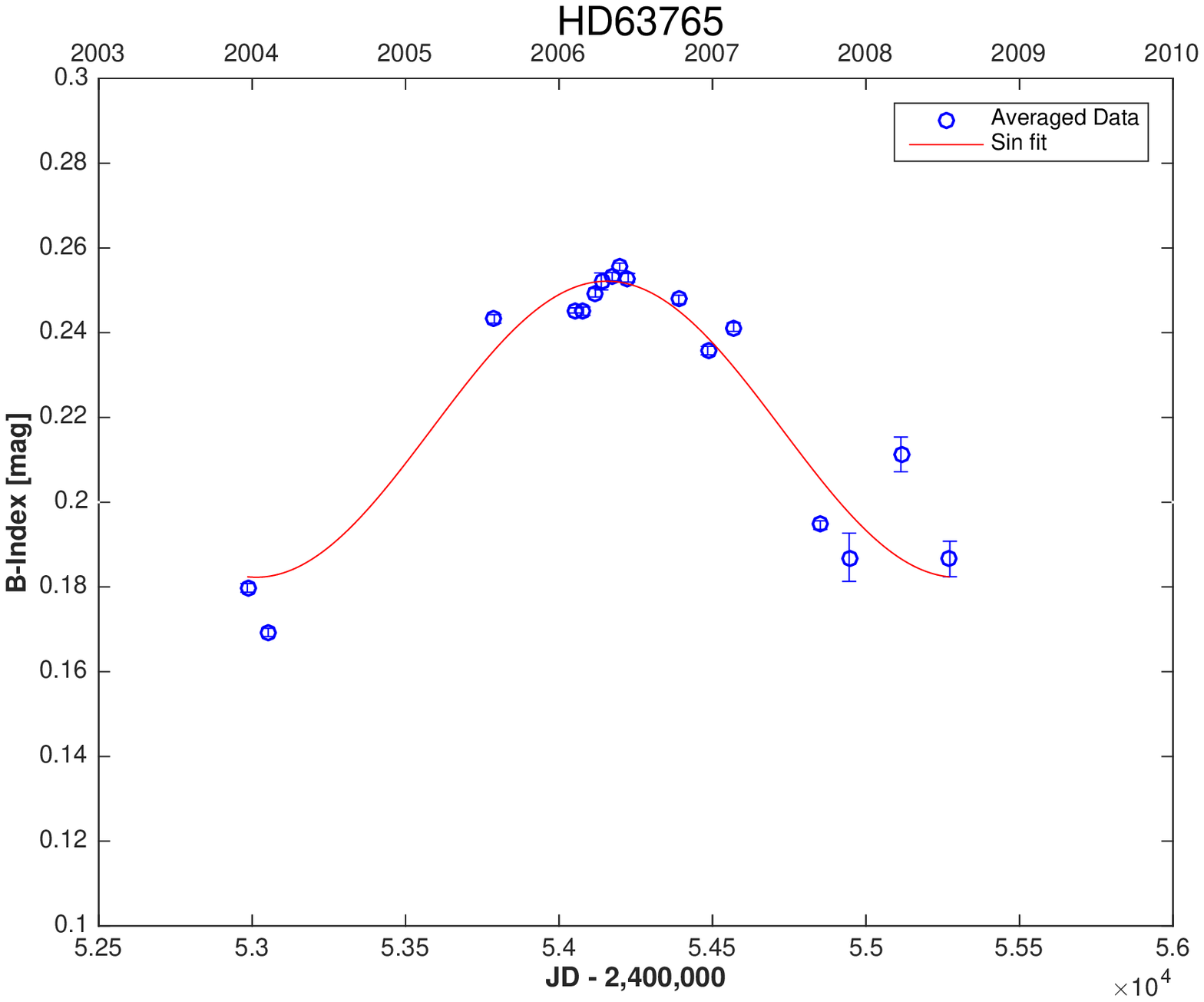}
        \end{subfigure} 
 \begin{subfigure}[b]{0.25\textwidth}
                \includegraphics[width=\linewidth]{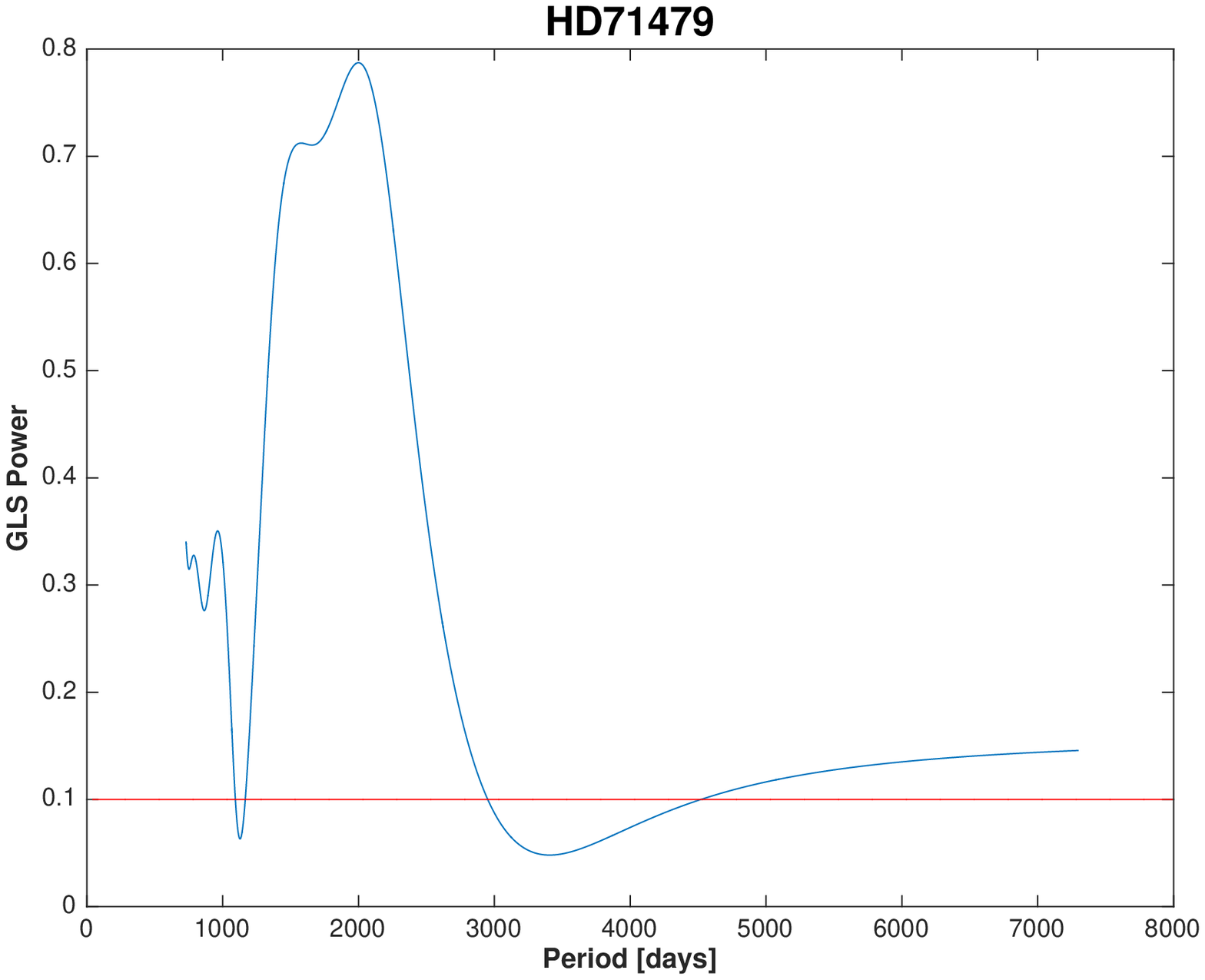}
        \end{subfigure}%
        \begin{subfigure}[b]{0.25\textwidth}
                \includegraphics[width=\linewidth]{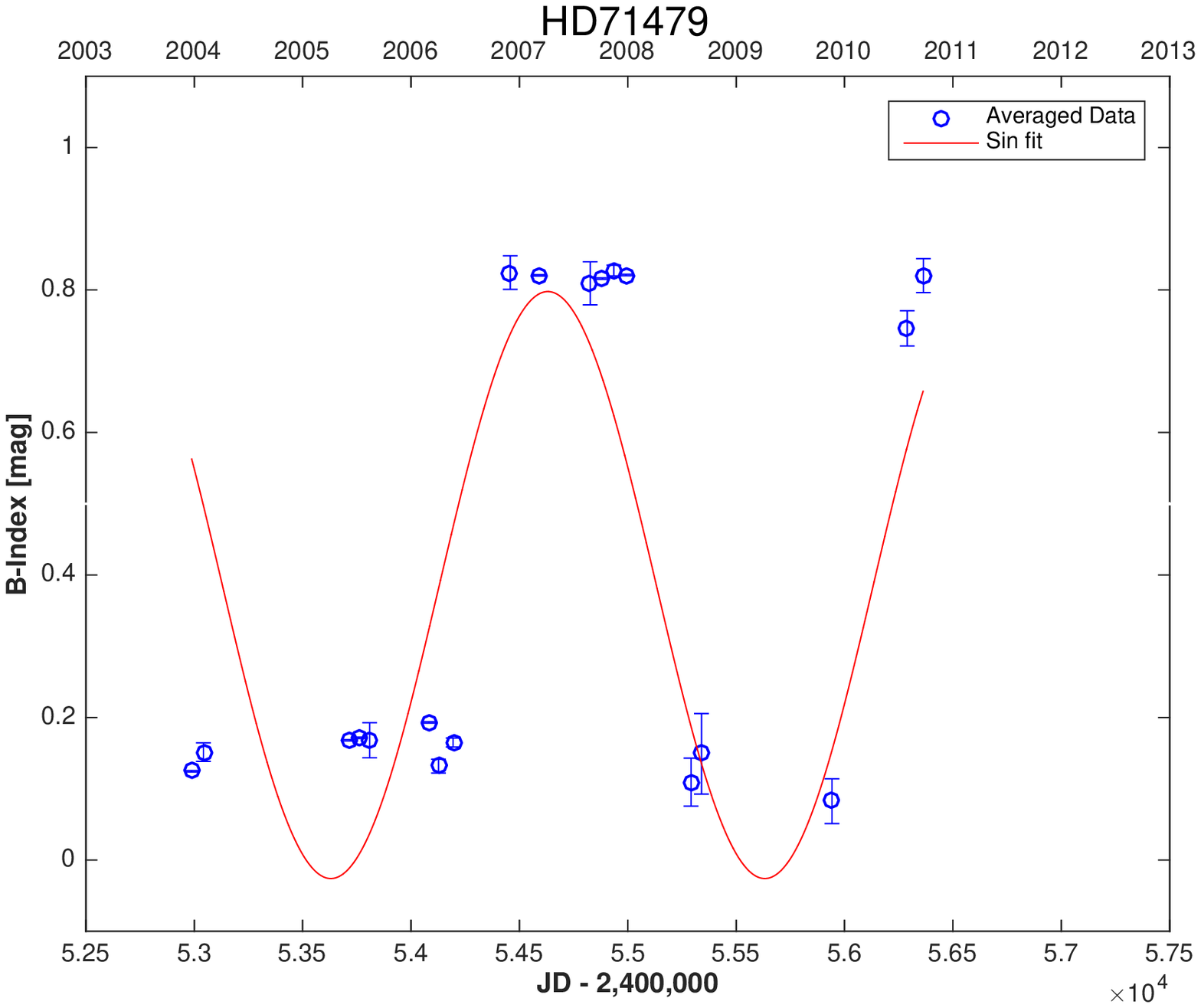}
        \end{subfigure} 
 \begin{subfigure}[b]{0.25\textwidth}
                \includegraphics[width=\linewidth]{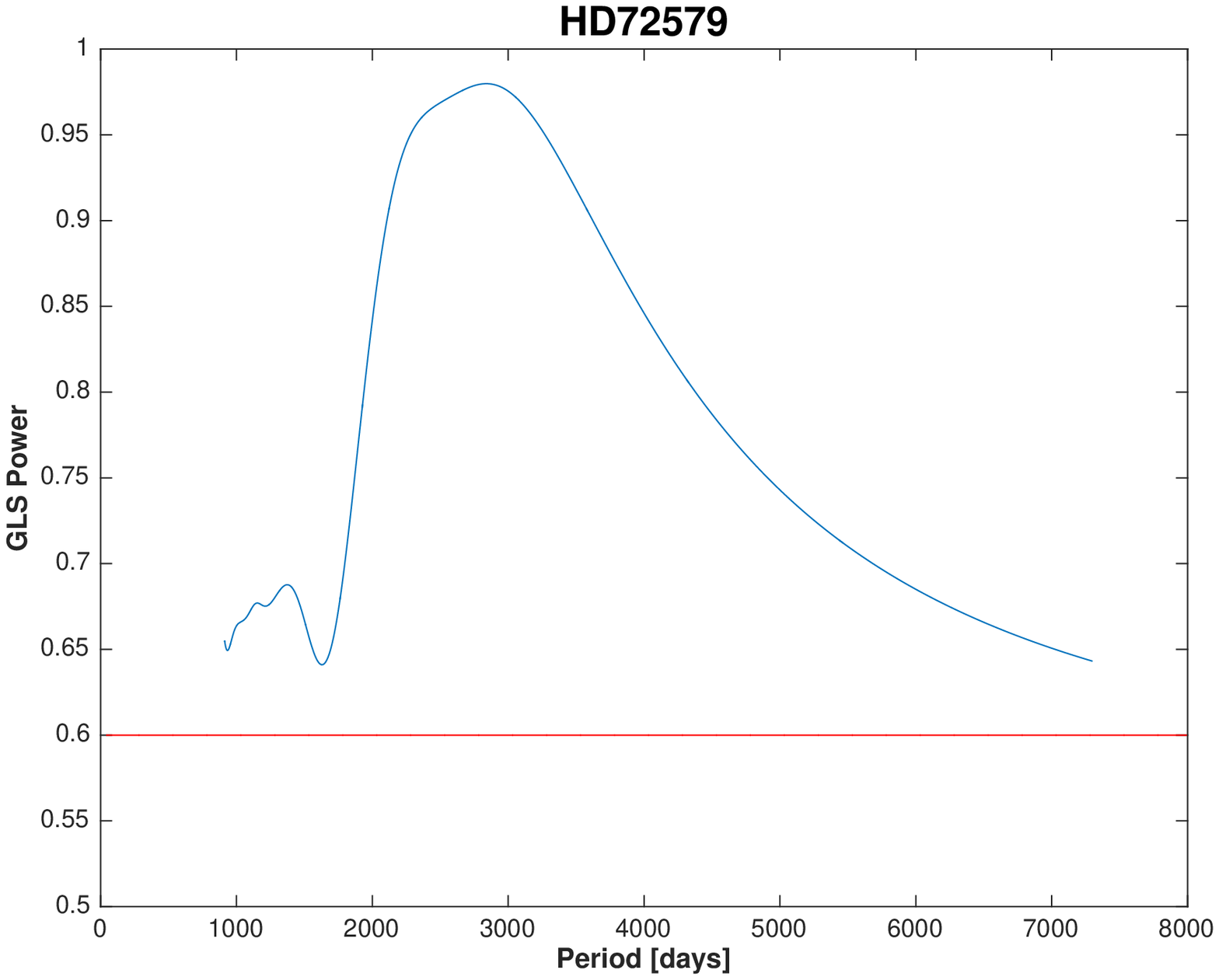}
        \end{subfigure}%
        \begin{subfigure}[b]{0.25\textwidth}
                \includegraphics[width=\linewidth]{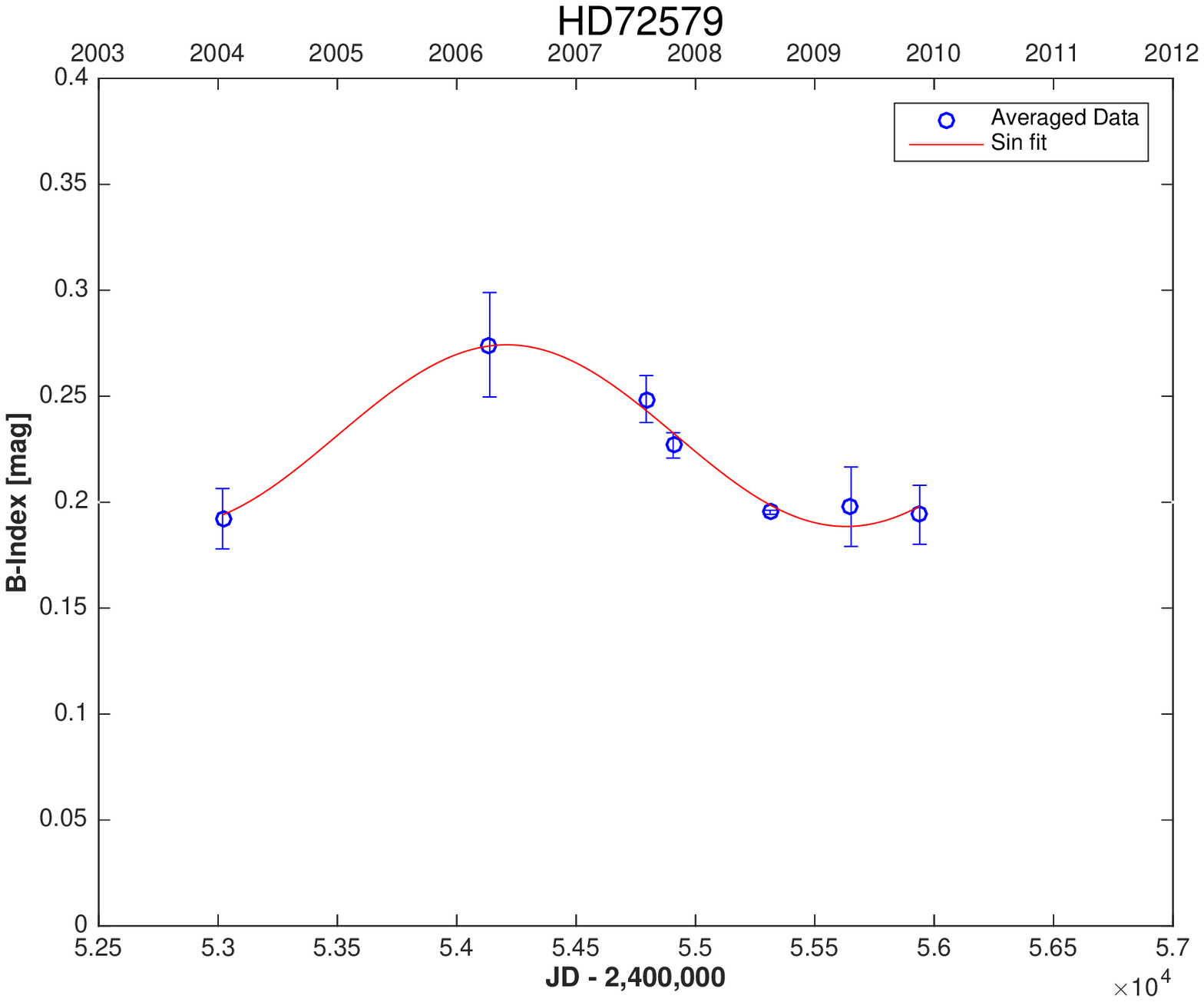}
        \end{subfigure} 
 \begin{subfigure}[b]{0.25\textwidth}
                \includegraphics[width=\linewidth]{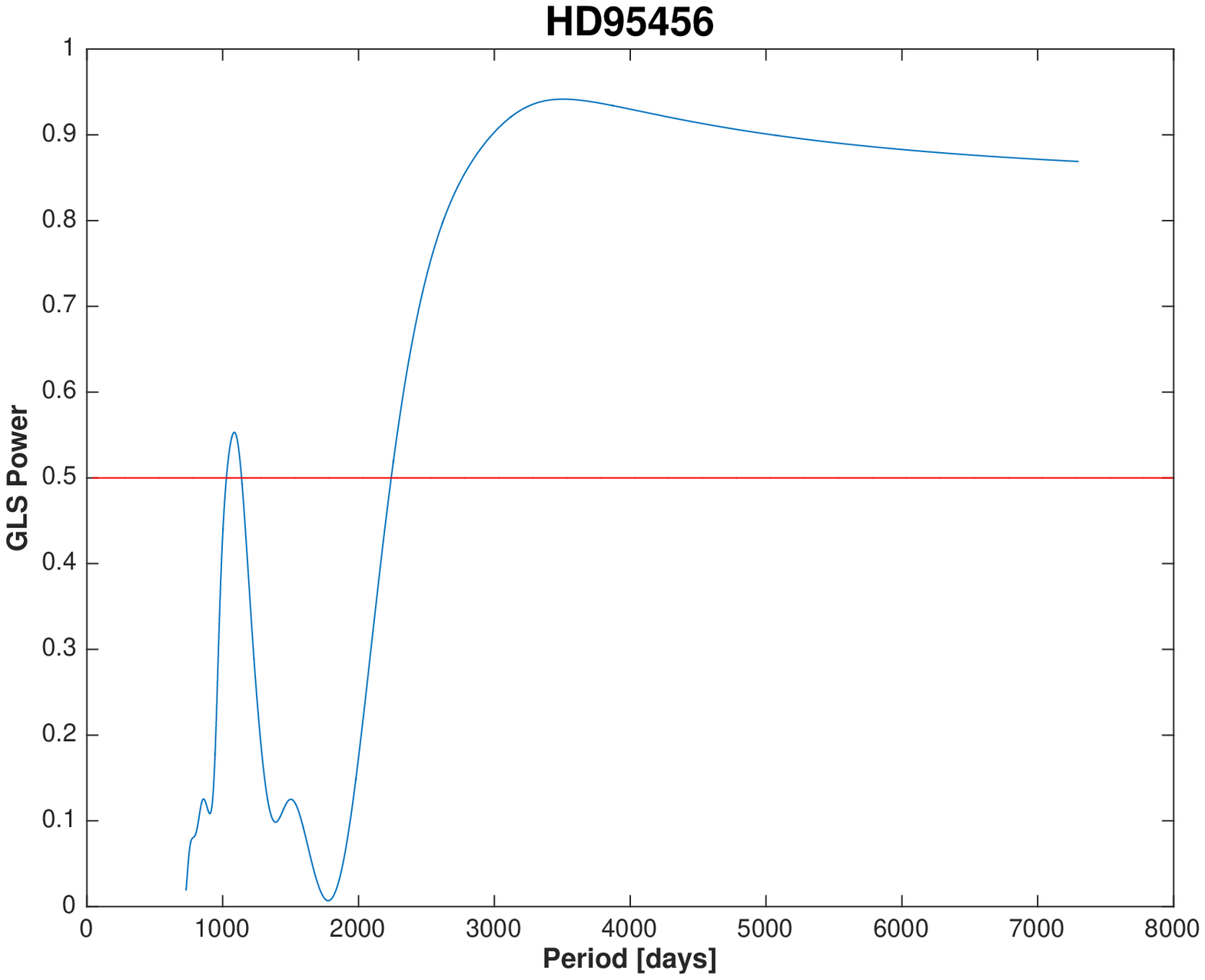}
        \end{subfigure}%
        \begin{subfigure}[b]{0.25\textwidth}
                \includegraphics[width=\linewidth]{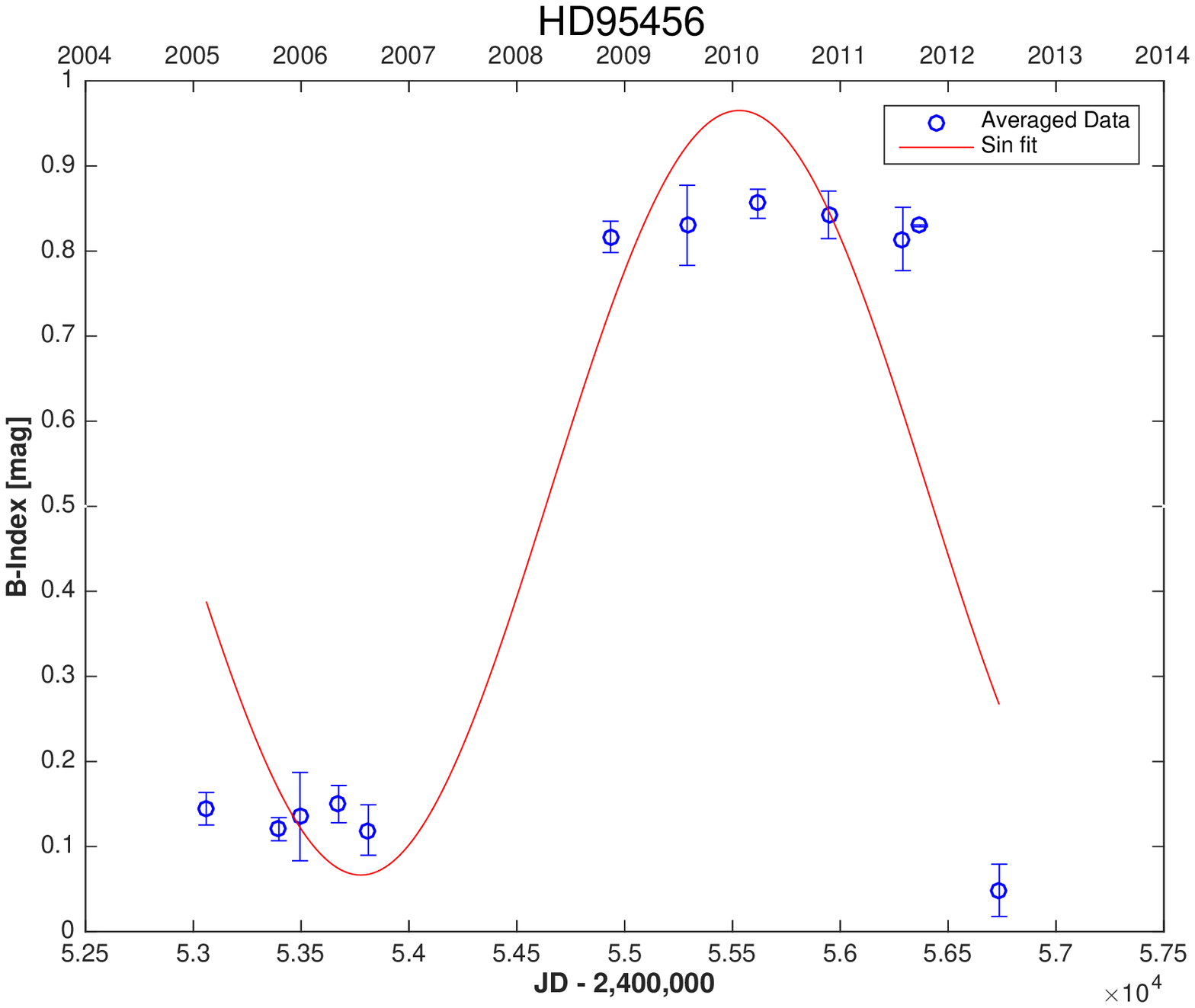}
        \end{subfigure} 
 \begin{subfigure}[b]{0.25\textwidth}
                \includegraphics[width=\linewidth]{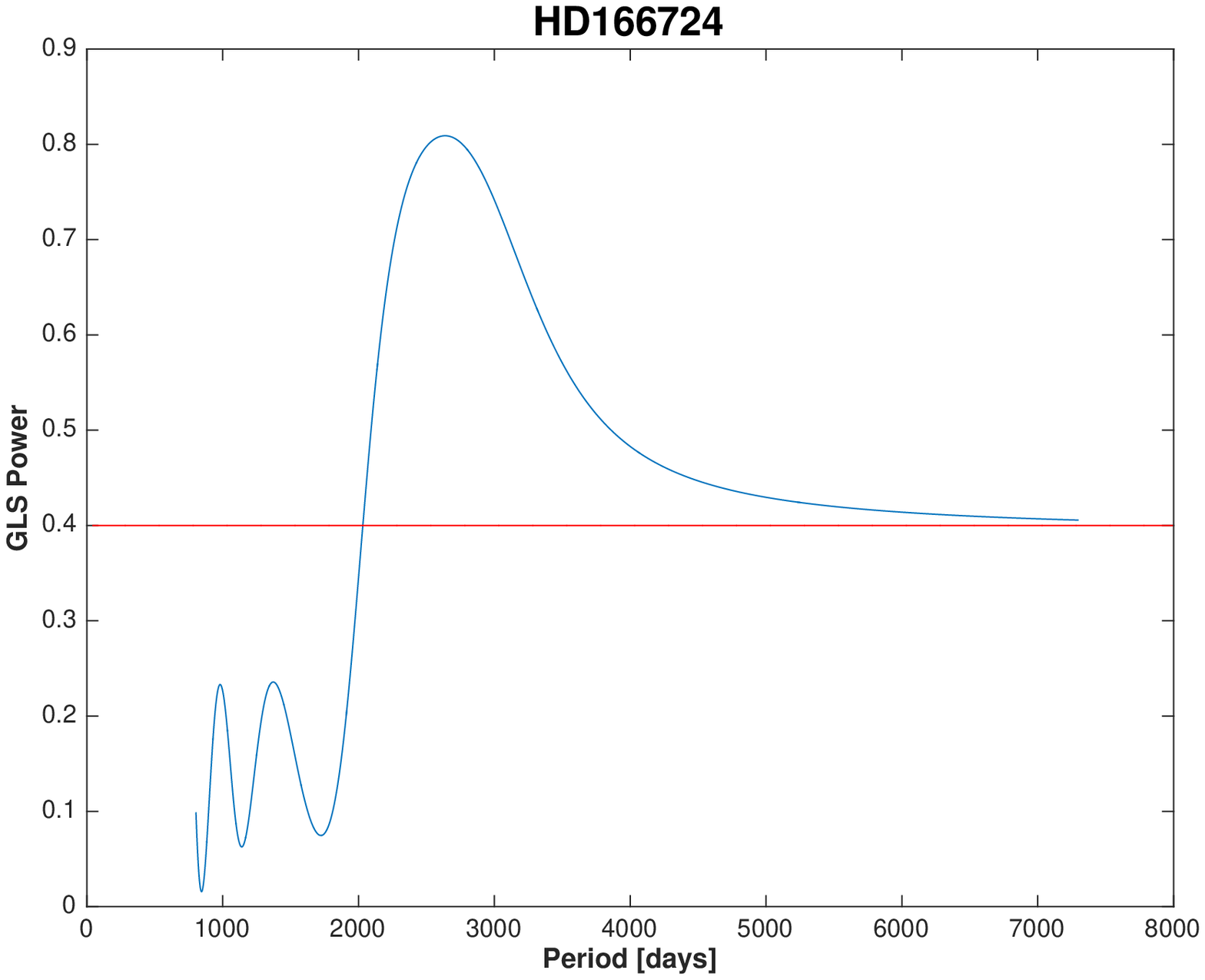}
        \end{subfigure}%
        \begin{subfigure}[b]{0.25\textwidth}
                \includegraphics[width=\linewidth]{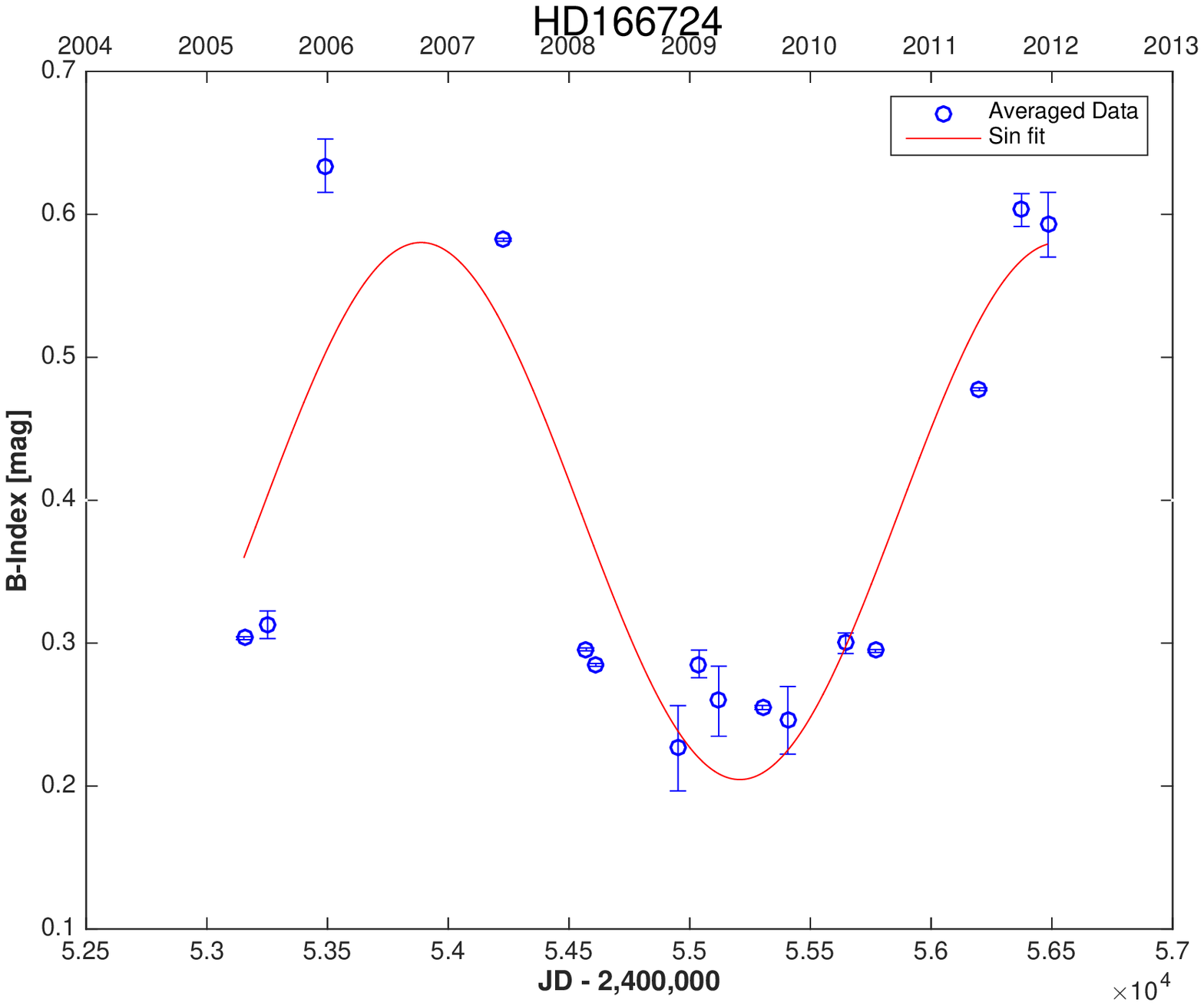}
        \end{subfigure} 
        
        \end{figure}
        
        \begin{figure}
                  \begin{subfigure}[b]{0.25\textwidth}
                \includegraphics[width=\linewidth]{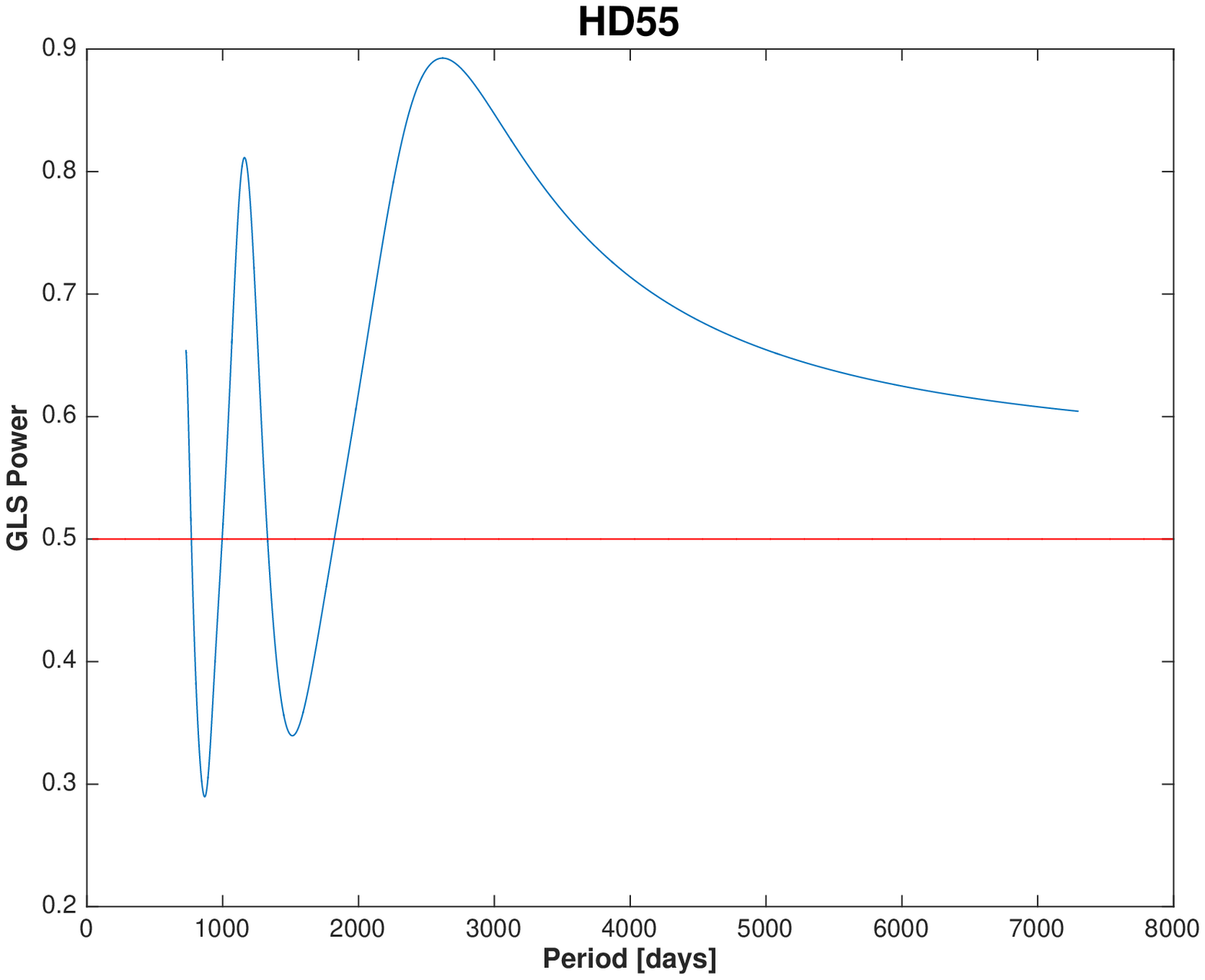}
        \end{subfigure}%
        \begin{subfigure}[b]{0.25\textwidth}
                \includegraphics[width=\linewidth]{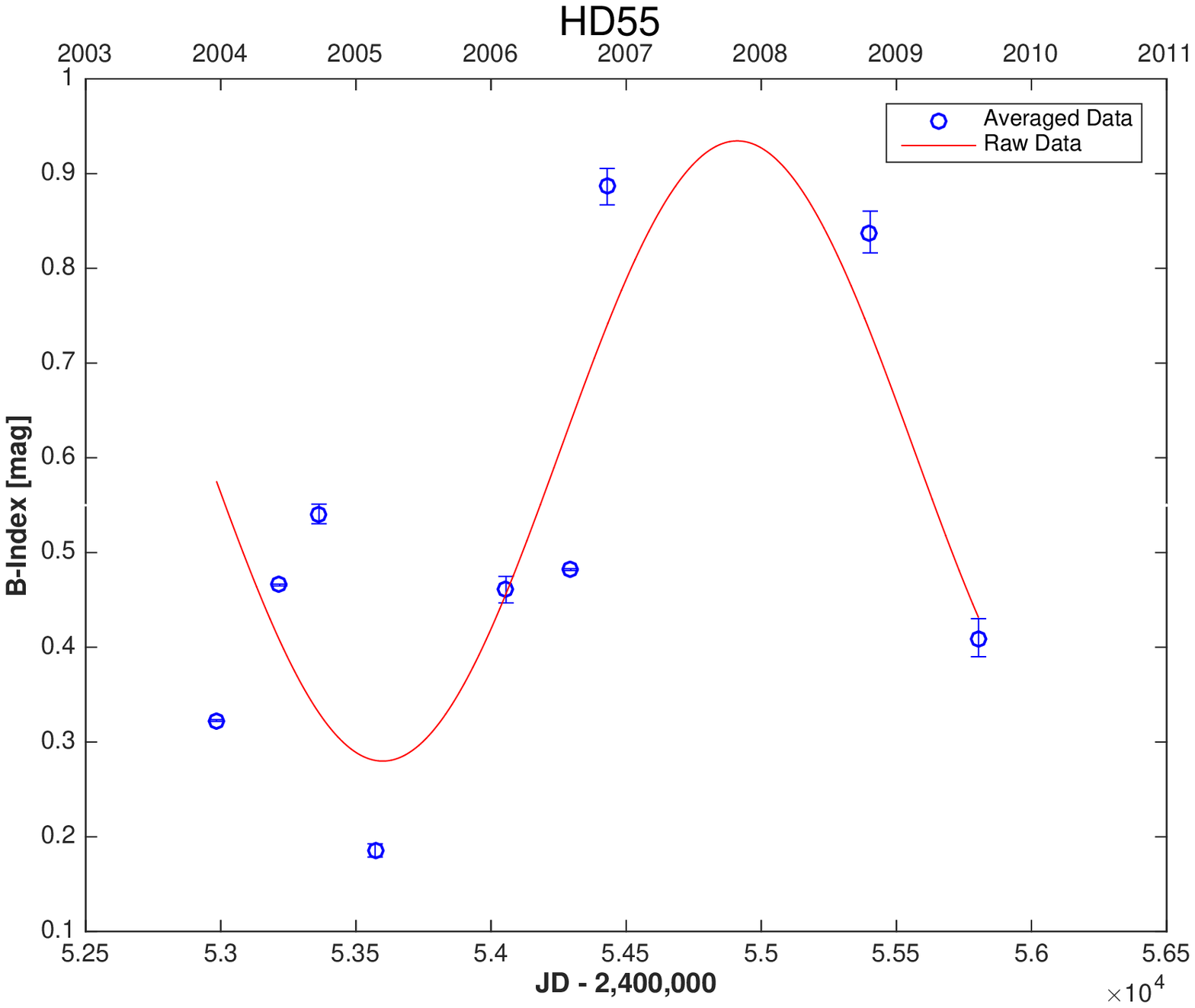}
        \end{subfigure} 
              
\end{figure}

\label{lastpage}

\end{document}